\nonstopmode
\documentclass[preliminary,copyright,creativecommons]{eptcs}

\newcommand{\AGDALATEX}[1]{#1}
\newcommand{\PLAINLATEX}[1]{}

\usepackage{amsfonts,amssymb,textgreek,stmaryrd}
\usepackage[postscript]{ucs}
\usepackage{pifont}
\usepackage[utf8x]{inputenc}

\AGDALATEX{%
\usepackage{latex/agda}
}
\PLAINLATEX{%
\usepackage{verbatim}
\newenvironment{code}{\verbatim}{\endverbatim}
\long\def\AgdaHide#1{} 
}
\DeclareUnicodeCharacter{8759}{::}
\DeclareUnicodeCharacter{9733}{$\star$}
\DeclareUnicodeCharacter{8718}{$\blacksquare$}
\DeclareUnicodeCharacter{10214}{$\llbracket$}
\DeclareUnicodeCharacter{10215}{$\rrbracket$}
\DeclareUnicodeCharacter{10218}{$\langle\!\langle$} 
\DeclareUnicodeCharacter{10219}{$\rangle\!\rangle$}

\newcommand{\DelayI}{\AgdaDatatype{Delay}~\AgdaBound{i}}
\newcommand{\DelayA}[1]{\AgdaDatatype{Delay}~\AgdaBound{#1}~\AgdaBound{A}}
\newcommand{\DelayIA}{\DelayA{i}}
\newcommand{\IDelayA}[1]{\AgdaDatatype{∞Delay}~\AgdaBound{#1}~\AgdaBound{A}}
\newcommand{\IDelayIA}{\IDelayA{i}}
\newcommand{\force}{\AgdaField{force}}
\newcommand{\ainf}{\AgdaBound{a∞}}
\newcommand{\aq}{\AgdaBound{a?}}
\newcommand{\aqp}{\AgdaBound{a?′}}
\newcommand{\jlti}{\ensuremath{\AgdaBound{j} < \AgdaBound{i}}}
\newcommand{\RawMonad}{\AgdaFunction{RawMonad}}
\newcommand{\RawFunctor}{\AgdaFunction{RawFunctor}}
\newcommand{\fmap}{\ensuremath{\AgdaFunction{\_\ensuremath{{<}\${>}}\_}}}
\newcommand{\return}{\AgdaFunction{return}}
\newcommand{\bind}{\ensuremath{\AgdaFunction{\ensuremath{\mathbin{{>}\mkern-8.5mu{>}\mkern-2mu{=}}}}}}
\newcommand{\prebind}{\ensuremath{\AgdaFunction{\_\ensuremath{\mathbin{{>}\mkern-8.5mu{>}\mkern-2mu{=}}}\_}}}
\newcommand{\preinfbind}{\ensuremath{\AgdaFunction{\_\ensuremath{\mathbin{\infty\mkern-3mu{>}\mkern-8.5mu{>}\mkern-2mu{=}}}\_}}}
\newcommand{\prebisim}{\ensuremath{\AgdaDatatype{\_∼\_}}}
\newcommand{\presizedbisim}[1]{\ensuremath{\AgdaDatatype{\_∼⟨~\AgdaBound{#1}~⟩∼\_}}}
\newcommand{\presizedinfbisim}[1]{\ensuremath{\AgdaDatatype{\_∞∼⟨~\AgdaBound{#1}~⟩∼\_}}}
\newcommand{\conv}{\ensuremath{\AgdaDatatype{⇓}}}
\newcommand{\ie}{i.e.}

\providecommand{\event}{MSFP 2014}
\title{Normalization by Evaluation in the Delay Monad \\
  \Large{\emph{A Case Study for Coinduction via Copatterns and Sized Types}}}
\author{Andreas Abel
\institute{Department of Computer Science and Engineering\\
Chalmers and Gothenburg University\\
Sweden}
\email{andreas.abel@gu.se}
\and
James Chapman
\institute{Institute of Cybernetics\\
Tallinn University of Technology\\
Estonia}
\email{james@cs.ioc.ee}
}
\def\titlerunning{Normalization by Evaluation in the Delay Monad}
\def\authorrunning{A. Abel \& J. Chapman}

\begin{document}

\maketitle
\begin{abstract}

In this paper, we present an Agda formalization of a normalizer for
simply-typed lambda terms.
The normalizer consists of two
coinductively defined functions in the delay monad:
One is a standard evaluator of lambda terms to closures, the other a
type-directed reifier from values to $\eta$-long $\beta$-normal forms.
Their composition,
normalization-by-evaluation, is shown to be a total function \emph{a
posteriori}, using a
standard logical-relations argument.

The successful formalization serves as a proof-of-concept for
coinductive programming and reasoning using sized types and copatterns,
a new and presently experimental feature of Agda.

\end{abstract}

\section{Introduction and Related Work}

It would be a great shame if dependently-typed programming (DTP)
restricted us to only writing very clever programs that were a priori
structurally recursive and hence obviously terminating. Put
another way, it is a lot to ask of the programmer to provide the
program and its termination proof in one go, programmers should also
be supported in working step-by-step. This paper champions a technique
that lowers the barrier of entry, from showing termination to only
showing productivity up front, and then later providing the
opportunity to show termination (convergence). In this paper, we write
a simple recursive normalizer for simply-typed lambda calculus which
as an intermediate step constructs closures
and finally constructs full $\eta$-long $\beta$-normal forms. The normalizer
is not structurally recursive and we represent it in Agda as a
potentially non-terminating but nonetheless productive corecursive
function targeting the coinductive delay monad. Later we show that the function
is indeed terminating as all such delayed computations converge (are
only finitely delayed) by a quite traditional strong computability
argument. The coinductive normalizer, when combined with the
termination proof, yields a terminating function returning undelayed
normal forms.

Our normalizer is an instance of \emph{normalization by evaluation} as
conceived by Danvy~\cite{danvy:tdpe}
and Abel, Coquand, and
Dybjer~\cite{abelCoquandDybjer:mpc08}:
Terms
are first evaluated into an applicative structure of values; herein, we
realize function values by closures, which can be seen as weak head
normal forms under explicit substitution.  The second phase goes in
the other direction: values are \emph{read back}
(terminology by Gr\'egoire and Leroy~\cite{gregoireLeroy:icfp02})
as terms in normal form.\footnote{%
  In a more strict terminology, normalization by evaluation must
  evaluate object-level functions as meta-level functions;  such is
  happening in Berger and Schwichtenberg's original work
  \cite{bergerSchwichtenberg:lics91}, but not here.
}
In contrast to the cited works, we employ
intrinsically well-typed representations of terms and values.  In
fact, our approach is closest to Altenkirch and Chapman's
\emph{big-step normalization}~\cite{altenkirchChapman:bigStepNormalisation,chapman:PhD};
this work can be consulted for more detailed descriptions of
well-typed terms and values.  Where Altenkirch and Chapman
represent partial functions via their inductively defined graphs, we
take the more direct route via the coinductive delay monad.  This is
the essential difference and contribution of the present work.

The delay monad has been used to implement evaluators before:
Danielsson's \emph{Operational Semantics Using the Partiality Monad}
~\cite{danielsson:icfp12} for untyped lambda terms is the model for
our evaluator.  However, we use a \emph{sized} delay monad, which
allows us to use the bind operation of the monad directly;
Danielsson, working with the previous version of Agda and its
coinduction, has to use a workaround to please Agda's guardedness
checker.

In spirit, evaluation into the delay monad is closely related to
\emph{continuous normalization} as implemented by Aehlig and
Joachimski~\cite{aehligJoachimski:continuousNormalization}.  Since
they compute possibly infinitely deep normal forms (from untyped
lambda terms), their type of terms is coinductive; further, our
\AgdaInductiveConstructor{later} constructor of the delay monad is one of their
constructors of lambda terms, called \emph{repetition constructor}.
They attribute this idea to Mints~\cite{mints:contNorm}.  In the
type-theoretic community, the delay monad has been popularized by
Capretta~\cite{capretta:generalRecursionViaCoinductiveTypes}, and it
is isomorphic to the trampolin type \cite{ganzFriedmanWand:icfp99}.
Escardo~\cite{escardo:metricPCF} describes a delay monad in the
context of a (ultra)metric model for PCF which allows
\emph{intensional functions} that can measure the termination speed of
their arguments.  Indeed, the coinductive delay monad is intensional
in the same sense as it makes the speed of convergence observable.

Using hereditary substitutions \cite{watkins:concurrentLFTR}, a
normalization function for the simply-typed lambda calculus can be
defined directly, by structural recursion on types.  This normalizer
has been formalized in Agda by Altenkirch and
Keller~\cite{kellerAltenkirch:msfp10}.  The idea of normalization by
induction on types is very old, see, e.g.,
Prawitz~\cite{prawitz:natDed}.  Note however, that normalization via
hereditary substitution implements a fixed strategy, bottom-up
normalization, which cannot be changed without losing the inductive
structure of the algorithm.  Our strategy, normalization via closures,
cannot be implemented directly by induction on types.  Further, the
simple induction on types also breaks down when switching to more
powerful lambda calculi like G\"odel's~T, while our approach scales
without effort.

To save paper and preserve precious forests,
we have only included the essential parts of the
Agda development; the full code is available online
\cite{abelChapman:msfp14lagda}.

\AgdaHide{
\begin{code}%
\>\AgdaSymbol{\{-\#} \AgdaKeyword{OPTIONS} --copatterns --sized-types \AgdaSymbol{\#-\}}\<%
\\
\>\AgdaKeyword{module} \AgdaSymbol{\_} \AgdaKeyword{where}\<%
\\
%
\\
\>\AgdaComment{-- Interface to standard library.}\<%
\\
%
\\
\>\AgdaKeyword{open} \AgdaKeyword{import} \AgdaModule{Level} \AgdaKeyword{public}\<%
\\
\>[0]\AgdaIndent{2}{}\<[2]%
\>[2]\AgdaKeyword{using} \AgdaSymbol{(}Level\AgdaSymbol{)} \AgdaKeyword{renaming} \AgdaSymbol{(}zero \AgdaSymbol{to} lzero\AgdaSymbol{;} suc \AgdaSymbol{to} lsuc\AgdaSymbol{)}\<%
\\
%
\\
\>\AgdaKeyword{open} \AgdaKeyword{import} \AgdaModule{Size} \AgdaKeyword{public}\<%
\\
%
\\
\>\AgdaKeyword{open} \AgdaKeyword{import} \AgdaModule{Category.Monad} \AgdaKeyword{public}\<%
\\
\>[0]\AgdaIndent{2}{}\<[2]%
\>[2]\AgdaKeyword{using} \AgdaSymbol{(}RawMonad\AgdaSymbol{;} \AgdaKeyword{module} RawMonad\AgdaSymbol{)}\<%
\\
%
\\
\>\AgdaKeyword{open} \AgdaKeyword{import} \AgdaModule{Data.Empty} \AgdaKeyword{public}\<%
\\
\>[0]\AgdaIndent{2}{}\<[2]%
\>[2]\AgdaKeyword{using} \AgdaSymbol{(}⊥\AgdaSymbol{;} ⊥-elim\AgdaSymbol{)}\<%
\\
%
\\
\>\AgdaKeyword{open} \AgdaKeyword{import} \AgdaModule{Data.List} \AgdaKeyword{public}\<%
\\
\>[0]\AgdaIndent{2}{}\<[2]%
\>[2]\AgdaKeyword{using} \AgdaSymbol{(}List\AgdaSymbol{;} []\AgdaSymbol{;} \_∷\_\AgdaSymbol{;} map\AgdaSymbol{)}\<%
\\
%
\\
\>\AgdaKeyword{open} \AgdaKeyword{import} \AgdaModule{Data.Maybe} \AgdaKeyword{public}\<%
\\
\>[0]\AgdaIndent{2}{}\<[2]%
\>[2]\AgdaKeyword{using} \AgdaSymbol{(}Maybe\AgdaSymbol{;} just\AgdaSymbol{;} nothing\AgdaSymbol{)} \AgdaKeyword{renaming} \AgdaSymbol{(}map \AgdaSymbol{to} fmap\AgdaSymbol{)}\<%
\\
%
\\
\>\AgdaKeyword{open} \AgdaKeyword{import} \AgdaModule{Data.Product} \AgdaKeyword{public}\<%
\\
\>[0]\AgdaIndent{2}{}\<[2]%
\>[2]\AgdaKeyword{using} \AgdaSymbol{(}∃\AgdaSymbol{;} \_×\_\AgdaSymbol{;} \_,\_\AgdaSymbol{)} \AgdaKeyword{renaming} \AgdaSymbol{(}proj₁ \AgdaSymbol{to} fst\AgdaSymbol{;} proj₂ \AgdaSymbol{to} snd\AgdaSymbol{)}\<%
\\
\>\AgdaKeyword{infixr} \AgdaNumber{1} \_,\_\<%
\\
%
\\
\>\AgdaKeyword{open} \AgdaKeyword{import} \AgdaModule{Data.Sum} \AgdaKeyword{public}\<%
\\
\>[0]\AgdaIndent{2}{}\<[2]%
\>[2]\AgdaKeyword{using} \AgdaSymbol{(}\_⊎\_\AgdaSymbol{;} [\_,\_]′\AgdaSymbol{)} \AgdaKeyword{renaming} \AgdaSymbol{(}inj₁ \AgdaSymbol{to} inl\AgdaSymbol{;} inj₂ \AgdaSymbol{to} inr\AgdaSymbol{)}\<%
\\
%
\\
\>\AgdaKeyword{open} \AgdaKeyword{import} \AgdaModule{Data.Unit} \<[23]%
\>[23]\AgdaKeyword{public}\<%
\\
\>[0]\AgdaIndent{2}{}\<[2]%
\>[2]\AgdaKeyword{using} \AgdaSymbol{(}⊤\AgdaSymbol{)}\<%
\\
%
\\
\>\AgdaKeyword{open} \AgdaKeyword{import} \AgdaModule{Function} \AgdaKeyword{public}\<%
\\
\>[0]\AgdaIndent{2}{}\<[2]%
\>[2]\AgdaKeyword{using} \AgdaSymbol{(}\_∘\_\AgdaSymbol{;} case\_of\_\AgdaSymbol{)}\<%
\\
%
\\
\>\AgdaKeyword{open} \AgdaKeyword{import} \AgdaModule{Relation.Nullary} \AgdaKeyword{public}\<%
\\
\>[0]\AgdaIndent{2}{}\<[2]%
\>[2]\AgdaKeyword{using} \AgdaSymbol{(}Dec\AgdaSymbol{;} yes\AgdaSymbol{;} no\AgdaSymbol{)}\<%
\\
%
\\
\>\AgdaKeyword{open} \AgdaKeyword{import} \AgdaModule{Relation.Binary} \AgdaKeyword{public}\<%
\\
\>[0]\AgdaIndent{2}{}\<[2]%
\>[2]\AgdaKeyword{using} \AgdaSymbol{(}Setoid\AgdaSymbol{;} \AgdaKeyword{module} Setoid\AgdaSymbol{)}\<%
\\
%
\\
\>\AgdaKeyword{import} \AgdaModule{Relation.Binary.PreorderReasoning}\<%
\\
\>\AgdaKeyword{module} \AgdaModule{Pre} \AgdaSymbol{=} \AgdaModule{Relation.Binary.PreorderReasoning}\<%
\\
%
\\
\>\AgdaKeyword{open} \AgdaKeyword{import} \AgdaModule{Relation.Binary.PropositionalEquality} \AgdaKeyword{public}\<%
\\
\>[0]\AgdaIndent{2}{}\<[2]%
\>[2]\AgdaKeyword{using} \AgdaSymbol{(}\_≡\_\AgdaSymbol{;} refl\AgdaSymbol{;} sym\AgdaSymbol{;} trans\AgdaSymbol{;} cong\AgdaSymbol{;} cong₂\AgdaSymbol{;} subst\AgdaSymbol{;} \AgdaKeyword{module} ≡-Reasoning\AgdaSymbol{)}\<%
\\
%
\\
\>\AgdaComment{--open ≡-Reasoning renaming (begin\_ to proof\_) public}\<%
\\
%
\\
\>\AgdaKeyword{open} \AgdaKeyword{import} \AgdaModule{Relation.Binary.HeterogeneousEquality} \AgdaKeyword{public}\<%
\\
\>[0]\AgdaIndent{2}{}\<[2]%
\>[2]\AgdaKeyword{using} \AgdaSymbol{(}\_≅\_\AgdaSymbol{;} refl\AgdaSymbol{;} ≡-to-≅\AgdaSymbol{;} \AgdaKeyword{module} ≅-Reasoning\AgdaSymbol{)}\<%
\\
\>[0]\AgdaIndent{2}{}\<[2]%
\>[2]\AgdaKeyword{renaming} \AgdaSymbol{(}sym \AgdaSymbol{to} hsym\AgdaSymbol{;} trans \AgdaSymbol{to} htrans\AgdaSymbol{;} cong \AgdaSymbol{to} hcong\AgdaSymbol{;}\<\\
\>            cong₂ \AgdaSymbol{to} hcong₂\AgdaSymbol{;} subst \AgdaSymbol{to} hsubst\AgdaSymbol{)}\<%
\\
%
\\
\>\AgdaFunction{hcong₃} \AgdaSymbol{:} \AgdaSymbol{\{}\AgdaBound{A} \AgdaSymbol{:} \AgdaPrimitiveType{Set}\AgdaSymbol{\}\{}\AgdaBound{B} \AgdaSymbol{:} \AgdaBound{A} \AgdaSymbol{→} \AgdaPrimitiveType{Set}\AgdaSymbol{\}\{}\AgdaBound{C} \AgdaSymbol{:} \AgdaSymbol{∀} \AgdaBound{a} \AgdaSymbol{→} \AgdaBound{B} \AgdaBound{a} \AgdaSymbol{→} \AgdaPrimitiveType{Set}\AgdaSymbol{\}\{}\AgdaBound{D} \AgdaSymbol{:} \AgdaSymbol{∀} \AgdaBound{a} \AgdaBound{b} \AgdaSymbol{→} \AgdaBound{C} \AgdaBound{a} \AgdaBound{b} \AgdaSymbol{→} \AgdaPrimitiveType{Set}\AgdaSymbol{\}}\<%
\\
\>[2]\AgdaIndent{9}{}\<[9]%
\>[9]\AgdaSymbol{(}\AgdaBound{\,f\,} \AgdaSymbol{:} \AgdaSymbol{∀} \AgdaBound{a} \AgdaBound{b} \AgdaBound{c} \AgdaSymbol{→} \AgdaBound{D} \AgdaBound{a} \AgdaBound{b} \AgdaBound{c}\AgdaSymbol{)}\<%
\\
\>[2]\AgdaIndent{9}{}\<[9]%
\>[9]\AgdaSymbol{\{}\AgdaBound{a} \AgdaBound{a′} \AgdaSymbol{:} \AgdaBound{A}\AgdaSymbol{\}} \AgdaSymbol{→} \AgdaBound{a} \AgdaDatatype{≅} \AgdaBound{a′} \AgdaSymbol{→}\<%
\\
\>[2]\AgdaIndent{9}{}\<[9]%
\>[9]\AgdaSymbol{\{}\AgdaBound{b} \AgdaSymbol{:} \AgdaBound{B} \AgdaBound{a}\AgdaSymbol{\}\{}\AgdaBound{b′} \AgdaSymbol{:} \AgdaBound{B} \AgdaBound{a′}\AgdaSymbol{\}} \AgdaSymbol{→} \AgdaBound{b} \AgdaDatatype{≅} \AgdaBound{b′} \AgdaSymbol{→}\<%
\\
\>[2]\AgdaIndent{9}{}\<[9]%
\>[9]\AgdaSymbol{\{}\AgdaBound{c} \AgdaSymbol{:} \AgdaBound{C} \AgdaBound{a} \AgdaBound{b}\AgdaSymbol{\}\{}\AgdaBound{c′} \AgdaSymbol{:} \AgdaBound{C} \AgdaBound{a′} \AgdaBound{b′}\AgdaSymbol{\}} \AgdaSymbol{→} \AgdaBound{c} \AgdaDatatype{≅} \AgdaBound{c′} \AgdaSymbol{→}\<%
\\
\>[2]\AgdaIndent{9}{}\<[9]%
\>[9]\AgdaBound{\,f\,} \AgdaBound{a} \AgdaBound{b} \AgdaBound{c} \AgdaDatatype{≅} \AgdaBound{\,f\,} \AgdaBound{a′} \AgdaBound{b′} \AgdaBound{c′}\<%
\\
\>\AgdaFunction{hcong₃} \AgdaBound{\,f\,} \AgdaInductiveConstructor{refl} \AgdaInductiveConstructor{refl} \AgdaInductiveConstructor{refl} \AgdaSymbol{=} \AgdaInductiveConstructor{refl}\<%
\\
%
\\
\>\AgdaFunction{≅-to-≡} \AgdaSymbol{:} \AgdaSymbol{∀} \AgdaSymbol{\{}\AgdaBound{a}\AgdaSymbol{\}} \AgdaSymbol{\{}\AgdaBound{A} \AgdaSymbol{:} \AgdaPrimitiveType{Set} \AgdaBound{a}\AgdaSymbol{\}} \AgdaSymbol{\{}\AgdaBound{x} \AgdaBound{y} \AgdaSymbol{:} \AgdaBound{A}\AgdaSymbol{\}} \AgdaSymbol{→} \AgdaBound{x} \AgdaDatatype{≅} \AgdaBound{y} \AgdaSymbol{→} \AgdaBound{x} \AgdaDatatype{≡} \AgdaBound{y}\<%
\\
\>\AgdaFunction{≅-to-≡} \AgdaInductiveConstructor{refl} \AgdaSymbol{=} \AgdaInductiveConstructor{refl}\<%
\end{code}
}
\section{Delay Monad}

The \AgdaDatatype{Delay} type is used to represent computations of type
\AgdaBound{A} whose result may be returned with some delay or never
be returned at all. A value available immediately is wrapped in the
\AgdaInductiveConstructor{now} constructor. A delayed value is wrapped in at least one
\AgdaInductiveConstructor{later} constructor. Each \AgdaInductiveConstructor{later} represents a single delay and an
infinite number of \AgdaInductiveConstructor{later} constructors wrapping a value
represents an infinite delay, i.e., a non-terminating computation.

It is interesting to compare the \AgdaDatatype{Delay} type with the
\AgdaDatatype{Maybe} type familiar from Haskell. Both are used to represent
partial values, but differ in the nature of partiality.
Pattern matching on an element of the
\AgdaDatatype{Maybe} type immediately yields either success (returning a value)
or failure (returning no value) whereas pattern matching on an element
of the \AgdaDatatype{Delay} type either yields success (returning a value) or
a delay after which one can pattern match again.  While \AgdaDatatype{Maybe}
lets us represent computation with error, possible \emph{non-termination} is
elegantly modeled by the \AgdaDatatype{Delay} type.
A definitely non-terminating value is represented by an infinite
succession of \AgdaInductiveConstructor{later} constructors, thus, \AgdaDatatype{Delay} must be a
coinductive type.  When analyzing a delayed value, we never know
whether after an initial succession of \AgdaInductiveConstructor{later} constructors we will
finally get a \AgdaInductiveConstructor{now} with a proper value---this reflects the
undecidability of termination in general.

In Agda, the \AgdaDatatype{Delay} type can be represented as a mutual
definition of an inductive datatype and a coinductive record.
The record \AgdaDatatype{∞Delay} is a coalgebra
and one interacts with it by using its single
observation (copattern) \force. Once forced we get an element of
the \AgdaDatatype{Delay} datatype which we can pattern match on to see if the
value is available \AgdaInductiveConstructor{now} or \AgdaInductiveConstructor{later}. If it is \AgdaInductiveConstructor{later}
then we get an element of \AgdaDatatype{∞Delay} which we can \force{}
again, and so forth.

\begin{code}%
\>\AgdaKeyword{mutual}\<%
\\
\>[0]\AgdaIndent{2}{}\<[2]%
\>[2]\AgdaKeyword{data} \AgdaDatatype{Delay} \AgdaSymbol{(}\AgdaBound{i} \AgdaSymbol{:} \AgdaPostulate{Size}\AgdaSymbol{)} \AgdaSymbol{(}\AgdaBound{A} \AgdaSymbol{:} \AgdaPrimitiveType{Set}\AgdaSymbol{)} \AgdaSymbol{:} \AgdaPrimitiveType{Set} \AgdaKeyword{where}\<%
\\
\>[2]\AgdaIndent{4}{}\<[4]%
\>[4]\AgdaInductiveConstructor{now} \<[11]%
\>[11]\AgdaSymbol{:} \<[14]%
\>[14]\AgdaBound{A} \<[26]%
\>[26]\AgdaSymbol{→} \AgdaDatatype{Delay} \AgdaBound{i} \AgdaBound{A}\<%
\\
\>[2]\AgdaIndent{4}{}\<[4]%
\>[4]\AgdaInductiveConstructor{later} \<[11]%
\>[11]\AgdaSymbol{:} \<[14]%
\>[14]\AgdaRecord{∞Delay} \AgdaBound{i} \AgdaBound{A} \<[26]%
\>[26]\AgdaSymbol{→} \AgdaDatatype{Delay} \AgdaBound{i} \AgdaBound{A}\<%
\\
%
\\
\>[0]\AgdaIndent{2}{}\<[2]%
\>[2]\AgdaKeyword{record} \AgdaRecord{∞Delay} \AgdaSymbol{(}\AgdaBound{i} \AgdaSymbol{:} \AgdaPostulate{Size}\AgdaSymbol{)} \AgdaSymbol{(}\AgdaBound{A} \AgdaSymbol{:} \AgdaPrimitiveType{Set}\AgdaSymbol{)} \AgdaSymbol{:} \AgdaPrimitiveType{Set} \AgdaKeyword{where}\<%
\\
\>[2]\AgdaIndent{4}{}\<[4]%
\>[4]\AgdaKeyword{coinductive}\<%
\\
\>[2]\AgdaIndent{4}{}\<[4]%
\>[4]\AgdaKeyword{field}\<%
\\
\>[4]\AgdaIndent{6}{}\<[6]%
\>[6]\AgdaField{force} \AgdaSymbol{:} \AgdaSymbol{\{}\AgdaBound{j} \AgdaSymbol{:} \AgdaPostulate{Size<} \AgdaBound{i}\AgdaSymbol{\}} \AgdaSymbol{→} \AgdaDatatype{Delay} \AgdaBound{j} \AgdaBound{A}\<%
\end{code}

\AgdaHide{
\begin{code}%
\>\AgdaKeyword{open} \AgdaModule{∞Delay} \AgdaKeyword{public}\<%
\end{code}}

Both types (\AgdaDatatype{Delay} and \AgdaDatatype{∞Delay}) are indexed by a size
\AgdaBound{i}.  This should be understood as \emph{observation depth}, i.e.,
a lower bound on the number of times we can iteratively \force{}
the delayed computation.   More precisely, forcing
$\ainf : \IDelayIA$
will result in a value $\ainf : \DelayA{j}$ of strictly smaller
observation depth \jlti.
An exception is a delayed value $\ainf : \IDelayA{∞}$ of infinite
observation depth, whose forcing $\force\;\ainf : \DelayA{∞}$
again has infinite observation depth.
The sizes (observation depths) are merely a means to establish
productivity of recursive definitions, in the end, we are only
interested in values $\aq : \DelayA{∞}$ of infinite depth.

If a corecursive function into \DelayIA{} only calls itself at
smaller depths \jlti{} it is guaranteed to be \emph{productive},
i.e., well-defined.  In the following definition of the
non-terminating value \AgdaFunction{never}, we make the hidden size arguments
explicit to demonstrate how they ensure productivity:

\begin{code}%
\>\AgdaFunction{never} \<[23]%
\>[23]\AgdaSymbol{:} \<[26]%
\>[26]\AgdaSymbol{∀\{}\AgdaBound{i} \AgdaBound{A}\AgdaSymbol{\}} \AgdaSymbol{→} \AgdaRecord{∞Delay} \AgdaBound{i} \AgdaBound{A}\<%
\\
\>\AgdaField{force} \AgdaSymbol{(}\AgdaFunction{never} \AgdaSymbol{\{}\AgdaBound{i}\AgdaSymbol{\})} \AgdaSymbol{\{}\AgdaBound{j}\AgdaSymbol{\}} \<[23]%
\>[23]\AgdaSymbol{=} \<[26]%
\>[26]\AgdaInductiveConstructor{later} \AgdaSymbol{(}\AgdaFunction{never} \AgdaSymbol{\{}\AgdaBound{j}\AgdaSymbol{\})}\<%
\end{code}

The value \AgdaFunction{never} is defined to be the thing that, if forced,
returns a postponed version of itself.  Formally, we have defined a
member of the record type \IDelayIA{} by giving the contents of
all of its fields, here only \force.  The use of a projection
like \force{} on the left hand side of a defining equation is
called a \emph{copattern}~\cite{abelPientkaThibodeauSetzer:popl13}.
Corecursive definitions by copatterns are the latest addition to Agda,
and can be activated since version 2.3.2 via the flag
\verb+--copatterns+.

The use of copatterns reduces productivity checking to termination
checking.  Agda simply checks that the size argument \AgdaBound{j} given in
the recursive call to \AgdaFunction{never} is smaller than the original
function parameter \AgdaBound{i}.  Indeed, \jlti{} is ensured by the
typing of projection \force.  A more detailed explanation and
theoretical foundations can be found in previous work of the first
author~\cite{abelPientka:icfp13}.  Agda can reconstruct size arguments
in programs if the sizes are declared in their type signature.  Thus,
we omit the hidden size arguments in the following.


At each observation depth \AgdaBound{i},
the functor \DelayI{} forms a monad. The \return{} of the monad is
given by \AgdaInductiveConstructor{now}, and bind \prebind{} is implemented below.
Notice that bind is size (observation depth) preserving; in other words, its
modulus of continuity is the identity.  The number of safe
observations on $\AgdaBound{a?}\;\bind\;\AgdaBound{\,f\,}$ is no less than those on both \aq{} and
$\AgdaBound{\,f\,}\;\AgdaBound{a}$ for any \AgdaBound{a}.
The implementation of bind follows a common scheme when working with
\AgdaDatatype{Delay}: we define two mutually recursive functions, the first by
pattern matching on \AgdaDatatype{Delay} and the second by copattern matching
on \AgdaDatatype{∞Delay}.

\begin{code}%
\>\AgdaKeyword{module} \AgdaModule{Bind} \AgdaKeyword{where}\<%
\\
\>[0]\AgdaIndent{2}{}\<[2]%
\>[2]\AgdaKeyword{mutual}\<%
\\
\>[2]\AgdaIndent{4}{}\<[4]%
\>[4]\AgdaFunction{\_\ensuremath{\mathbin{{>}\mkern-8.5mu{>}\mkern-2mu{=}}}\_} \<[23]%
\>[23]\AgdaSymbol{:} \<[26]%
\>[26]\AgdaSymbol{∀} \AgdaSymbol{\{}\AgdaBound{i} \AgdaBound{A} \AgdaBound{B}\AgdaSymbol{\}} \AgdaSymbol{→} \AgdaDatatype{Delay} \AgdaBound{i} \AgdaBound{A} \AgdaSymbol{→} \AgdaSymbol{(}\AgdaBound{A} \AgdaSymbol{→} \AgdaDatatype{Delay} \AgdaBound{i} \AgdaBound{B}\AgdaSymbol{)} \AgdaSymbol{→} \AgdaDatatype{Delay} \AgdaBound{i} \AgdaBound{B}\<%
\\
\>[2]\AgdaIndent{4}{}\<[4]%
\>[4]\AgdaInductiveConstructor{now} \<[11]%
\>[11]\AgdaBound{a} \<[15]%
\>[15]\AgdaFunction{\ensuremath{\mathbin{{>}\mkern-8.5mu{>}\mkern-2mu{=}}}} \<[20]%
\>[20]\AgdaBound{\,f\,} \<[23]%
\>[23]\AgdaSymbol{=} \<[26]%
\>[26]\AgdaBound{\,f\,} \AgdaBound{a}\<%
\\
\>[2]\AgdaIndent{4}{}\<[4]%
\>[4]\AgdaInductiveConstructor{later} \<[11]%
\>[11]\AgdaBound{a∞} \<[15]%
\>[15]\AgdaFunction{\ensuremath{\mathbin{{>}\mkern-8.5mu{>}\mkern-2mu{=}}}} \<[20]%
\>[20]\AgdaBound{\,f\,} \<[23]%
\>[23]\AgdaSymbol{=} \<[26]%
\>[26]\AgdaInductiveConstructor{later} \AgdaSymbol{(}\AgdaBound{a∞} \AgdaFunction{\ensuremath{\mathbin{\infty\mkern-3mu{>}\mkern-8.5mu{>}\mkern-2mu{=}}}} \AgdaBound{\,f\,}\AgdaSymbol{)}\<%
\\
%
\\
\>[2]\AgdaIndent{4}{}\<[4]%
\>[4]\AgdaFunction{\_\ensuremath{\mathbin{\infty\mkern-3mu{>}\mkern-8.5mu{>}\mkern-2mu{=}}}\_} \<[23]%
\>[23]\AgdaSymbol{:} \<[26]%
\>[26]\AgdaSymbol{∀} \AgdaSymbol{\{}\AgdaBound{i} \AgdaBound{A} \AgdaBound{B}\AgdaSymbol{\}} \AgdaSymbol{→} \AgdaRecord{∞Delay} \AgdaBound{i} \AgdaBound{A} \AgdaSymbol{→} \AgdaSymbol{(}\AgdaBound{A} \AgdaSymbol{→} \AgdaDatatype{Delay} \AgdaBound{i} \AgdaBound{B}\AgdaSymbol{)} \AgdaSymbol{→} \AgdaRecord{∞Delay} \AgdaBound{i} \AgdaBound{B}\<%
\\
\>[2]\AgdaIndent{4}{}\<[4]%
\>[4]\AgdaField{force} \AgdaSymbol{(}\AgdaBound{a∞} \AgdaFunction{\ensuremath{\mathbin{\infty\mkern-3mu{>}\mkern-8.5mu{>}\mkern-2mu{=}}}} \AgdaBound{\,f\,}\AgdaSymbol{)} \<[23]%
\>[23]\AgdaSymbol{=} \<[26]%
\>[26]\AgdaField{force} \AgdaBound{a∞} \AgdaFunction{\ensuremath{\mathbin{{>}\mkern-8.5mu{>}\mkern-2mu{=}}}} \AgdaBound{\,f\,}\<%
\end{code}

\noindent
We make \DelayI{} an instance of \RawMonad{} (it is called `raw' as it does not enforce the laws) as defined in
the Agda standard library.  This provides us automatically with a
\RawFunctor{} instance, with map function \fmap{} written
infix as in Haskell's base library.

\begin{code}%
\>\AgdaFunction{delayMonad} \AgdaSymbol{:} \AgdaSymbol{∀} \AgdaSymbol{\{}\AgdaBound{i}\AgdaSymbol{\}} \AgdaSymbol{→} \AgdaFunction{RawMonad} \AgdaSymbol{(}\AgdaDatatype{Delay} \AgdaBound{i}\AgdaSymbol{)}\<%
\\
\>\AgdaFunction{delayMonad} \AgdaSymbol{\{}\AgdaBound{i}\AgdaSymbol{\}} \AgdaSymbol{=} \AgdaKeyword{record}\<%
\\
\>[0]\AgdaIndent{2}{}\<[2]%
\>[2]\AgdaSymbol{\{} \AgdaField{return} \<[12]%
\>[12]\AgdaSymbol{=} \<[15]%
\>[15]\AgdaInductiveConstructor{now}\<%
\\
\>[0]\AgdaIndent{2}{}\<[2]%
\>[2]\AgdaSymbol{;} \AgdaField{\_\ensuremath{\mathbin{{>}\mkern-8.5mu{>}\mkern-2mu{=}}}\_} \<[12]%
\>[12]\AgdaSymbol{=} \<[15]%
\>[15]\AgdaFunction{\_\ensuremath{\mathbin{{>}\mkern-8.5mu{>}\mkern-2mu{=}}}\_} \AgdaSymbol{\{}\AgdaBound{i}\AgdaSymbol{\}}\<%
\\
\>[0]\AgdaIndent{2}{}\<[2]%
\>[2]\AgdaSymbol{\}} \AgdaKeyword{where} \AgdaKeyword{open} \AgdaModule{Bind}\<%
\end{code}

\AgdaHide{
\begin{code}%
\>\AgdaKeyword{module} \AgdaModule{\_} \AgdaSymbol{\{}\AgdaBound{i} \AgdaSymbol{:} \AgdaPostulate{Size}\AgdaSymbol{\}} \AgdaKeyword{where}\<%
\\
\>[0]\AgdaIndent{2}{}\<[2]%
\>[2]\AgdaKeyword{open} \AgdaKeyword{module} \AgdaModule{DelayMonad} \AgdaSymbol{=} \AgdaModule{RawMonad} \AgdaSymbol{(}\AgdaFunction{delayMonad} \AgdaSymbol{\{}\AgdaSymbol{i} \AgdaSymbol{=} \AgdaBound{i}\AgdaSymbol{\})}\<%
\\
\>[2]\AgdaIndent{27}{}\<[27]%
\>[27]\AgdaKeyword{public} \AgdaKeyword{renaming} \AgdaSymbol{(}\_⊛\_ \AgdaSymbol{to} \_<*>\_\AgdaSymbol{)}\<%
\\
\>\AgdaKeyword{open} \AgdaModule{Bind} \AgdaKeyword{public} \AgdaKeyword{using} \AgdaSymbol{(}\_\ensuremath{\mathbin{\infty\mkern-3mu{>}\mkern-8.5mu{>}\mkern-2mu{=}}}\_\AgdaSymbol{)}\<%
\\
%
\\
\>\AgdaComment{-- Map for ∞Delay}\<%
\\
%
\\
\>\AgdaFunction{\_∞\ensuremath{{<}\${>}}\_} \AgdaSymbol{:} \AgdaSymbol{∀} \AgdaSymbol{\{}\AgdaBound{i} \AgdaBound{A} \AgdaBound{B}\AgdaSymbol{\}} \AgdaSymbol{(}\AgdaBound{\,f\,} \AgdaSymbol{:} \AgdaBound{A} \AgdaSymbol{→} \AgdaBound{B}\AgdaSymbol{)} \AgdaSymbol{(}\AgdaBound{∞a} \AgdaSymbol{:} \AgdaRecord{∞Delay} \AgdaBound{i} \AgdaBound{A}\AgdaSymbol{)} \AgdaSymbol{→} \AgdaRecord{∞Delay} \AgdaBound{i} \AgdaBound{B}\<%
\\
\>\AgdaBound{\,f\,} \AgdaFunction{∞\ensuremath{{<}\${>}}} \AgdaBound{∞a} \AgdaSymbol{=} \AgdaBound{∞a} \AgdaFunction{\ensuremath{\mathbin{\infty\mkern-3mu{>}\mkern-8.5mu{>}\mkern-2mu{=}}}} \AgdaSymbol{λ} \AgdaBound{a} \AgdaSymbol{→} \AgdaFunction{return} \AgdaSymbol{(}\AgdaBound{\,f\,} \AgdaBound{a}\AgdaSymbol{)}\<%
\\
\>\AgdaComment{-- force (f ∞\ensuremath{{<}\${>}} ∞a) = f \ensuremath{{<}\${>}} force ∞a}\<%
\\
%
\\
\>\AgdaComment{-- Double bind}\<%
\\
%
\\
\>\AgdaFunction{\_\ensuremath{\mathbin{{=}\mkern-2mu{<}\mkern-8.5mu{<}}}2\_,\_} \AgdaSymbol{:} \AgdaSymbol{∀} \AgdaSymbol{\{}\AgdaBound{i} \AgdaBound{A} \AgdaBound{B} \AgdaBound{C}\AgdaSymbol{\}} \AgdaSymbol{→} \AgdaSymbol{(}\AgdaBound{A} \AgdaSymbol{→} \AgdaBound{B} \AgdaSymbol{→} \AgdaDatatype{Delay} \AgdaBound{i} \AgdaBound{C}\AgdaSymbol{)} \AgdaSymbol{→} \AgdaDatatype{Delay} \AgdaBound{i} \AgdaBound{A} \AgdaSymbol{→} \AgdaDatatype{Delay} \AgdaBound{i} \AgdaBound{B} \AgdaSymbol{→} \AgdaDatatype{Delay} \AgdaBound{i} \AgdaBound{C}\<%
\\
\>\AgdaBound{\,f\,} \AgdaFunction{\ensuremath{\mathbin{{=}\mkern-2mu{<}\mkern-8.5mu{<}}}2} \AgdaBound{x} \AgdaFunction{,} \AgdaBound{y} \AgdaSymbol{=} \AgdaBound{x} \AgdaFunction{\ensuremath{\mathbin{{>}\mkern-8.5mu{>}\mkern-2mu{=}}}} \AgdaSymbol{λ} \AgdaBound{a} \AgdaSymbol{→} \AgdaBound{y} \AgdaFunction{\ensuremath{\mathbin{{>}\mkern-8.5mu{>}\mkern-2mu{=}}}} \AgdaSymbol{λ} \AgdaBound{b} \AgdaSymbol{→} \AgdaBound{\,f\,} \AgdaBound{a} \AgdaBound{b}\<%
\end{code}
}

\subsection{Strong Bisimilarity}

We can define the coinductive strong bisimilarity relation \prebisim{}
for \DelayA{∞}
following the same pattern as for \AgdaDatatype{Delay} itself. Two finite
computations are \emph{strongly bisimilar} if they contain the same value and
the same amount of delay (number of \AgdaInductiveConstructor{later}s).  Non-terminating
computations are also identified.\footnote{One could also consider other
relations such as \emph{weak bisimilarity} which identifies finite
computations containing the same value but different numbers of
\AgdaInductiveConstructor{later}s.}

\begin{code}%
\>\AgdaKeyword{mutual}\<%
\\
\>[0]\AgdaIndent{2}{}\<[2]%
\>[2]\AgdaKeyword{data} \AgdaDatatype{\_∼\_} \AgdaSymbol{\{}\AgdaBound{i} \AgdaSymbol{:} \AgdaPostulate{Size}\AgdaSymbol{\}} \AgdaSymbol{\{}\AgdaBound{A} \AgdaSymbol{:} \AgdaPrimitiveType{Set}\AgdaSymbol{\}} \AgdaSymbol{:} \AgdaSymbol{(}\AgdaBound{a?} \AgdaBound{b?} \AgdaSymbol{:} \AgdaDatatype{Delay} \AgdaPostulate{∞} \AgdaBound{A}\AgdaSymbol{)} \AgdaSymbol{→} \AgdaPrimitiveType{Set} \AgdaKeyword{where}\<%
\\
\>[2]\AgdaIndent{4}{}\<[4]%
\>[4]\AgdaInductiveConstructor{∼now} \<[12]%
\>[12]\AgdaSymbol{:} \<[15]%
\>[15]\AgdaSymbol{∀} \AgdaBound{a} \<[48]%
\>[48]\AgdaSymbol{→} \<[51]%
\>[51]\AgdaInductiveConstructor{now} \AgdaBound{a} \<[61]%
\>[61]\AgdaDatatype{∼} \AgdaInductiveConstructor{now} \AgdaBound{a}\<%
\\
\>[2]\AgdaIndent{4}{}\<[4]%
\>[4]\AgdaInductiveConstructor{∼later} \<[12]%
\>[12]\AgdaSymbol{:} \<[15]%
\>[15]\AgdaSymbol{∀} \AgdaSymbol{\{}\AgdaBound{a∞} \AgdaBound{b∞}\AgdaSymbol{\}} \AgdaSymbol{(}\AgdaBound{eq} \AgdaSymbol{:} \AgdaBound{a∞} \AgdaRecord{∞∼⟨} \AgdaBound{i} \AgdaRecord{⟩∼} \AgdaBound{b∞}\AgdaSymbol{)} \<[48]%
\>[48]\AgdaSymbol{→} \<[51]%
\>[51]\AgdaInductiveConstructor{later} \AgdaBound{a∞} \<[61]%
\>[61]\AgdaDatatype{∼} \AgdaInductiveConstructor{later} \AgdaBound{b∞}\<%
\\
%
\\
\>[0]\AgdaIndent{2}{}\<[2]%
\>[2]\AgdaFunction{\_∼⟨\_⟩∼\_} \AgdaSymbol{=} \AgdaSymbol{λ} \AgdaSymbol{\{}\AgdaBound{A}\AgdaSymbol{\}} \AgdaBound{a?} \AgdaBound{i} \AgdaBound{b?} \AgdaSymbol{→} \AgdaDatatype{\_∼\_} \AgdaSymbol{\{}\AgdaBound{i}\AgdaSymbol{\}\{}\AgdaBound{A}\AgdaSymbol{\}} \AgdaBound{a?} \AgdaBound{b?}\<%
\\
%
\\
\>[0]\AgdaIndent{2}{}\<[2]%
\>[2]\AgdaKeyword{record} \AgdaRecord{\_∞∼⟨\_⟩∼\_} \AgdaSymbol{\{}\AgdaBound{A}\AgdaSymbol{\}} \AgdaSymbol{(}\AgdaBound{a∞} \AgdaSymbol{:} \AgdaRecord{∞Delay} \AgdaPostulate{∞} \AgdaBound{A}\AgdaSymbol{)} \AgdaBound{i} \AgdaSymbol{(}\AgdaBound{b∞} \AgdaSymbol{:} \AgdaRecord{∞Delay} \AgdaPostulate{∞} \AgdaBound{A}\AgdaSymbol{)} \AgdaSymbol{:} \AgdaPrimitiveType{Set} \AgdaKeyword{where}\<%
\\
\>[2]\AgdaIndent{4}{}\<[4]%
\>[4]\AgdaKeyword{coinductive}\<%
\\
\>[2]\AgdaIndent{4}{}\<[4]%
\>[4]\AgdaKeyword{field}\<%
\\
\>[4]\AgdaIndent{6}{}\<[6]%
\>[6]\AgdaField{∼force} \AgdaSymbol{:} \AgdaSymbol{\{}\AgdaBound{j} \AgdaSymbol{:} \AgdaPostulate{Size<} \AgdaBound{i}\AgdaSymbol{\}} \AgdaSymbol{→} \AgdaField{force} \AgdaBound{a∞} \AgdaFunction{∼⟨} \AgdaBound{j} \AgdaFunction{⟩∼} \AgdaField{force} \AgdaBound{b∞}\<%
\\
%
\\
\>\AgdaFunction{\_∞∼\_} \AgdaSymbol{=} \AgdaSymbol{λ} \AgdaSymbol{\{}\AgdaBound{i}\AgdaSymbol{\}} \AgdaSymbol{\{}\AgdaBound{A}\AgdaSymbol{\}} \AgdaBound{a∞} \AgdaBound{b∞} \AgdaSymbol{→} \AgdaRecord{\_∞∼⟨\_⟩∼\_} \AgdaSymbol{\{}\AgdaBound{A}\AgdaSymbol{\}} \AgdaBound{a∞} \AgdaBound{i} \AgdaBound{b∞}\<%
\end{code}

\noindent
The definition includes the two sized relations
\presizedbisim{i} on \DelayA{∞} and
\presizedinfbisim{i} on \IDelayA{∞} that exist for the purpose of
recursively constructing derivations (proofs) of bisimilarity in a way
that convinces Agda of their productivity.  These are approximations
of bisimilarity in the sense that they are intermediate, partially
defined relations needed for the construction
of the fully defined relations
\presizedbisim{∞} and
\presizedinfbisim{∞}.
They are subtly different to the approximations $\cong_n$ of
strong bisimilarity $\cong$ in the context of ultrametric spaces
\cite[Sec.~2.2] {aehligJoachimski:continuousNormalization}.
Those approximations are fully defined relations that approximate the
concept of equality, for instance at stage $n=0$ all values are equal,
at $n=1$ they are equal if observations of depth one coincide, until
at stage $n=\omega$ observation of arbitrary depth must yield the same
result.

\AgdaHide{
\begin{code}%
\>\AgdaKeyword{open} \AgdaModule{\_∞∼⟨\_⟩∼\_} \AgdaKeyword{public}\<%
\\
%
\\
\>\AgdaFunction{∼never} \AgdaSymbol{:} \AgdaSymbol{∀\{}\AgdaBound{i} \AgdaBound{A}\AgdaSymbol{\}} \AgdaSymbol{→} \AgdaSymbol{(}\AgdaFunction{never} \AgdaSymbol{\{}\AgdaSymbol{A} \AgdaSymbol{=} \AgdaBound{A}\AgdaSymbol{\})} \AgdaRecord{∞∼⟨} \AgdaBound{i} \AgdaRecord{⟩∼} \AgdaFunction{never}\<%
\\
\>\AgdaField{∼force} \AgdaFunction{∼never} \AgdaSymbol{=} \AgdaInductiveConstructor{∼later} \AgdaFunction{∼never}\<%
\end{code}}

All bisimilarity relations \presizedbisim{i} and
\presizedinfbisim{i} are equivalences.  The proofs by coinduction are
straightforward and omitted here.

\begin{code}%
\>\AgdaFunction{∼refl} \<[9]%
\>[9]\AgdaSymbol{:} \<[12]%
\>[12]\AgdaSymbol{∀\{}\AgdaBound{i} \AgdaBound{A}\AgdaSymbol{\}} \AgdaSymbol{(}\AgdaBound{a?} \<[24]%
\>[24]\AgdaSymbol{:} \AgdaDatatype{Delay} \AgdaPostulate{∞} \AgdaBound{A}\AgdaSymbol{)} \<[39]%
\>[39]\AgdaSymbol{→} \<[42]%
\>[42]\AgdaBound{a?} \AgdaFunction{∼⟨} \AgdaBound{i} \AgdaFunction{⟩∼} \AgdaBound{a?}\<%
\\
\>\AgdaFunction{∞∼refl} \<[9]%
\>[9]\AgdaSymbol{:} \<[12]%
\>[12]\AgdaSymbol{∀\{}\AgdaBound{i} \AgdaBound{A}\AgdaSymbol{\}} \AgdaSymbol{(}\AgdaBound{a∞} \<[24]%
\>[24]\AgdaSymbol{:} \AgdaRecord{∞Delay} \AgdaPostulate{∞} \AgdaBound{A}\AgdaSymbol{)} \<[39]%
\>[39]\AgdaSymbol{→} \<[42]%
\>[42]\AgdaBound{a∞} \AgdaRecord{∞∼⟨} \AgdaBound{i} \AgdaRecord{⟩∼} \AgdaBound{a∞}\<%
\\
%
\\
\>\AgdaFunction{∼sym} \<[9]%
\>[9]\AgdaSymbol{:} \<[12]%
\>[12]\AgdaSymbol{∀\{}\AgdaBound{i} \AgdaBound{A}\AgdaSymbol{\}\{}\AgdaBound{a?} \<[23]%
\>[23]\AgdaBound{b?} \<[27]%
\>[27]\AgdaSymbol{:} \AgdaDatatype{Delay} \AgdaPostulate{∞} \AgdaBound{A} \AgdaSymbol{\}} \<[42]%
\>[42]\AgdaSymbol{→} \<[45]%
\>[45]\AgdaBound{a?} \AgdaFunction{∼⟨} \AgdaBound{i} \AgdaFunction{⟩∼} \AgdaBound{b?} \<[61]%
\>[61]\AgdaSymbol{→} \<[64]%
\>[64]\AgdaBound{b?} \AgdaFunction{∼⟨} \AgdaBound{i} \AgdaFunction{⟩∼} \AgdaBound{a?}\<%
\\
\>\AgdaFunction{∞∼sym} \<[9]%
\>[9]\AgdaSymbol{:} \<[12]%
\>[12]\AgdaSymbol{∀\{}\AgdaBound{i} \AgdaBound{A}\AgdaSymbol{\}\{}\AgdaBound{a∞} \<[23]%
\>[23]\AgdaBound{b∞} \<[27]%
\>[27]\AgdaSymbol{:} \AgdaRecord{∞Delay} \AgdaPostulate{∞} \AgdaBound{A}\AgdaSymbol{\}} \<[42]%
\>[42]\AgdaSymbol{→} \<[45]%
\>[45]\AgdaBound{a∞} \AgdaRecord{∞∼⟨} \AgdaBound{i} \AgdaRecord{⟩∼} \AgdaBound{b∞} \<[61]%
\>[61]\AgdaSymbol{→} \<[64]%
\>[64]\AgdaBound{b∞} \AgdaRecord{∞∼⟨} \AgdaBound{i} \AgdaRecord{⟩∼} \AgdaBound{a∞}\<%
\\
%
\\
\>\AgdaFunction{∼trans} \<[9]%
\>[9]\AgdaSymbol{:} \<[12]%
\>[12]\AgdaSymbol{∀\{}\AgdaBound{i} \AgdaBound{A}\AgdaSymbol{\}\{}\AgdaBound{a?} \AgdaBound{b?} \AgdaBound{c?} \AgdaSymbol{:} \AgdaDatatype{Delay} \AgdaPostulate{∞} \AgdaBound{A}\AgdaSymbol{\}} \AgdaSymbol{→}\<%
\\
\>[6]\AgdaIndent{12}{}\<[12]%
\>[12]\AgdaBound{a?} \AgdaFunction{∼⟨} \AgdaBound{i} \AgdaFunction{⟩∼} \AgdaBound{b?} \AgdaSymbol{→} \<[29]%
\>[29]\AgdaBound{b?} \AgdaFunction{∼⟨} \AgdaBound{i} \AgdaFunction{⟩∼} \AgdaBound{c?} \AgdaSymbol{→} \AgdaBound{a?} \AgdaFunction{∼⟨} \AgdaBound{i} \AgdaFunction{⟩∼} \AgdaBound{c?}\<%
\\
\>\AgdaFunction{∞∼trans} \<[9]%
\>[9]\AgdaSymbol{:} \<[12]%
\>[12]\AgdaSymbol{∀\{}\AgdaBound{i} \AgdaBound{A}\AgdaSymbol{\}\{}\AgdaBound{a∞} \AgdaBound{b∞} \AgdaBound{c∞} \AgdaSymbol{:} \AgdaRecord{∞Delay} \AgdaPostulate{∞} \AgdaBound{A}\AgdaSymbol{\}} \AgdaSymbol{→}\<%
\\
\>[6]\AgdaIndent{12}{}\<[12]%
\>[12]\AgdaBound{a∞} \AgdaRecord{∞∼⟨} \AgdaBound{i} \AgdaRecord{⟩∼} \AgdaBound{b∞} \AgdaSymbol{→} \<[30]%
\>[30]\AgdaBound{b∞} \AgdaRecord{∞∼⟨} \AgdaBound{i} \AgdaRecord{⟩∼} \AgdaBound{c∞} \AgdaSymbol{→} \AgdaBound{a∞} \AgdaRecord{∞∼⟨} \AgdaBound{i} \AgdaRecord{⟩∼} \AgdaBound{c∞}\<%
\end{code}

\AgdaHide{
\begin{code}%
\>\AgdaFunction{∼refl} \AgdaSymbol{(}\AgdaInductiveConstructor{now} \AgdaBound{a}\AgdaSymbol{)} \<[17]%
\>[17]\AgdaSymbol{=} \AgdaInductiveConstructor{∼now} \AgdaBound{a}\<%
\\
\>\AgdaFunction{∼refl} \AgdaSymbol{(}\AgdaInductiveConstructor{later} \AgdaBound{a∞}\AgdaSymbol{)} \AgdaSymbol{=} \AgdaInductiveConstructor{∼later} \AgdaSymbol{(}\AgdaFunction{∞∼refl} \AgdaBound{a∞}\AgdaSymbol{)}\<%
\\
%
\\
\>\AgdaField{∼force} \AgdaSymbol{(}\AgdaFunction{∞∼refl} \AgdaBound{a∞}\AgdaSymbol{)} \AgdaSymbol{=} \AgdaFunction{∼refl} \AgdaSymbol{(}\AgdaField{force} \AgdaBound{a∞}\AgdaSymbol{)}\<%
\\
%
\\
\>\AgdaFunction{∼sym} \AgdaSymbol{(}\AgdaInductiveConstructor{∼now} \AgdaBound{a}\AgdaSymbol{)} \<[17]%
\>[17]\AgdaSymbol{=} \AgdaInductiveConstructor{∼now} \AgdaBound{a}\<%
\\
\>\AgdaFunction{∼sym} \AgdaSymbol{(}\AgdaInductiveConstructor{∼later} \AgdaBound{eq}\AgdaSymbol{)} \AgdaSymbol{=} \AgdaInductiveConstructor{∼later} \AgdaSymbol{(}\AgdaFunction{∞∼sym} \AgdaBound{eq}\AgdaSymbol{)}\<%
\\
%
\\
\>\AgdaField{∼force} \AgdaSymbol{(}\AgdaFunction{∞∼sym} \AgdaBound{eq}\AgdaSymbol{)} \AgdaSymbol{=} \AgdaFunction{∼sym} \AgdaSymbol{(}\AgdaField{∼force} \AgdaBound{eq}\AgdaSymbol{)}\<%
\\
%
\\
\>\AgdaFunction{∼trans} \AgdaSymbol{(}\AgdaInductiveConstructor{∼now} \AgdaBound{a}\AgdaSymbol{)} \<[19]%
\>[19]\AgdaSymbol{(}\AgdaInductiveConstructor{∼now} \AgdaSymbol{.}\AgdaBound{a}\AgdaSymbol{)} \<[32]%
\>[32]\AgdaSymbol{=} \AgdaInductiveConstructor{∼now} \AgdaBound{a}\<%
\\
\>\AgdaFunction{∼trans} \AgdaSymbol{(}\AgdaInductiveConstructor{∼later} \AgdaBound{eq}\AgdaSymbol{)} \AgdaSymbol{(}\AgdaInductiveConstructor{∼later} \AgdaBound{eq′}\AgdaSymbol{)} \AgdaSymbol{=} \AgdaInductiveConstructor{∼later} \AgdaSymbol{(}\AgdaFunction{∞∼trans} \AgdaBound{eq} \AgdaBound{eq′}\AgdaSymbol{)}\<%
\\
%
\\
\>\AgdaField{∼force} \AgdaSymbol{(}\AgdaFunction{∞∼trans} \AgdaBound{eq} \AgdaBound{eq′}\AgdaSymbol{)} \AgdaSymbol{=} \AgdaFunction{∼trans} \AgdaSymbol{(}\AgdaField{∼force} \AgdaBound{eq}\AgdaSymbol{)} \AgdaSymbol{(}\AgdaField{∼force} \AgdaBound{eq′}\AgdaSymbol{)}\<%
\\
%
\\
\>\AgdaComment{-- Equality reasoning}\<%
\\
%
\\
\>\AgdaFunction{∼setoid} \AgdaSymbol{:} \AgdaSymbol{(}\AgdaBound{i} \AgdaSymbol{:} \AgdaPostulate{Size}\AgdaSymbol{)} \AgdaSymbol{(}\AgdaBound{A} \AgdaSymbol{:} \AgdaPrimitiveType{Set}\AgdaSymbol{)} \AgdaSymbol{→} \AgdaRecord{Setoid} \AgdaPrimitive{lzero} \AgdaPrimitive{lzero}\<%
\\
\>\AgdaFunction{∼setoid} \AgdaBound{i} \AgdaBound{A} \AgdaSymbol{=} \AgdaKeyword{record}\<%
\\
\>[0]\AgdaIndent{2}{}\<[2]%
\>[2]\AgdaSymbol{\{} \AgdaField{Carrier} \<[18]%
\>[18]\AgdaSymbol{=} \AgdaDatatype{Delay} \AgdaPostulate{∞} \AgdaBound{A}\<%
\\
\>[0]\AgdaIndent{2}{}\<[2]%
\>[2]\AgdaSymbol{;} \AgdaField{\_≈\_} \<[18]%
\>[18]\AgdaSymbol{=} \AgdaDatatype{\_∼\_} \AgdaSymbol{\{}\AgdaBound{i}\AgdaSymbol{\}}\<%
\\
\>[0]\AgdaIndent{2}{}\<[2]%
\>[2]\AgdaSymbol{;} \AgdaField{isEquivalence} \AgdaSymbol{=} \AgdaKeyword{record}\<%
\\
\>[2]\AgdaIndent{4}{}\<[4]%
\>[4]\AgdaSymbol{\{} \AgdaField{refl} \<[12]%
\>[12]\AgdaSymbol{=} \AgdaSymbol{λ} \AgdaSymbol{\{}\AgdaBound{a?}\AgdaSymbol{\}} \AgdaSymbol{→} \AgdaFunction{∼refl} \AgdaBound{a?}\<%
\\
\>[2]\AgdaIndent{4}{}\<[4]%
\>[4]\AgdaSymbol{;} \AgdaField{sym} \<[12]%
\>[12]\AgdaSymbol{=} \AgdaFunction{∼sym}\<%
\\
\>[2]\AgdaIndent{4}{}\<[4]%
\>[4]\AgdaSymbol{;} \AgdaField{trans} \AgdaSymbol{=} \AgdaFunction{∼trans}\<%
\\
\>[2]\AgdaIndent{4}{}\<[4]%
\>[4]\AgdaSymbol{\}}\<%
\\
\>[0]\AgdaIndent{2}{}\<[2]%
\>[2]\AgdaSymbol{\}}\<%
\\
%
\\
\>\AgdaFunction{∞∼setoid} \AgdaSymbol{:} \AgdaSymbol{(}\AgdaBound{i} \AgdaSymbol{:} \AgdaPostulate{Size}\AgdaSymbol{)} \AgdaSymbol{(}\AgdaBound{A} \AgdaSymbol{:} \AgdaPrimitiveType{Set}\AgdaSymbol{)} \AgdaSymbol{→} \AgdaRecord{Setoid} \AgdaPrimitive{lzero} \AgdaPrimitive{lzero}\<%
\\
\>\AgdaFunction{∞∼setoid} \AgdaBound{i} \AgdaBound{A} \AgdaSymbol{=} \AgdaKeyword{record}\<%
\\
\>[0]\AgdaIndent{2}{}\<[2]%
\>[2]\AgdaSymbol{\{} \AgdaField{Carrier} \<[18]%
\>[18]\AgdaSymbol{=} \AgdaRecord{∞Delay} \AgdaPostulate{∞} \AgdaBound{A}\<%
\\
\>[0]\AgdaIndent{2}{}\<[2]%
\>[2]\AgdaSymbol{;} \AgdaField{\_≈\_} \<[18]%
\>[18]\AgdaSymbol{=} \AgdaFunction{\_∞∼\_} \AgdaSymbol{\{}\AgdaBound{i}\AgdaSymbol{\}}\<%
\\
\>[0]\AgdaIndent{2}{}\<[2]%
\>[2]\AgdaSymbol{;} \AgdaField{isEquivalence} \AgdaSymbol{=} \AgdaKeyword{record}\<%
\\
\>[2]\AgdaIndent{4}{}\<[4]%
\>[4]\AgdaSymbol{\{} \AgdaField{refl} \<[12]%
\>[12]\AgdaSymbol{=} \AgdaSymbol{λ} \AgdaSymbol{\{}\AgdaBound{a?}\AgdaSymbol{\}} \AgdaSymbol{→} \AgdaFunction{∞∼refl} \AgdaBound{a?}\<%
\\
\>[2]\AgdaIndent{4}{}\<[4]%
\>[4]\AgdaSymbol{;} \AgdaField{sym} \<[12]%
\>[12]\AgdaSymbol{=} \AgdaFunction{∞∼sym}\<%
\\
\>[2]\AgdaIndent{4}{}\<[4]%
\>[4]\AgdaSymbol{;} \AgdaField{trans} \AgdaSymbol{=} \AgdaFunction{∞∼trans}\<%
\\
\>[2]\AgdaIndent{4}{}\<[4]%
\>[4]\AgdaSymbol{\}}\<%
\\
\>[0]\AgdaIndent{2}{}\<[2]%
\>[2]\AgdaSymbol{\}}\<%
\\
%
\\
\>\AgdaKeyword{module} \AgdaModule{∼-Reasoning} \AgdaSymbol{\{}\AgdaBound{i} \AgdaSymbol{:} \AgdaPostulate{Size}\AgdaSymbol{\}} \AgdaSymbol{\{}\AgdaBound{A} \AgdaSymbol{:} \AgdaPrimitiveType{Set}\AgdaSymbol{\}} \AgdaKeyword{where}\<%
\\
\>[0]\AgdaIndent{2}{}\<[2]%
\>[2]\AgdaKeyword{open} \AgdaModule{Pre} \AgdaSymbol{(}\AgdaFunction{Setoid.preorder} \AgdaSymbol{(}\AgdaFunction{∼setoid} \AgdaBound{i} \AgdaBound{A}\AgdaSymbol{))} \AgdaKeyword{public}\<%
\\
\>\AgdaComment{--    using (begin\_; \_∎) (\_≈⟨⟩\_ to \_∼⟨⟩\_; \_≈⟨\_⟩\_ to \_∼⟨\_⟩\_)}\<%
\\
\>[2]\AgdaIndent{4}{}\<[4]%
\>[4]\AgdaKeyword{renaming} \AgdaSymbol{(}\_≈⟨⟩\_ \AgdaSymbol{to} \_≡⟨⟩\_\AgdaSymbol{;} \_≈⟨\_⟩\_ \AgdaSymbol{to} \_≡⟨\_⟩\_\AgdaSymbol{;} \_∼⟨\_⟩\_ \AgdaSymbol{to} \_∼⟨\_⟩\_\AgdaSymbol{;} begin\_ \AgdaSymbol{to} proof\_\AgdaSymbol{)}\<%
\\
%
\\
\>\AgdaKeyword{module} \AgdaModule{∞∼-Reasoning} \AgdaSymbol{\{}\AgdaBound{i} \AgdaSymbol{:} \AgdaPostulate{Size}\AgdaSymbol{\}} \AgdaSymbol{\{}\AgdaBound{A} \AgdaSymbol{:} \AgdaPrimitiveType{Set}\AgdaSymbol{\}} \AgdaKeyword{where}\<%
\\
\>[0]\AgdaIndent{2}{}\<[2]%
\>[2]\AgdaKeyword{open} \AgdaModule{Pre} \AgdaSymbol{(}\AgdaFunction{Setoid.preorder} \AgdaSymbol{(}\AgdaFunction{∞∼setoid} \AgdaBound{i} \AgdaBound{A}\AgdaSymbol{))} \AgdaKeyword{public}\<%
\\
\>\AgdaComment{--    using (begin\_; \_∎) (\_≈⟨⟩\_ to \_∼⟨⟩\_; \_≈⟨\_⟩\_ to \_∼⟨\_⟩\_)}\<%
\\
\>[2]\AgdaIndent{4}{}\<[4]%
\>[4]\AgdaKeyword{renaming} \AgdaSymbol{(}\_≈⟨⟩\_ \AgdaSymbol{to} \_≡⟨⟩\_\AgdaSymbol{;} \_≈⟨\_⟩\_ \AgdaSymbol{to} \_≡⟨\_⟩\_\AgdaSymbol{;} \_∼⟨\_⟩\_ \AgdaSymbol{to} \_∞∼⟨\_⟩\_\AgdaSymbol{;} begin\_ \AgdaSymbol{to} proof\_\AgdaSymbol{)}\<%
\end{code}
}

\noindent
The associativity law of the delay monad holds up to strong
bisimilarity.  Here, we spell out the proof by coinduction:

\begin{code}%
\>\AgdaKeyword{mutual}\<%
\\
\>[0]\AgdaIndent{2}{}\<[2]%
\>[2]\AgdaFunction{bind-assoc} \<[27]%
\>[27]\AgdaSymbol{:} \<[30]%
\>[30]\AgdaSymbol{∀\{}\AgdaBound{i} \AgdaBound{A} \AgdaBound{B} \AgdaBound{C}\AgdaSymbol{\}} \AgdaSymbol{(}\AgdaBound{m} \AgdaSymbol{:} \AgdaDatatype{Delay} \AgdaPostulate{∞} \AgdaBound{A}\AgdaSymbol{)}\<%
\\
\>[2]\AgdaIndent{30}{}\<[30]%
\>[30]\AgdaSymbol{\{}\AgdaBound{k} \AgdaSymbol{:} \AgdaBound{A} \AgdaSymbol{→} \AgdaDatatype{Delay} \AgdaPostulate{∞} \AgdaBound{B}\AgdaSymbol{\}} \AgdaSymbol{\{}\AgdaBound{l} \AgdaSymbol{:} \AgdaBound{B} \AgdaSymbol{→} \AgdaDatatype{Delay} \AgdaPostulate{∞} \AgdaBound{C}\AgdaSymbol{\}} \AgdaSymbol{→}\<%
\\
\>[2]\AgdaIndent{30}{}\<[30]%
\>[30]\AgdaSymbol{((}\AgdaBound{m} \AgdaFunction{\ensuremath{\mathbin{{>}\mkern-8.5mu{>}\mkern-2mu{=}}}} \AgdaBound{k}\AgdaSymbol{)} \AgdaFunction{\ensuremath{\mathbin{{>}\mkern-8.5mu{>}\mkern-2mu{=}}}} \AgdaBound{l}\AgdaSymbol{)} \AgdaFunction{∼⟨} \AgdaBound{i} \AgdaFunction{⟩∼} \AgdaSymbol{(}\AgdaBound{m} \AgdaFunction{\ensuremath{\mathbin{{>}\mkern-8.5mu{>}\mkern-2mu{=}}}} \AgdaSymbol{λ} \AgdaBound{a} \AgdaSymbol{→} \AgdaSymbol{(}\AgdaBound{k} \AgdaBound{a} \AgdaFunction{\ensuremath{\mathbin{{>}\mkern-8.5mu{>}\mkern-2mu{=}}}} \AgdaBound{l}\AgdaSymbol{))}\<%
\\
\>[0]\AgdaIndent{2}{}\<[2]%
\>[2]\AgdaFunction{bind-assoc} \AgdaSymbol{(}\AgdaInductiveConstructor{now} \AgdaBound{a}\AgdaSymbol{)} \<[27]%
\>[27]\AgdaSymbol{=} \<[30]%
\>[30]\AgdaFunction{∼refl} \AgdaSymbol{\_}\<%
\\
\>[0]\AgdaIndent{2}{}\<[2]%
\>[2]\AgdaFunction{bind-assoc} \AgdaSymbol{(}\AgdaInductiveConstructor{later} \AgdaBound{a∞}\AgdaSymbol{)} \<[27]%
\>[27]\AgdaSymbol{=} \<[30]%
\>[30]\AgdaInductiveConstructor{∼later} \AgdaSymbol{(}\AgdaFunction{∞bind-assoc} \AgdaBound{a∞}\AgdaSymbol{)}\<%
\\
%
\\
\>[0]\AgdaIndent{2}{}\<[2]%
\>[2]\AgdaFunction{∞bind-assoc} \<[27]%
\>[27]\AgdaSymbol{:} \<[30]%
\>[30]\AgdaSymbol{∀\{}\AgdaBound{i} \AgdaBound{A} \AgdaBound{B} \AgdaBound{C}\AgdaSymbol{\}} \AgdaSymbol{(}\AgdaBound{a∞} \AgdaSymbol{:} \AgdaRecord{∞Delay} \AgdaPostulate{∞} \AgdaBound{A}\AgdaSymbol{)}\<%
\\
\>[2]\AgdaIndent{30}{}\<[30]%
\>[30]\AgdaSymbol{\{}\AgdaBound{k} \AgdaSymbol{:} \AgdaBound{A} \AgdaSymbol{→} \AgdaDatatype{Delay} \AgdaPostulate{∞} \AgdaBound{B}\AgdaSymbol{\}} \AgdaSymbol{\{}\AgdaBound{l} \AgdaSymbol{:} \AgdaBound{B} \AgdaSymbol{→} \AgdaDatatype{Delay} \AgdaPostulate{∞} \AgdaBound{C}\AgdaSymbol{\}} \AgdaSymbol{→}\<%
\\
\>[2]\AgdaIndent{30}{}\<[30]%
\>[30]\AgdaSymbol{((}\AgdaBound{a∞} \AgdaFunction{\ensuremath{\mathbin{\infty\mkern-3mu{>}\mkern-8.5mu{>}\mkern-2mu{=}}}} \AgdaBound{k}\AgdaSymbol{)} \AgdaFunction{\ensuremath{\mathbin{\infty\mkern-3mu{>}\mkern-8.5mu{>}\mkern-2mu{=}}}} \AgdaBound{l}\AgdaSymbol{)} \AgdaRecord{∞∼⟨} \AgdaBound{i} \AgdaRecord{⟩∼} \AgdaSymbol{(}\AgdaBound{a∞} \AgdaFunction{\ensuremath{\mathbin{\infty\mkern-3mu{>}\mkern-8.5mu{>}\mkern-2mu{=}}}} \AgdaSymbol{λ} \AgdaBound{a} \AgdaSymbol{→} \AgdaSymbol{(}\AgdaBound{k} \AgdaBound{a} \AgdaFunction{\ensuremath{\mathbin{{>}\mkern-8.5mu{>}\mkern-2mu{=}}}} \AgdaBound{l}\AgdaSymbol{))}\<%
\\
\>[0]\AgdaIndent{2}{}\<[2]%
\>[2]\AgdaField{∼force} \AgdaSymbol{(}\AgdaFunction{∞bind-assoc} \AgdaBound{a∞}\AgdaSymbol{)} \<[27]%
\>[27]\AgdaSymbol{=} \<[30]%
\>[30]\AgdaFunction{bind-assoc} \AgdaSymbol{(}\AgdaField{force} \AgdaBound{a∞}\AgdaSymbol{)}\<%
\end{code}

\noindent
Further, bind (\prebind{} and \preinfbind{}) and is a
congruence in both arguments (proofs omitted here).
\begin{code}%
\>\AgdaFunction{bind-cong-l} \<[14]%
\>[14]\AgdaSymbol{:} \<[17]%
\>[17]\AgdaSymbol{∀\{}\AgdaBound{i} \AgdaBound{A} \AgdaBound{B}\AgdaSymbol{\}\{}\AgdaBound{a?} \AgdaBound{b?} \AgdaSymbol{:} \AgdaDatatype{Delay} \AgdaPostulate{∞} \AgdaBound{A}\AgdaSymbol{\}} \AgdaSymbol{→} \<[48]%
\>[48]\AgdaBound{a?} \AgdaFunction{∼⟨} \AgdaBound{i} \AgdaFunction{⟩∼} \AgdaBound{b?} \AgdaSymbol{→}\<%
\\
\>[2]\AgdaIndent{17}{}\<[17]%
\>[17]\AgdaSymbol{(}\AgdaBound{k} \AgdaSymbol{:} \AgdaBound{A} \AgdaSymbol{→} \AgdaDatatype{Delay} \AgdaPostulate{∞} \AgdaBound{B}\AgdaSymbol{)} \AgdaSymbol{→} \AgdaSymbol{(}\AgdaBound{a?} \AgdaFunction{\ensuremath{\mathbin{{>}\mkern-8.5mu{>}\mkern-2mu{=}}}} \AgdaBound{k}\AgdaSymbol{)} \AgdaFunction{∼⟨} \AgdaBound{i} \AgdaFunction{⟩∼} \AgdaSymbol{(}\AgdaBound{b?} \AgdaFunction{\ensuremath{\mathbin{{>}\mkern-8.5mu{>}\mkern-2mu{=}}}} \AgdaBound{k}\AgdaSymbol{)}\<%
\\
%
\\
\>\AgdaFunction{∞bind-cong-l} \<[14]%
\>[14]\AgdaSymbol{:} \<[17]%
\>[17]\AgdaSymbol{∀\{}\AgdaBound{i} \AgdaBound{A} \AgdaBound{B}\AgdaSymbol{\}\{}\AgdaBound{a∞} \AgdaBound{b∞} \AgdaSymbol{:} \AgdaRecord{∞Delay} \AgdaPostulate{∞} \AgdaBound{A}\AgdaSymbol{\}} \AgdaSymbol{→} \AgdaBound{a∞} \AgdaRecord{∞∼⟨} \AgdaBound{i} \AgdaRecord{⟩∼} \AgdaBound{b∞} \AgdaSymbol{→}\<%
\\
\>[2]\AgdaIndent{17}{}\<[17]%
\>[17]\AgdaSymbol{(}\AgdaBound{k} \AgdaSymbol{:} \AgdaBound{A} \AgdaSymbol{→} \AgdaDatatype{Delay} \AgdaPostulate{∞} \AgdaBound{B}\AgdaSymbol{)} \AgdaSymbol{→} \AgdaSymbol{(}\AgdaBound{a∞} \AgdaFunction{\ensuremath{\mathbin{\infty\mkern-3mu{>}\mkern-8.5mu{>}\mkern-2mu{=}}}} \AgdaBound{k}\AgdaSymbol{)} \AgdaRecord{∞∼⟨} \AgdaBound{i} \AgdaRecord{⟩∼} \AgdaSymbol{(}\AgdaBound{b∞} \AgdaFunction{\ensuremath{\mathbin{\infty\mkern-3mu{>}\mkern-8.5mu{>}\mkern-2mu{=}}}} \AgdaBound{k}\AgdaSymbol{)}\<%
\\
%
\\
\>\AgdaFunction{bind-cong-r} \<[14]%
\>[14]\AgdaSymbol{:} \<[17]%
\>[17]\AgdaSymbol{∀\{}\AgdaBound{i} \AgdaBound{A} \AgdaBound{B}\AgdaSymbol{\}(}\AgdaBound{a?} \AgdaSymbol{:} \AgdaDatatype{Delay} \AgdaPostulate{∞} \AgdaBound{A}\AgdaSymbol{)\{}\AgdaBound{k} \AgdaBound{l} \AgdaSymbol{:} \AgdaBound{A} \AgdaSymbol{→} \AgdaDatatype{Delay} \AgdaPostulate{∞} \AgdaBound{B}\AgdaSymbol{\}} \AgdaSymbol{→}\<%
\\
\>[2]\AgdaIndent{17}{}\<[17]%
\>[17]\AgdaSymbol{(∀} \AgdaBound{a} \AgdaSymbol{→} \AgdaSymbol{(}\AgdaBound{k} \AgdaBound{a}\AgdaSymbol{)} \AgdaFunction{∼⟨} \AgdaBound{i} \AgdaFunction{⟩∼} \AgdaSymbol{(}\AgdaBound{l} \AgdaBound{a}\AgdaSymbol{))} \AgdaSymbol{→} \AgdaSymbol{(}\AgdaBound{a?} \AgdaFunction{\ensuremath{\mathbin{{>}\mkern-8.5mu{>}\mkern-2mu{=}}}} \AgdaBound{k}\AgdaSymbol{)} \AgdaFunction{∼⟨} \AgdaBound{i} \AgdaFunction{⟩∼} \AgdaSymbol{(}\AgdaBound{a?} \AgdaFunction{\ensuremath{\mathbin{{>}\mkern-8.5mu{>}\mkern-2mu{=}}}} \AgdaBound{l}\AgdaSymbol{)}\<%
\\
%
\\
\>\AgdaFunction{∞bind-cong-r} \<[14]%
\>[14]\AgdaSymbol{:} \<[17]%
\>[17]\AgdaSymbol{∀\{}\AgdaBound{i} \AgdaBound{A} \AgdaBound{B}\AgdaSymbol{\}(}\AgdaBound{a∞} \AgdaSymbol{:} \AgdaRecord{∞Delay} \AgdaPostulate{∞} \AgdaBound{A}\AgdaSymbol{)\{}\AgdaBound{k} \AgdaBound{l} \AgdaSymbol{:} \AgdaBound{A} \AgdaSymbol{→} \AgdaDatatype{Delay} \AgdaPostulate{∞} \AgdaBound{B}\AgdaSymbol{\}} \AgdaSymbol{→}\<%
\\
\>[2]\AgdaIndent{17}{}\<[17]%
\>[17]\AgdaSymbol{(∀} \AgdaBound{a} \AgdaSymbol{→} \AgdaSymbol{(}\AgdaBound{k} \AgdaBound{a}\AgdaSymbol{)} \AgdaFunction{∼⟨} \AgdaBound{i} \AgdaFunction{⟩∼} \AgdaSymbol{(}\AgdaBound{l} \AgdaBound{a}\AgdaSymbol{))} \AgdaSymbol{→} \AgdaSymbol{(}\AgdaBound{a∞} \AgdaFunction{\ensuremath{\mathbin{\infty\mkern-3mu{>}\mkern-8.5mu{>}\mkern-2mu{=}}}} \AgdaBound{k}\AgdaSymbol{)} \AgdaRecord{∞∼⟨} \AgdaBound{i} \AgdaRecord{⟩∼} \AgdaSymbol{(}\AgdaBound{a∞} \AgdaFunction{\ensuremath{\mathbin{\infty\mkern-3mu{>}\mkern-8.5mu{>}\mkern-2mu{=}}}} \AgdaBound{l}\AgdaSymbol{)}\<%
\end{code}

\AgdaHide{
\begin{code}%
\>\AgdaFunction{bind-cong-l} \AgdaSymbol{(}\AgdaInductiveConstructor{∼now} \AgdaBound{a}\AgdaSymbol{)} \<[24]%
\>[24]\AgdaBound{k} \AgdaSymbol{=} \AgdaFunction{∼refl} \AgdaSymbol{\_}\<%
\\
\>\AgdaFunction{bind-cong-l} \AgdaSymbol{(}\AgdaInductiveConstructor{∼later} \AgdaBound{eq}\AgdaSymbol{)} \AgdaBound{k} \AgdaSymbol{=} \AgdaInductiveConstructor{∼later} \AgdaSymbol{(}\AgdaFunction{∞bind-cong-l} \AgdaBound{eq} \AgdaBound{k}\AgdaSymbol{)}\<%
\\
%
\\
\>\AgdaField{∼force} \AgdaSymbol{(}\AgdaFunction{∞bind-cong-l} \AgdaBound{eq} \AgdaBound{k}\AgdaSymbol{)} \AgdaSymbol{=} \AgdaFunction{bind-cong-l} \AgdaSymbol{(}\AgdaField{∼force} \AgdaBound{eq}\AgdaSymbol{)} \AgdaBound{k}\<%
\\
%
\\
\>\AgdaFunction{bind-cong-r} \AgdaSymbol{(}\AgdaInductiveConstructor{now} \AgdaBound{a}\AgdaSymbol{)} \<[23]%
\>[23]\AgdaBound{h} \AgdaSymbol{=} \AgdaBound{h} \AgdaBound{a}\<%
\\
\>\AgdaFunction{bind-cong-r} \AgdaSymbol{(}\AgdaInductiveConstructor{later} \AgdaBound{a∞}\AgdaSymbol{)} \AgdaBound{h} \AgdaSymbol{=} \AgdaInductiveConstructor{∼later} \AgdaSymbol{(}\AgdaFunction{∞bind-cong-r} \AgdaBound{a∞} \AgdaBound{h}\AgdaSymbol{)}\<%
\\
%
\\
\>\AgdaField{∼force} \AgdaSymbol{(}\AgdaFunction{∞bind-cong-r} \AgdaBound{a∞} \AgdaBound{h}\AgdaSymbol{)} \AgdaSymbol{=} \AgdaFunction{bind-cong-r} \AgdaSymbol{(}\AgdaField{force} \AgdaBound{a∞}\AgdaSymbol{)} \AgdaBound{h}\<%
\end{code}}

\noindent
As map (\fmap{}) is defined in terms of bind and return, laws for
map are instances of the monad laws:
\begin{code}%
\>\AgdaFunction{map-compose} \<[16]%
\>[16]\AgdaSymbol{:} \<[19]%
\>[19]\AgdaSymbol{∀\{}\AgdaBound{i} \AgdaBound{A} \AgdaBound{B} \AgdaBound{C}\AgdaSymbol{\}} \AgdaSymbol{(}\AgdaBound{a?} \AgdaSymbol{:} \AgdaDatatype{Delay} \AgdaPostulate{∞} \AgdaBound{A}\AgdaSymbol{)} \AgdaSymbol{\{}\AgdaBound{\,f\,} \AgdaSymbol{:} \AgdaBound{A} \AgdaSymbol{→} \AgdaBound{B}\AgdaSymbol{\}} \AgdaSymbol{\{}\AgdaBound{g} \AgdaSymbol{:} \AgdaBound{B} \AgdaSymbol{→} \AgdaBound{C}\AgdaSymbol{\}} \AgdaSymbol{→}\<%
\\
\>[17]\AgdaIndent{19}{}\<[19]%
\>[19]\AgdaSymbol{(}\AgdaBound{g} \AgdaFunction{\ensuremath{{<}\${>}}} \AgdaSymbol{(}\AgdaBound{\,f\,} \AgdaFunction{\ensuremath{{<}\${>}}} \AgdaBound{a?}\AgdaSymbol{))} \AgdaFunction{∼⟨} \AgdaBound{i} \AgdaFunction{⟩∼} \AgdaSymbol{((}\AgdaBound{g} \AgdaFunction{∘} \AgdaBound{\,f\,}\AgdaSymbol{)} \AgdaFunction{\ensuremath{{<}\${>}}} \AgdaBound{a?}\AgdaSymbol{)}\<%
\\
\>\AgdaFunction{map-compose} \AgdaBound{a?} \<[16]%
\>[16]\AgdaSymbol{=} \<[19]%
\>[19]\AgdaFunction{bind-assoc} \AgdaBound{a?}\<%
\\
%
\\
\>\AgdaFunction{map-cong} \<[16]%
\>[16]\AgdaSymbol{:} \<[19]%
\>[19]\AgdaSymbol{∀\{}\AgdaBound{i} \AgdaBound{A} \AgdaBound{B}\AgdaSymbol{\}\{}\AgdaBound{a?} \AgdaBound{b?} \AgdaSymbol{:} \AgdaDatatype{Delay} \AgdaPostulate{∞} \AgdaBound{A}\AgdaSymbol{\}} \AgdaSymbol{(}\AgdaBound{\,f\,} \AgdaSymbol{:} \AgdaBound{A} \AgdaSymbol{→} \AgdaBound{B}\AgdaSymbol{)} \AgdaSymbol{→}\<%
\\
\>[17]\AgdaIndent{19}{}\<[19]%
\>[19]\AgdaBound{a?} \AgdaFunction{∼⟨} \AgdaBound{i} \AgdaFunction{⟩∼} \AgdaBound{b?} \AgdaSymbol{→} \AgdaSymbol{(}\AgdaBound{\,f\,} \AgdaFunction{\ensuremath{{<}\${>}}} \AgdaBound{a?}\AgdaSymbol{)} \AgdaFunction{∼⟨} \AgdaBound{i} \AgdaFunction{⟩∼} \AgdaSymbol{(}\AgdaBound{\,f\,} \AgdaFunction{\ensuremath{{<}\${>}}} \AgdaBound{b?}\AgdaSymbol{)}\<%
\\
\>\AgdaFunction{map-cong} \AgdaBound{\,f\,} \AgdaBound{eq} \<[16]%
\>[16]\AgdaSymbol{=} \<[19]%
\>[19]\AgdaFunction{bind-cong-l} \AgdaBound{eq} \AgdaSymbol{(}\AgdaInductiveConstructor{now} \AgdaFunction{∘} \AgdaBound{\,f\,}\AgdaSymbol{)}\<%
\end{code} 

\subsection{Convergence}

We define convergence as a relation between delayed computations of
type \DelayA{∞} and values of type~\AgdaBound{A}.  If $\AgdaBound{a?}\;\conv\;\AgdaBound{a}$,
then the delayed computation \aq{} eventually yields the value
\AgdaBound{a}. This is a central concept in this paper as we will write a
(productive) normalizer that produces delayed normal forms and then
prove that all such delayed normal forms converge to a value yielding
termination of the normalizer. Notice that convergence is an \emph{inductive}
relation defined on coinductive data.

\begin{code}%
\>\AgdaKeyword{data} \AgdaDatatype{\_⇓\_} \AgdaSymbol{\{}\AgdaBound{A} \AgdaSymbol{:} \AgdaPrimitiveType{Set}\AgdaSymbol{\}} \AgdaSymbol{:} \AgdaSymbol{(}\AgdaBound{a?} \AgdaSymbol{:} \AgdaDatatype{Delay} \AgdaPostulate{∞} \AgdaBound{A}\AgdaSymbol{)} \AgdaSymbol{(}\AgdaBound{a} \AgdaSymbol{:} \AgdaBound{A}\AgdaSymbol{)} \AgdaSymbol{→} \AgdaPrimitiveType{Set} \AgdaKeyword{where}\<%
\\
\>[0]\AgdaIndent{2}{}\<[2]%
\>[2]\AgdaInductiveConstructor{now⇓} \<[10]%
\>[10]\AgdaSymbol{:} \<[13]%
\>[13]\AgdaSymbol{∀\{}\AgdaBound{a}\AgdaSymbol{\}} \<[52]%
\>[52]\AgdaSymbol{→} \AgdaInductiveConstructor{now} \AgdaBound{a} \AgdaDatatype{⇓} \AgdaBound{a}\<%
\\
\>[0]\AgdaIndent{2}{}\<[2]%
\>[2]\AgdaInductiveConstructor{later⇓} \<[10]%
\>[10]\AgdaSymbol{:} \<[13]%
\>[13]\AgdaSymbol{∀\{}\AgdaBound{a}\AgdaSymbol{\}} \AgdaSymbol{\{}\AgdaBound{a∞} \AgdaSymbol{:} \AgdaRecord{∞Delay} \AgdaPostulate{∞} \AgdaBound{A}\AgdaSymbol{\}} \AgdaSymbol{→} \AgdaField{force} \AgdaBound{a∞} \AgdaDatatype{⇓} \AgdaBound{a} \<[52]%
\>[52]\AgdaSymbol{→} \AgdaInductiveConstructor{later} \AgdaBound{a∞} \AgdaDatatype{⇓} \AgdaBound{a}\<%
\\
%
\\
\>\AgdaFunction{\_⇓} \<[5]%
\>[5]\AgdaSymbol{:} \<[8]%
\>[8]\AgdaSymbol{\{}\AgdaBound{A} \AgdaSymbol{:} \AgdaPrimitiveType{Set}\AgdaSymbol{\}} \AgdaSymbol{(}\AgdaBound{x} \AgdaSymbol{:} \AgdaDatatype{Delay} \AgdaPostulate{∞} \AgdaBound{A}\AgdaSymbol{)} \AgdaSymbol{→} \AgdaPrimitiveType{Set}\<%
\\
\>\AgdaBound{x} \AgdaFunction{⇓} \<[5]%
\>[5]\AgdaSymbol{=} \<[8]%
\>[8]\AgdaFunction{∃} \AgdaSymbol{λ} \AgdaBound{a} \AgdaSymbol{→} \AgdaBound{x} \AgdaDatatype{⇓} \AgdaBound{a}\<%
\end{code}

\noindent
We define some useful utilities about convergence: We can map
functions on values over a convergence relation (see \AgdaFunction{map⇓}).
If a delayed computation \aq{} converges
to a value \AgdaBound{a} then so does any strongly bisimilar computation
\aqp{} (see \AgdaFunction{subst∼⇓}).
If we apply a function \AgdaBound{\,f\,} to a delayed value
\aq{} using bind and we know that the delayed value converges to a
value \AgdaBound{a} then we can replace the bind with an ordinary
application $\AgdaBound{\,f\,}\;\AgdaBound{a}$ (see \AgdaFunction{bind⇓}).

\begin{code}%
\>\AgdaFunction{map⇓} \<[9]%
\>[9]\AgdaSymbol{:} \<[12]%
\>[12]\AgdaSymbol{∀\{}\AgdaBound{A} \AgdaBound{B}\AgdaSymbol{\}\{}\AgdaBound{a} \AgdaSymbol{:} \AgdaBound{A}\AgdaSymbol{\}\{}\AgdaBound{a?} \AgdaSymbol{:} \AgdaDatatype{Delay} \AgdaPostulate{∞} \AgdaBound{A}\AgdaSymbol{\}(}\AgdaBound{\,f\,} \AgdaSymbol{:} \AgdaBound{A} \AgdaSymbol{→} \AgdaBound{B}\AgdaSymbol{)} \AgdaSymbol{→} \AgdaBound{a?} \AgdaDatatype{⇓} \AgdaBound{a} \AgdaSymbol{→} \AgdaSymbol{(}\AgdaBound{\,f\,} \AgdaFunction{\ensuremath{{<}\${>}}} \AgdaBound{a?}\AgdaSymbol{)} \AgdaDatatype{⇓} \AgdaBound{\,f\,} \AgdaBound{a}\<%
\\
%
\\
\>\AgdaFunction{subst∼⇓} \<[9]%
\>[9]\AgdaSymbol{:} \<[12]%
\>[12]\AgdaSymbol{∀\{}\AgdaBound{A}\AgdaSymbol{\}\{}\AgdaBound{a?} \AgdaBound{a?′} \AgdaSymbol{:} \AgdaDatatype{Delay} \AgdaPostulate{∞} \AgdaBound{A}\AgdaSymbol{\}\{}\AgdaBound{a} \AgdaSymbol{:} \AgdaBound{A}\AgdaSymbol{\}} \AgdaSymbol{→} \AgdaBound{a?} \AgdaDatatype{⇓} \AgdaBound{a} \AgdaSymbol{→} \AgdaBound{a?} \AgdaDatatype{∼} \AgdaBound{a?′} \AgdaSymbol{→} \AgdaBound{a?′} \AgdaDatatype{⇓} \AgdaBound{a}\<%
\\
%
\\
\>\AgdaFunction{bind⇓} \<[9]%
\>[9]\AgdaSymbol{:} \<[12]%
\>[12]\AgdaSymbol{∀\{}\AgdaBound{A} \AgdaBound{B}\AgdaSymbol{\}(}\AgdaBound{\,f\,} \AgdaSymbol{:} \AgdaBound{A} \AgdaSymbol{→} \AgdaDatatype{Delay} \AgdaPostulate{∞} \AgdaBound{B}\AgdaSymbol{)\{}\AgdaBound{?a} \AgdaSymbol{:} \AgdaDatatype{Delay} \AgdaPostulate{∞} \AgdaBound{A}\AgdaSymbol{\}\{}\AgdaBound{a} \AgdaSymbol{:} \AgdaBound{A}\AgdaSymbol{\}\{}\AgdaBound{b} \AgdaSymbol{:} \AgdaBound{B}\AgdaSymbol{\}} \AgdaSymbol{→}\<%
\\
\>[2]\AgdaIndent{12}{}\<[12]%
\>[12]\AgdaBound{?a} \AgdaDatatype{⇓} \AgdaBound{a} \AgdaSymbol{→} \AgdaBound{\,f\,} \AgdaBound{a} \AgdaDatatype{⇓} \AgdaBound{b} \AgdaSymbol{→} \AgdaSymbol{(}\AgdaBound{?a} \AgdaFunction{\ensuremath{\mathbin{{>}\mkern-8.5mu{>}\mkern-2mu{=}}}} \AgdaBound{\,f\,}\AgdaSymbol{)} \AgdaDatatype{⇓} \AgdaBound{b}\<%
\end{code} 
\AgdaHide{%
\begin{code}%
\>\AgdaFunction{map⇓} \AgdaBound{\,f\,} \AgdaInductiveConstructor{now⇓} \<[19]%
\>[19]\AgdaSymbol{=} \AgdaInductiveConstructor{now⇓}\<%
\\
\>\AgdaFunction{map⇓} \AgdaBound{\,f\,} \AgdaSymbol{(}\AgdaInductiveConstructor{later⇓} \AgdaBound{a⇓}\AgdaSymbol{)} \AgdaSymbol{=} \AgdaInductiveConstructor{later⇓} \AgdaSymbol{(}\AgdaFunction{map⇓} \AgdaBound{\,f\,} \AgdaBound{a⇓}\AgdaSymbol{)}\<%
\\
%
\\
\>\AgdaFunction{subst∼⇓} \AgdaInductiveConstructor{now⇓} \AgdaSymbol{(}\AgdaInductiveConstructor{∼now} \AgdaBound{a}\AgdaSymbol{)} \AgdaSymbol{=} \AgdaInductiveConstructor{now⇓}\<%
\\
\>\AgdaFunction{subst∼⇓} \AgdaSymbol{(}\AgdaInductiveConstructor{later⇓} \AgdaBound{p}\AgdaSymbol{)} \AgdaSymbol{(}\AgdaInductiveConstructor{∼later} \AgdaBound{eq}\AgdaSymbol{)} \AgdaSymbol{=} \AgdaInductiveConstructor{later⇓} \AgdaSymbol{(}\AgdaFunction{subst∼⇓} \AgdaBound{p} \AgdaSymbol{(}\AgdaField{∼force} \AgdaBound{eq}\AgdaSymbol{))}\<%
\\
%
\\
\>\AgdaFunction{bind⇓} \AgdaBound{\,f\,} \AgdaInductiveConstructor{now⇓} \AgdaBound{q} \AgdaSymbol{=} \AgdaBound{q}\<%
\\
\>\AgdaFunction{bind⇓} \AgdaBound{\,f\,} \AgdaSymbol{(}\AgdaInductiveConstructor{later⇓} \AgdaBound{p}\AgdaSymbol{)} \AgdaBound{q} \AgdaSymbol{=} \AgdaInductiveConstructor{later⇓} \AgdaSymbol{(}\AgdaFunction{bind⇓} \AgdaBound{\,f\,} \AgdaBound{p} \AgdaBound{q}\AgdaSymbol{)}\<%
\end{code}}

\noindent
That completes our discussion of the delay infrastructure.

\section{Well-typed terms, values, and coinductive normalization}

\AgdaHide{
\begin{code}%
\>\<%
\\
\>\AgdaKeyword{infixr} \AgdaNumber{6} \_⇒\_\<%
\\
\>\AgdaComment{--infixl 1 \_,\_}\<%
\end{code}}

We present the syntax of the well-typed lambda terms, which is
Altenkirch and Chapman's
\cite{altenkirchChapman:bigStepNormalisation} without explicit
substitutions.
First we
introduce simple types \AgdaDatatype{Ty} with one base type \AgdaInductiveConstructor{★} and function types
$\AgdaBound{a}\;\AgdaInductiveConstructor{⇒}\;\AgdaBound{b}$.

\begin{samepage}
\begin{code}%
\>\AgdaKeyword{data} \AgdaDatatype{Ty} \<[9]%
\>[9]\AgdaSymbol{:} \AgdaPrimitiveType{Set} \AgdaKeyword{where}\<%
\\
\>[0]\AgdaIndent{2}{}\<[2]%
\>[2]\AgdaInductiveConstructor{★} \<[9]%
\>[9]\AgdaSymbol{:} \AgdaDatatype{Ty}\<%
\\
\>[0]\AgdaIndent{2}{}\<[2]%
\>[2]\AgdaInductiveConstructor{\_⇒\_} \<[9]%
\>[9]\AgdaSymbol{:} \AgdaSymbol{(}\AgdaBound{a} \AgdaBound{b} \AgdaSymbol{:} \AgdaDatatype{Ty}\AgdaSymbol{)} \AgdaSymbol{→} \<[25]%
\>[25]\AgdaDatatype{Ty}\<%
\end{code}
\end{samepage}

\noindent
We use de Bruijn indices to
represent variables, so contexts \AgdaDatatype{Cxt} are just lists of (unnamed) types.

\begin{code}%
\>\AgdaKeyword{data} \AgdaDatatype{Cxt} \<[10]%
\>[10]\AgdaSymbol{:} \AgdaPrimitiveType{Set} \AgdaKeyword{where}\<%
\\
\>[0]\AgdaIndent{2}{}\<[2]%
\>[2]\AgdaInductiveConstructor{ε} \<[10]%
\>[10]\AgdaSymbol{:} \AgdaDatatype{Cxt}\<%
\\
\>[0]\AgdaIndent{2}{}\<[2]%
\>[2]\AgdaInductiveConstructor{\_,\_} \<[10]%
\>[10]\AgdaSymbol{:} \AgdaSymbol{(}\AgdaBound{Γ} \AgdaSymbol{:} \AgdaDatatype{Cxt}\AgdaSymbol{)} \AgdaSymbol{(}\AgdaBound{a} \AgdaSymbol{:} \AgdaDatatype{Ty}\AgdaSymbol{)} \AgdaSymbol{→} \<[34]%
\>[34]\AgdaDatatype{Cxt}\<%
\end{code}

\noindent
Variables are de Bruijn indices, just natural numbers. They are
indexed by context and type which guarantees that they are well-scoped
and well-typed. Notice that only non-empty contexts can have variables,
since none of the constructors targets the empty context. The \AgdaInductiveConstructor{zero}th
variable has the same type as the type at the end of the context.

\begin{code}%
\>\AgdaKeyword{data} \AgdaDatatype{Var} \AgdaSymbol{:} \AgdaSymbol{(}\AgdaBound{Γ} \AgdaSymbol{:} \AgdaDatatype{Cxt}\AgdaSymbol{)} \AgdaSymbol{(}\AgdaBound{a} \AgdaSymbol{:} \AgdaDatatype{Ty}\AgdaSymbol{)} \AgdaSymbol{→} \AgdaPrimitiveType{Set} \AgdaKeyword{where}\<%
\\
\>[0]\AgdaIndent{2}{}\<[2]%
\>[2]\AgdaInductiveConstructor{zero} \<[8]%
\>[8]\AgdaSymbol{:} \AgdaSymbol{∀\{}\AgdaBound{Γ} \AgdaBound{a}\AgdaSymbol{\}} \<[34]%
\>[34]\AgdaSymbol{→} \AgdaDatatype{Var} \AgdaSymbol{(}\AgdaBound{Γ} \AgdaInductiveConstructor{,} \AgdaBound{a}\AgdaSymbol{)} \AgdaBound{a}\<%
\\
\>[0]\AgdaIndent{2}{}\<[2]%
\>[2]\AgdaInductiveConstructor{suc} \<[8]%
\>[8]\AgdaSymbol{:} \AgdaSymbol{∀\{}\AgdaBound{Γ} \AgdaBound{a} \AgdaBound{b}\AgdaSymbol{\}} \AgdaSymbol{(}\AgdaBound{x} \AgdaSymbol{:} \AgdaDatatype{Var} \AgdaBound{Γ} \AgdaBound{a}\AgdaSymbol{)} \<[34]%
\>[34]\AgdaSymbol{→} \AgdaDatatype{Var} \AgdaSymbol{(}\AgdaBound{Γ} \AgdaInductiveConstructor{,} \AgdaBound{b}\AgdaSymbol{)} \AgdaBound{a}\<%
\end{code}

\noindent
Terms are also indexed by context and type, guaranteeing
well-typedness and well-scopedness. Terms are either variables, lambda
abstractions, or applications. Notice that the context index in the
body of the lambda tracks that one more variable has been bound. Further,
applications are guaranteed to be well-typed.

\begin{code}%
\>\AgdaKeyword{data} \AgdaDatatype{Tm} \AgdaSymbol{(}\AgdaBound{Γ} \AgdaSymbol{:} \AgdaDatatype{Cxt}\AgdaSymbol{)} \AgdaSymbol{:} \AgdaSymbol{(}\AgdaBound{a} \AgdaSymbol{:} \AgdaDatatype{Ty}\AgdaSymbol{)} \AgdaSymbol{→} \AgdaPrimitiveType{Set} \AgdaKeyword{where}\<%
\\
\>[0]\AgdaIndent{2}{}\<[2]%
\>[2]\AgdaInductiveConstructor{var} \<[7]%
\>[7]\AgdaSymbol{:} \AgdaSymbol{∀\{}\AgdaBound{a}\AgdaSymbol{\}} \<[17]%
\>[17]\AgdaSymbol{(}\AgdaBound{x} \AgdaSymbol{:} \AgdaDatatype{Var} \AgdaBound{Γ} \AgdaBound{a}\AgdaSymbol{)} \<[50]%
\>[50]\AgdaSymbol{→} \AgdaDatatype{Tm} \AgdaBound{Γ} \AgdaBound{a}\<%
\\
\>[0]\AgdaIndent{2}{}\<[2]%
\>[2]\AgdaInductiveConstructor{abs} \<[7]%
\>[7]\AgdaSymbol{:} \AgdaSymbol{∀\{}\AgdaBound{a} \AgdaBound{b}\AgdaSymbol{\}} \<[17]%
\>[17]\AgdaSymbol{(}\AgdaBound{t} \AgdaSymbol{:} \AgdaDatatype{Tm} \AgdaSymbol{(}\AgdaBound{Γ} \AgdaInductiveConstructor{,} \AgdaBound{a}\AgdaSymbol{)} \AgdaBound{b}\AgdaSymbol{)} \<[50]%
\>[50]\AgdaSymbol{→} \AgdaDatatype{Tm} \AgdaBound{Γ} \AgdaSymbol{(}\AgdaBound{a} \AgdaInductiveConstructor{⇒} \AgdaBound{b}\AgdaSymbol{)}\<%
\\
\>[0]\AgdaIndent{2}{}\<[2]%
\>[2]\AgdaInductiveConstructor{app} \<[7]%
\>[7]\AgdaSymbol{:} \AgdaSymbol{∀\{}\AgdaBound{a} \AgdaBound{b}\AgdaSymbol{\}} \<[17]%
\>[17]\AgdaSymbol{(}\AgdaBound{t} \AgdaSymbol{:} \AgdaDatatype{Tm} \AgdaBound{Γ} \AgdaSymbol{(}\AgdaBound{a} \AgdaInductiveConstructor{⇒} \AgdaBound{b}\AgdaSymbol{))} \AgdaSymbol{(}\AgdaBound{u} \AgdaSymbol{:} \AgdaDatatype{Tm} \AgdaBound{Γ} \AgdaBound{a}\AgdaSymbol{)} \<[50]%
\>[50]\AgdaSymbol{→} \AgdaDatatype{Tm} \AgdaBound{Γ} \AgdaBound{b}\<%
\end{code}

\noindent
We introduce neutral terms, parametric in the argument type \AgdaBound{Ξ} of
application as we will need both neutral weak-head normal and beta-eta
normal forms. Intuitively, neutrals are \emph{stuck}.
In plain lambda calculus, they are either
variables, or applications that cannot compute as there is a neutral
term in the function position.

\begin{code}%
\>\AgdaKeyword{data} \AgdaDatatype{Ne} \AgdaSymbol{(}\AgdaBound{Ξ} \AgdaSymbol{:} \AgdaDatatype{Cxt} \AgdaSymbol{→} \AgdaDatatype{Ty} \AgdaSymbol{→} \AgdaPrimitiveType{Set}\AgdaSymbol{)(}\AgdaBound{Γ} \AgdaSymbol{:} \AgdaDatatype{Cxt}\AgdaSymbol{)} \AgdaSymbol{:} \AgdaDatatype{Ty} \AgdaSymbol{→} \AgdaPrimitiveType{Set} \AgdaKeyword{where}\<%
\\
\>[0]\AgdaIndent{2}{}\<[2]%
\>[2]\AgdaInductiveConstructor{var} \<[7]%
\>[7]\AgdaSymbol{:} \AgdaSymbol{∀\{}\AgdaBound{a}\AgdaSymbol{\}} \<[17]%
\>[17]\AgdaSymbol{→} \AgdaDatatype{Var} \AgdaBound{Γ} \AgdaBound{a} \<[43]%
\>[43]\AgdaSymbol{→} \AgdaDatatype{Ne} \AgdaBound{Ξ} \AgdaBound{Γ} \AgdaBound{a}\<%
\\
\>[0]\AgdaIndent{2}{}\<[2]%
\>[2]\AgdaInductiveConstructor{app} \<[7]%
\>[7]\AgdaSymbol{:} \AgdaSymbol{∀\{}\AgdaBound{a} \AgdaBound{b}\AgdaSymbol{\}} \<[17]%
\>[17]\AgdaSymbol{→} \AgdaDatatype{Ne} \AgdaBound{Ξ} \AgdaBound{Γ} \AgdaSymbol{(}\AgdaBound{a} \AgdaInductiveConstructor{⇒} \AgdaBound{b}\AgdaSymbol{)} \AgdaSymbol{→} \AgdaBound{Ξ} \AgdaBound{Γ} \AgdaBound{a} \<[43]%
\>[43]\AgdaSymbol{→} \AgdaDatatype{Ne} \AgdaBound{Ξ} \AgdaBound{Γ} \AgdaBound{b}\<%
\end{code}

\noindent
Weak head normal forms (\AgdaDatatype{Val}ues) are either neutral terms or
closures of a body of a lambda and an environment containing values
for the all the variables except the lambda bound variable. Once a
value for the lambda bound variable is available the body of the
lambda may be evaluated in the now complete environment. \AgdaDatatype{Val}ues
are defined mutually with \AgdaDatatype{Env}ironments which are just lists of
values. We also provide a \AgdaFunction{lookup} function that looks
variables up in the environment. Notice that the typing ensures that
\AgdaFunction{lookup} never tries to access a variable that is out of scope
and, indeed, never encounters an empty environment as no variables can
exist there.

\begin{samepage}
\begin{code}%
\>\AgdaKeyword{mutual}\<%
\\
\>[0]\AgdaIndent{2}{}\<[2]%
\>[2]\AgdaKeyword{data} \AgdaDatatype{Val} \AgdaSymbol{(}\AgdaBound{Δ} \AgdaSymbol{:} \AgdaDatatype{Cxt}\AgdaSymbol{)} \AgdaSymbol{:} \AgdaSymbol{(}\AgdaBound{a} \AgdaSymbol{:} \AgdaDatatype{Ty}\AgdaSymbol{)} \AgdaSymbol{→} \AgdaPrimitiveType{Set} \AgdaKeyword{where}\<%
\\
\>[2]\AgdaIndent{4}{}\<[4]%
\>[4]\AgdaInductiveConstructor{ne} \<[9]%
\>[9]\AgdaSymbol{:} \AgdaSymbol{∀\{}\AgdaBound{a}\AgdaSymbol{\}} \<[21]%
\>[21]\AgdaSymbol{(}\AgdaBound{w} \AgdaSymbol{:} \AgdaDatatype{Ne} \AgdaDatatype{Val} \AgdaBound{Δ} \AgdaBound{a}\AgdaSymbol{)} \<[55]%
\>[55]\AgdaSymbol{→} \AgdaDatatype{Val} \AgdaBound{Δ} \AgdaBound{a}\<%
\\
\>[2]\AgdaIndent{4}{}\<[4]%
\>[4]\AgdaInductiveConstructor{lam} \<[9]%
\>[9]\AgdaSymbol{:} \AgdaSymbol{∀\{}\AgdaBound{Γ} \AgdaBound{a} \AgdaBound{b}\AgdaSymbol{\}} \<[21]%
\>[21]\AgdaSymbol{(}\AgdaBound{t} \AgdaSymbol{:} \AgdaDatatype{Tm} \AgdaSymbol{(}\AgdaBound{Γ} \AgdaInductiveConstructor{,} \AgdaBound{a}\AgdaSymbol{)} \AgdaBound{b}\AgdaSymbol{)} \AgdaSymbol{(}\AgdaBound{ρ} \AgdaSymbol{:} \AgdaDatatype{Env} \AgdaBound{Δ} \AgdaBound{Γ}\AgdaSymbol{)} \<[55]%
\>[55]\AgdaSymbol{→} \AgdaDatatype{Val} \AgdaBound{Δ} \AgdaSymbol{(}\AgdaBound{a} \AgdaInductiveConstructor{⇒} \AgdaBound{b}\AgdaSymbol{)}\<%
\\
%
\\
\>[0]\AgdaIndent{2}{}\<[2]%
\>[2]\AgdaKeyword{data} \AgdaDatatype{Env} \AgdaSymbol{(}\AgdaBound{Δ} \AgdaSymbol{:} \AgdaDatatype{Cxt}\AgdaSymbol{)} \AgdaSymbol{:} \AgdaSymbol{(}\AgdaBound{Γ} \AgdaSymbol{:} \AgdaDatatype{Cxt}\AgdaSymbol{)} \AgdaSymbol{→} \AgdaPrimitiveType{Set} \AgdaKeyword{where}\<%
\\
\>[2]\AgdaIndent{4}{}\<[4]%
\>[4]\AgdaInductiveConstructor{ε} \<[9]%
\>[9]\AgdaSymbol{:} \<[50]%
\>[50]\AgdaDatatype{Env} \AgdaBound{Δ} \AgdaInductiveConstructor{ε}\<%
\\
\>[2]\AgdaIndent{4}{}\<[4]%
\>[4]\AgdaInductiveConstructor{\_,\_} \<[9]%
\>[9]\AgdaSymbol{:} \AgdaSymbol{∀} \AgdaSymbol{\{}\AgdaBound{Γ} \AgdaBound{a}\AgdaSymbol{\}} \AgdaSymbol{(}\AgdaBound{ρ} \AgdaSymbol{:} \AgdaDatatype{Env} \AgdaBound{Δ} \AgdaBound{Γ}\AgdaSymbol{)} \AgdaSymbol{(}\AgdaBound{v} \AgdaSymbol{:} \AgdaDatatype{Val} \AgdaBound{Δ} \AgdaBound{a}\AgdaSymbol{)} \AgdaSymbol{→} \<[50]%
\>[50]\AgdaDatatype{Env} \AgdaBound{Δ} \AgdaSymbol{(}\AgdaBound{Γ} \AgdaInductiveConstructor{,} \AgdaBound{a}\AgdaSymbol{)}\<%
\end{code}
\end{samepage}

\begin{code}%
\>\AgdaFunction{lookup} \<[25]%
\>[25]\AgdaSymbol{:} \<[28]%
\>[28]\AgdaSymbol{∀} \AgdaSymbol{\{}\AgdaBound{Γ} \AgdaBound{Δ} \AgdaBound{a}\AgdaSymbol{\}} \AgdaSymbol{→} \AgdaDatatype{Var} \AgdaBound{Γ} \AgdaBound{a} \AgdaSymbol{→} \AgdaDatatype{Env} \AgdaBound{Δ} \AgdaBound{Γ} \AgdaSymbol{→} \AgdaDatatype{Val} \AgdaBound{Δ} \AgdaBound{a}\<%
\\
\>\AgdaFunction{lookup} \AgdaInductiveConstructor{zero} \<[16]%
\>[16]\AgdaSymbol{(}\AgdaBound{ρ} \AgdaInductiveConstructor{,} \AgdaBound{v}\AgdaSymbol{)} \<[25]%
\>[25]\AgdaSymbol{=} \<[28]%
\>[28]\AgdaBound{v}\<%
\\
\>\AgdaFunction{lookup} \AgdaSymbol{(}\AgdaInductiveConstructor{suc} \AgdaBound{x}\AgdaSymbol{)} \<[16]%
\>[16]\AgdaSymbol{(}\AgdaBound{ρ} \AgdaInductiveConstructor{,} \AgdaBound{v}\AgdaSymbol{)} \<[25]%
\>[25]\AgdaSymbol{=} \<[28]%
\>[28]\AgdaFunction{lookup} \AgdaBound{x} \AgdaBound{ρ}\<%
\end{code}

\noindent Evaluation \AgdaFunction{eval} takes a term and a suitable
environment and returns a delayed value.  It is defined mutually with
an \AgdaFunction{apply} function that applies function values to
argument values, and a function \AgdaFunction{beta} that reduces a
β-redex, i.e., a closure applied to a value.  While
\AgdaFunction{eval} and \AgdaFunction{beta} are recursively invoked
only on subterms, \AgdaFunction{apply} is called with arguments
\AgdaBound{\,f\,} and \AgdaBound{v} which are results of evaluating terms
\AgdaBound{t} and \AgdaBound{u} and not structurally smaller than the
arguments of caller \AgdaFunction{eval}.  Thus, the three functions are
not defined by structural induction but by mutual \emph{coinduction}.

\begin{code}%
\>\AgdaFunction{eval} \<[7]%
\>[7]\AgdaSymbol{:} \<[10]%
\>[10]\AgdaSymbol{∀\{}\AgdaBound{i} \AgdaBound{Γ} \AgdaBound{Δ} \AgdaBound{b}\AgdaSymbol{\}} \<[24]%
\>[24]\AgdaSymbol{→} \AgdaDatatype{Tm} \AgdaBound{Γ} \AgdaBound{b} \<[41]%
\>[41]\AgdaSymbol{→} \AgdaDatatype{Env} \AgdaBound{Δ} \AgdaBound{Γ} \<[63]%
\>[63]\AgdaSymbol{→} \AgdaDatatype{Delay} \AgdaBound{i} \AgdaSymbol{(}\AgdaDatatype{Val} \AgdaBound{Δ} \AgdaBound{b}\AgdaSymbol{)}\<%
\\
\>\AgdaFunction{apply} \<[7]%
\>[7]\AgdaSymbol{:} \<[10]%
\>[10]\AgdaSymbol{∀\{}\AgdaBound{i} \AgdaBound{Δ} \AgdaBound{a} \AgdaBound{b}\AgdaSymbol{\}} \<[24]%
\>[24]\AgdaSymbol{→} \AgdaDatatype{Val} \AgdaBound{Δ} \AgdaSymbol{(}\AgdaBound{a} \AgdaInductiveConstructor{⇒} \AgdaBound{b}\AgdaSymbol{)} \<[52]%
\>[52]\AgdaSymbol{→} \AgdaDatatype{Val} \AgdaBound{Δ} \AgdaBound{a} \<[63]%
\>[63]\AgdaSymbol{→} \AgdaDatatype{Delay} \AgdaBound{i} \AgdaSymbol{(}\AgdaDatatype{Val} \AgdaBound{Δ} \AgdaBound{b}\AgdaSymbol{)}\<%
\\
\>\AgdaFunction{beta} \<[7]%
\>[7]\AgdaSymbol{:} \<[10]%
\>[10]\AgdaSymbol{∀\{}\AgdaBound{i} \AgdaBound{Γ} \AgdaBound{Δ} \AgdaBound{a} \AgdaBound{b}\AgdaSymbol{\}} \<[24]%
\>[24]\AgdaSymbol{→} \AgdaDatatype{Tm} \AgdaSymbol{(}\AgdaBound{Γ} \AgdaInductiveConstructor{,} \AgdaBound{a}\AgdaSymbol{)} \AgdaBound{b} \<[41]%
\>[41]\AgdaSymbol{→} \AgdaDatatype{Env} \AgdaBound{Δ} \AgdaBound{Γ} \<[52]%
\>[52]\AgdaSymbol{→} \AgdaDatatype{Val} \AgdaBound{Δ} \AgdaBound{a} \<[63]%
\>[63]\AgdaSymbol{→} \AgdaRecord{∞Delay} \AgdaBound{i} \AgdaSymbol{(}\AgdaDatatype{Val} \AgdaBound{Δ} \AgdaBound{b}\AgdaSymbol{)}\<%
\\
%
\\
\>\AgdaFunction{eval} \AgdaSymbol{(}\AgdaInductiveConstructor{var} \AgdaBound{x}\AgdaSymbol{)} \<[17]%
\>[17]\AgdaBound{ρ} \<[21]%
\>[21]\AgdaSymbol{=} \AgdaInductiveConstructor{now} \AgdaSymbol{(}\AgdaFunction{lookup} \AgdaBound{x} \AgdaBound{ρ}\AgdaSymbol{)}\<%
\\
\>\AgdaFunction{eval} \AgdaSymbol{(}\AgdaInductiveConstructor{abs} \AgdaBound{t}\AgdaSymbol{)} \<[17]%
\>[17]\AgdaBound{ρ} \<[21]%
\>[21]\AgdaSymbol{=} \AgdaInductiveConstructor{now} \AgdaSymbol{(}\AgdaInductiveConstructor{lam} \AgdaBound{t} \AgdaBound{ρ}\AgdaSymbol{)}\<%
\\
\>\AgdaFunction{eval} \AgdaSymbol{(}\AgdaInductiveConstructor{app} \AgdaBound{t} \AgdaBound{u}\AgdaSymbol{)} \<[17]%
\>[17]\AgdaBound{ρ} \<[21]%
\>[21]\AgdaSymbol{=} \AgdaFunction{eval} \AgdaBound{t} \AgdaBound{ρ} \AgdaFunction{\ensuremath{\mathbin{{>}\mkern-8.5mu{>}\mkern-2mu{=}}}} \AgdaSymbol{λ} \AgdaBound{\,f\,} \AgdaSymbol{→} \AgdaFunction{eval} \AgdaBound{u} \AgdaBound{ρ} \AgdaFunction{\ensuremath{\mathbin{{>}\mkern-8.5mu{>}\mkern-2mu{=}}}} \AgdaSymbol{λ} \AgdaBound{v} \AgdaSymbol{→} \AgdaFunction{apply} \AgdaBound{\,f\,} \AgdaBound{v}\<%
\\
%
\\
\>\AgdaFunction{apply} \AgdaSymbol{(}\AgdaInductiveConstructor{ne} \AgdaBound{w}\AgdaSymbol{)} \<[17]%
\>[17]\AgdaBound{v} \<[21]%
\>[21]\AgdaSymbol{=} \AgdaInductiveConstructor{now} \AgdaSymbol{(}\AgdaInductiveConstructor{ne} \AgdaSymbol{(}\AgdaInductiveConstructor{app} \AgdaBound{w} \AgdaBound{v}\AgdaSymbol{))}\<%
\\
\>\AgdaFunction{apply} \AgdaSymbol{(}\AgdaInductiveConstructor{lam} \AgdaBound{t} \AgdaBound{ρ}\AgdaSymbol{)} \<[17]%
\>[17]\AgdaBound{v} \<[21]%
\>[21]\AgdaSymbol{=} \AgdaInductiveConstructor{later} \AgdaSymbol{(}\AgdaFunction{beta} \AgdaBound{t} \AgdaBound{ρ} \AgdaBound{v}\AgdaSymbol{)}\<%
\\
%
\\
\>\AgdaField{force} \AgdaSymbol{(}\AgdaFunction{beta} \AgdaBound{t} \AgdaBound{ρ} \AgdaBound{v}\AgdaSymbol{)} \<[21]%
\>[21]\AgdaSymbol{=} \AgdaFunction{eval} \AgdaBound{t} \AgdaSymbol{(}\AgdaBound{ρ} \AgdaInductiveConstructor{,} \AgdaBound{v}\AgdaSymbol{)}\<%
\end{code}

\noindent
To justify the coinductive definition, the recursive calls must be
\emph{guarded}.  Immediately guarded is only \AgdaFunction{beta} which
only unfolds if \AgdaField{force}d.  The \AgdaFunction{apply} function
only calls
\AgdaFunction{beta}, and this call is under constructor
\AgdaInductiveConstructor{later}, \ie, not under any elimination,
thus, the code for \AgdaFunction{apply} is also not endangering
productivity.   Yet \AgdaFunction{eval} makes three recursive calls as
arguments to the elimination \prebind{}, violating the syntactic
guardedness condition \cite{coquand:infiniteobjects,gimenez:guardeddefinitions}
as implemented, e.g., in Coq~\cite{inria:coq84} and previous Agda~\cite{altenkirchDanielsson:par10}.  Sized types come to the
rescue here!  The typing of bind
\begin{center}
\AgdaCodeStyle
\begin{ptboxed}
\>[4]\prebind{} \<[23]%
\>[23]\AgdaSymbol{:} \<[26]%
\>[26]\AgdaSymbol{∀} \AgdaSymbol{\{}\AgdaBound{i} \AgdaBound{A} \AgdaBound{B}\AgdaSymbol{\}} \AgdaSymbol{→} \AgdaDatatype{Delay} \AgdaBound{i} \AgdaBound{A} \AgdaSymbol{→} \AgdaSymbol{(}\AgdaBound{A} \AgdaSymbol{→} \AgdaDatatype{Delay} \AgdaBound{i} \AgdaBound{B}\AgdaSymbol{)} \AgdaSymbol{→} \AgdaDatatype{Delay} \AgdaBound{i} \AgdaBound{B}\<%
\end{ptboxed}  
\end{center}
guarantees that its two arguments are observed no deeper than its
result; thus, guardedness is not destroyed by a use of bind.  Finally,
\AgdaFunction{eval} calls itself only on subterms, thus, these two
recursive calls, while not guarded by explicit delays,  
can be justified by a local structural induction on
\AgdaDatatype{Tm}.  Agda's termination checker is able to recognize
lexicographic termination measures \cite{abelAltenkirch:jfp02}, in this case
it is a lexicographic recursion first on observation depth in the
\AgdaDatatype{Delay} monad and second on the height of
\AgdaDatatype{Tm} trees.

Beta-eta normal forms are either of function type, in which case they
must be a lambda term, or of base type, in which case they must be a
neutral term, meaning, a variable applied to normal forms.

\begin{code}%
\>\AgdaKeyword{data} \AgdaDatatype{Nf} \AgdaSymbol{(}\AgdaBound{Γ} \AgdaSymbol{:} \AgdaDatatype{Cxt}\AgdaSymbol{)} \AgdaSymbol{:} \AgdaDatatype{Ty} \AgdaSymbol{→} \AgdaPrimitiveType{Set} \AgdaKeyword{where}\<%
\\
\>[0]\AgdaIndent{2}{}\<[2]%
\>[2]\AgdaInductiveConstructor{lam} \<[7]%
\>[7]\AgdaSymbol{:} \AgdaSymbol{∀\{}\AgdaBound{a} \AgdaBound{b}\AgdaSymbol{\}} \<[17]%
\>[17]\AgdaSymbol{(}\AgdaBound{n} \<[21]%
\>[21]\AgdaSymbol{:} \AgdaDatatype{Nf} \AgdaSymbol{(}\AgdaBound{Γ} \AgdaInductiveConstructor{,} \AgdaBound{a}\AgdaSymbol{)} \AgdaBound{b}\AgdaSymbol{)} \<[38]%
\>[38]\AgdaSymbol{→} \AgdaDatatype{Nf} \AgdaBound{Γ} \AgdaSymbol{(}\AgdaBound{a} \AgdaInductiveConstructor{⇒} \AgdaBound{b}\AgdaSymbol{)}\<%
\\
\>[0]\AgdaIndent{2}{}\<[2]%
\>[2]\AgdaInductiveConstructor{ne} \<[7]%
\>[7]\AgdaSymbol{:} \<[17]%
\>[17]\AgdaSymbol{(}\AgdaBound{m} \<[21]%
\>[21]\AgdaSymbol{:} \AgdaDatatype{Ne} \AgdaDatatype{Nf} \AgdaBound{Γ} \AgdaInductiveConstructor{★}\AgdaSymbol{)} \<[38]%
\>[38]\AgdaSymbol{→} \AgdaDatatype{Nf} \AgdaBound{Γ} \AgdaInductiveConstructor{★}\<%
\end{code}

\noindent
To turn values into normal forms we must be able to apply functional
values to fresh variables.  We need an operation on values that
introduces a fresh variable into the context:

\begin{code}%
\>\AgdaFunction{weakVal} \AgdaSymbol{:} \AgdaSymbol{∀\{}\AgdaBound{Δ} \AgdaBound{a} \AgdaBound{c}\AgdaSymbol{\}} \AgdaSymbol{→} \AgdaDatatype{Val} \AgdaBound{Δ} \AgdaBound{c} \AgdaSymbol{→} \AgdaDatatype{Val} \AgdaSymbol{(}\AgdaBound{Δ} \AgdaInductiveConstructor{,} \AgdaBound{a}\AgdaSymbol{)} \AgdaBound{c}\<%
\end{code}

\noindent
We take the approach of implementing this operation using so-called
order preserving embeddings (OPEs) which represent weakenings in
arbitrary positions in the context. Order preserving embeddings can be
represented in a first order way which simplifies reasoning about
them.

\begin{samepage}
\begin{code}%
\>\AgdaKeyword{data} \AgdaDatatype{\_≤\_} \AgdaSymbol{:} \AgdaSymbol{(}\AgdaBound{Γ} \AgdaBound{Δ} \AgdaSymbol{:} \AgdaDatatype{Cxt}\AgdaSymbol{)} \AgdaSymbol{→} \AgdaPrimitiveType{Set} \AgdaKeyword{where}\<%
\\
\>[0]\AgdaIndent{2}{}\<[2]%
\>[2]\AgdaInductiveConstructor{id} \<[8]%
\>[8]\AgdaSymbol{:} \AgdaSymbol{∀\{}\AgdaBound{Γ}\AgdaSymbol{\}} \<[28]%
\>[28]\AgdaSymbol{→} \AgdaBound{Γ} \AgdaDatatype{≤} \AgdaBound{Γ}\<%
\\
\>[0]\AgdaIndent{2}{}\<[2]%
\>[2]\AgdaInductiveConstructor{weak} \<[8]%
\>[8]\AgdaSymbol{:} \AgdaSymbol{∀\{}\AgdaBound{Γ} \AgdaBound{Δ} \AgdaBound{a}\AgdaSymbol{\}} \AgdaSymbol{→} \AgdaBound{Γ} \AgdaDatatype{≤} \AgdaBound{Δ} \<[28]%
\>[28]\AgdaSymbol{→} \AgdaSymbol{(}\AgdaBound{Γ} \AgdaInductiveConstructor{,} \AgdaBound{a}\AgdaSymbol{)} \AgdaDatatype{≤} \AgdaBound{Δ}\<%
\\
\>[0]\AgdaIndent{2}{}\<[2]%
\>[2]\AgdaInductiveConstructor{lift} \<[8]%
\>[8]\AgdaSymbol{:} \AgdaSymbol{∀\{}\AgdaBound{Γ} \AgdaBound{Δ} \AgdaBound{a}\AgdaSymbol{\}} \AgdaSymbol{→} \AgdaBound{Γ} \AgdaDatatype{≤} \AgdaBound{Δ} \<[28]%
\>[28]\AgdaSymbol{→} \AgdaSymbol{(}\AgdaBound{Γ} \AgdaInductiveConstructor{,} \AgdaBound{a}\AgdaSymbol{)} \AgdaDatatype{≤} \AgdaSymbol{(}\AgdaBound{Δ} \AgdaInductiveConstructor{,} \AgdaBound{a}\AgdaSymbol{)}\<%
\end{code}
\end{samepage}

\noindent
We implement composition of OPEs and prove that \AgdaFunction{id} is the right
unit of composition (proof suppressed). The left unit property holds
definitionally. We could additionally prove associativity and observe
that OPEs form a category but this is not required in this paper.

\begin{code}%
\>\AgdaFunction{\_•\_} \<[19]%
\>[19]\AgdaSymbol{:} \<[22]%
\>[22]\AgdaSymbol{∀} \AgdaSymbol{\{}\AgdaBound{Γ} \AgdaBound{Δ} \AgdaBound{Δ\kern-1pt′}\AgdaSymbol{\}} \AgdaSymbol{(}\AgdaBound{η} \AgdaSymbol{:} \AgdaBound{Γ} \AgdaDatatype{≤} \AgdaBound{Δ}\AgdaSymbol{)} \AgdaSymbol{(}\AgdaBound{η′} \AgdaSymbol{:} \AgdaBound{Δ} \AgdaDatatype{≤} \AgdaBound{Δ\kern-1pt′}\AgdaSymbol{)} \AgdaSymbol{→} \AgdaBound{Γ} \AgdaDatatype{≤} \AgdaBound{Δ\kern-1pt′}\<%
\\
\>\AgdaInductiveConstructor{id} \<[8]%
\>[8]\AgdaFunction{•} \AgdaBound{η′} \<[19]%
\>[19]\AgdaSymbol{=} \<[22]%
\>[22]\AgdaBound{η′}\<%
\\
\>\AgdaInductiveConstructor{weak} \AgdaBound{η} \<[8]%
\>[8]\AgdaFunction{•} \AgdaBound{η′} \<[19]%
\>[19]\AgdaSymbol{=} \<[22]%
\>[22]\AgdaInductiveConstructor{weak} \<[28]%
\>[28]\AgdaSymbol{(}\AgdaBound{η} \AgdaFunction{•} \AgdaBound{η′}\AgdaSymbol{)}\<%
\\
\>\AgdaInductiveConstructor{lift} \AgdaBound{η} \<[8]%
\>[8]\AgdaFunction{•} \AgdaInductiveConstructor{id} \<[19]%
\>[19]\AgdaSymbol{=} \<[22]%
\>[22]\AgdaInductiveConstructor{lift} \<[28]%
\>[28]\AgdaBound{η}\<%
\\
\>\AgdaInductiveConstructor{lift} \AgdaBound{η} \<[8]%
\>[8]\AgdaFunction{•} \AgdaInductiveConstructor{weak} \AgdaBound{η′} \<[19]%
\>[19]\AgdaSymbol{=} \<[22]%
\>[22]\AgdaInductiveConstructor{weak} \<[28]%
\>[28]\AgdaSymbol{(}\AgdaBound{η} \AgdaFunction{•} \AgdaBound{η′}\AgdaSymbol{)}\<%
\\
\>\AgdaInductiveConstructor{lift} \AgdaBound{η} \<[8]%
\>[8]\AgdaFunction{•} \AgdaInductiveConstructor{lift} \AgdaBound{η′} \<[19]%
\>[19]\AgdaSymbol{=} \<[22]%
\>[22]\AgdaInductiveConstructor{lift} \<[28]%
\>[28]\AgdaSymbol{(}\AgdaBound{η} \AgdaFunction{•} \AgdaBound{η′}\AgdaSymbol{)}\<%
\\
%
\\
\>\AgdaFunction{η•id} \<[19]%
\>[19]\AgdaSymbol{:} \<[22]%
\>[22]\AgdaSymbol{∀} \AgdaSymbol{\{}\AgdaBound{Γ} \AgdaBound{Δ}\AgdaSymbol{\}} \AgdaSymbol{(}\AgdaBound{η} \AgdaSymbol{:} \AgdaBound{Γ} \AgdaDatatype{≤} \AgdaBound{Δ}\AgdaSymbol{)} \AgdaSymbol{→} \AgdaBound{η} \AgdaFunction{•} \AgdaInductiveConstructor{id} \AgdaDatatype{≡} \AgdaBound{η}\<%
\end{code}

\AgdaHide{
\begin{code}%
\>\AgdaFunction{η•id} \AgdaInductiveConstructor{id} \<[14]%
\>[14]\AgdaSymbol{=} \AgdaInductiveConstructor{refl}\<%
\\
\>\AgdaFunction{η•id} \AgdaSymbol{(}\AgdaInductiveConstructor{weak} \AgdaBound{η}\AgdaSymbol{)} \AgdaSymbol{=} \AgdaFunction{cong} \AgdaInductiveConstructor{weak} \AgdaSymbol{(}\AgdaFunction{η•id} \AgdaBound{η}\AgdaSymbol{)}\<%
\\
\>\AgdaFunction{η•id} \AgdaSymbol{(}\AgdaInductiveConstructor{lift} \AgdaBound{η}\AgdaSymbol{)} \AgdaSymbol{=} \AgdaInductiveConstructor{refl}\<%
\end{code}}

\noindent
We define a map operation that weakens variables, values,
environments, normal forms and neutral terms by OPEs.

\begin{code}%
\>\AgdaFunction{var≤} \<[6]%
\>[6]\AgdaSymbol{:} \AgdaSymbol{∀\{}\AgdaBound{Γ} \AgdaBound{Δ}\AgdaSymbol{\}} \AgdaSymbol{→} \AgdaBound{Γ} \AgdaDatatype{≤} \AgdaBound{Δ} \AgdaSymbol{→} \AgdaSymbol{∀\{}\AgdaBound{a}\AgdaSymbol{\}} \<[31]%
\>[31]\AgdaSymbol{→} \AgdaDatatype{Var} \AgdaBound{Δ} \AgdaBound{a} \<[45]%
\>[45]\AgdaSymbol{→} \AgdaDatatype{Var} \AgdaBound{Γ} \AgdaBound{a}\<%
\\
\>\AgdaFunction{val≤} \<[6]%
\>[6]\AgdaSymbol{:} \AgdaSymbol{∀\{}\AgdaBound{Γ} \AgdaBound{Δ}\AgdaSymbol{\}} \AgdaSymbol{→} \AgdaBound{Γ} \AgdaDatatype{≤} \AgdaBound{Δ} \AgdaSymbol{→} \AgdaSymbol{∀\{}\AgdaBound{a}\AgdaSymbol{\}} \<[31]%
\>[31]\AgdaSymbol{→} \AgdaDatatype{Val} \AgdaBound{Δ} \AgdaBound{a} \<[45]%
\>[45]\AgdaSymbol{→} \AgdaDatatype{Val} \AgdaBound{Γ} \AgdaBound{a}\<%
\\
\>\AgdaFunction{env≤} \<[6]%
\>[6]\AgdaSymbol{:} \AgdaSymbol{∀\{}\AgdaBound{Γ} \AgdaBound{Δ}\AgdaSymbol{\}} \AgdaSymbol{→} \AgdaBound{Γ} \AgdaDatatype{≤} \AgdaBound{Δ} \AgdaSymbol{→} \AgdaSymbol{∀\{}\AgdaBound{E}\AgdaSymbol{\}} \<[31]%
\>[31]\AgdaSymbol{→} \AgdaDatatype{Env} \AgdaBound{Δ} \AgdaBound{E} \<[45]%
\>[45]\AgdaSymbol{→} \AgdaDatatype{Env} \AgdaBound{Γ} \AgdaBound{E}\<%
\\
\>\AgdaFunction{nev≤} \<[6]%
\>[6]\AgdaSymbol{:} \AgdaSymbol{∀\{}\AgdaBound{Γ} \AgdaBound{Δ}\AgdaSymbol{\}} \AgdaSymbol{→} \AgdaBound{Γ} \AgdaDatatype{≤} \AgdaBound{Δ} \AgdaSymbol{→} \AgdaSymbol{∀\{}\AgdaBound{a}\AgdaSymbol{\}} \<[31]%
\>[31]\AgdaSymbol{→} \AgdaDatatype{Ne} \AgdaDatatype{Val} \AgdaBound{Δ} \AgdaBound{a} \<[45]%
\>[45]\AgdaSymbol{→} \AgdaDatatype{Ne} \AgdaDatatype{Val} \AgdaBound{Γ} \AgdaBound{a}\<%
\\
\>\AgdaFunction{nf≤} \<[6]%
\>[6]\AgdaSymbol{:} \AgdaSymbol{∀\{}\AgdaBound{Γ} \AgdaBound{Δ}\AgdaSymbol{\}} \AgdaSymbol{→} \AgdaBound{Γ} \AgdaDatatype{≤} \AgdaBound{Δ} \AgdaSymbol{→} \AgdaSymbol{∀\{}\AgdaBound{a}\AgdaSymbol{\}} \<[31]%
\>[31]\AgdaSymbol{→} \AgdaDatatype{Nf} \AgdaBound{Δ} \AgdaBound{a} \<[45]%
\>[45]\AgdaSymbol{→} \AgdaDatatype{Nf} \AgdaBound{Γ} \AgdaBound{a}\<%
\\
\>\AgdaFunction{nen≤} \<[6]%
\>[6]\AgdaSymbol{:} \AgdaSymbol{∀\{}\AgdaBound{Γ} \AgdaBound{Δ}\AgdaSymbol{\}} \AgdaSymbol{→} \AgdaBound{Γ} \AgdaDatatype{≤} \AgdaBound{Δ} \AgdaSymbol{→} \AgdaSymbol{∀\{}\AgdaBound{a}\AgdaSymbol{\}} \<[31]%
\>[31]\AgdaSymbol{→} \AgdaDatatype{Ne} \AgdaDatatype{Nf} \AgdaBound{Δ} \AgdaBound{a} \<[45]%
\>[45]\AgdaSymbol{→} \AgdaDatatype{Ne} \AgdaDatatype{Nf} \AgdaBound{Γ} \AgdaBound{a}\<%
\end{code}
\AgdaHide{
\begin{code}%
\>\AgdaFunction{var≤} \AgdaInductiveConstructor{id} \<[15]%
\>[15]\AgdaBound{x} \<[22]%
\>[22]\AgdaSymbol{=} \AgdaBound{x}\<%
\\
\>\AgdaFunction{var≤} \AgdaSymbol{(}\AgdaInductiveConstructor{weak} \AgdaBound{η}\AgdaSymbol{)} \<[15]%
\>[15]\AgdaBound{x} \<[22]%
\>[22]\AgdaSymbol{=} \AgdaInductiveConstructor{suc} \AgdaSymbol{(}\AgdaFunction{var≤} \AgdaBound{η} \AgdaBound{x}\AgdaSymbol{)}\<%
\\
\>\AgdaFunction{var≤} \AgdaSymbol{(}\AgdaInductiveConstructor{lift} \AgdaBound{η}\AgdaSymbol{)} \<[15]%
\>[15]\AgdaInductiveConstructor{zero} \<[22]%
\>[22]\AgdaSymbol{=} \AgdaInductiveConstructor{zero}\<%
\\
\>\AgdaFunction{var≤} \AgdaSymbol{(}\AgdaInductiveConstructor{lift} \AgdaBound{η}\AgdaSymbol{)} \AgdaSymbol{(}\AgdaInductiveConstructor{suc} \AgdaBound{x}\AgdaSymbol{)} \AgdaSymbol{=} \AgdaInductiveConstructor{suc} \AgdaSymbol{(}\AgdaFunction{var≤} \AgdaBound{η} \AgdaBound{x}\AgdaSymbol{)}\<%
\\
%
\\
\>\AgdaFunction{val≤} \AgdaBound{η} \AgdaSymbol{(}\AgdaInductiveConstructor{ne} \AgdaBound{w}\AgdaSymbol{)} \<[18]%
\>[18]\AgdaSymbol{=} \AgdaInductiveConstructor{ne} \AgdaSymbol{(}\AgdaFunction{nev≤} \AgdaBound{η} \AgdaBound{w}\AgdaSymbol{)}\<%
\\
\>\AgdaFunction{val≤} \AgdaBound{η} \AgdaSymbol{(}\AgdaInductiveConstructor{lam} \AgdaBound{t} \AgdaBound{ρ}\AgdaSymbol{)} \<[18]%
\>[18]\AgdaSymbol{=} \AgdaInductiveConstructor{lam} \AgdaBound{t} \AgdaSymbol{(}\AgdaFunction{env≤} \AgdaBound{η} \AgdaBound{ρ}\AgdaSymbol{)}\<%
\\
%
\\
\>\AgdaFunction{env≤} \AgdaBound{η} \AgdaInductiveConstructor{ε} \<[15]%
\>[15]\AgdaSymbol{=} \AgdaInductiveConstructor{ε}\<%
\\
\>\AgdaFunction{env≤} \AgdaBound{η} \AgdaSymbol{(}\AgdaBound{ρ} \AgdaInductiveConstructor{,} \AgdaBound{v}\AgdaSymbol{)} \AgdaSymbol{=} \AgdaFunction{env≤} \AgdaBound{η} \AgdaBound{ρ} \AgdaInductiveConstructor{,} \AgdaFunction{val≤} \AgdaBound{η} \AgdaBound{v}\<%
\\
%
\\
\>\AgdaFunction{nev≤} \AgdaBound{η} \AgdaSymbol{(}\AgdaInductiveConstructor{var} \AgdaBound{x}\AgdaSymbol{)} \<[17]%
\>[17]\AgdaSymbol{=} \AgdaInductiveConstructor{var} \AgdaSymbol{(}\AgdaFunction{var≤} \AgdaBound{η} \AgdaBound{x}\AgdaSymbol{)}\<%
\\
\>\AgdaFunction{nev≤} \AgdaBound{η} \AgdaSymbol{(}\AgdaInductiveConstructor{app} \AgdaBound{w} \AgdaBound{v}\AgdaSymbol{)} \AgdaSymbol{=} \AgdaInductiveConstructor{app} \AgdaSymbol{(}\AgdaFunction{nev≤} \AgdaBound{η} \AgdaBound{w}\AgdaSymbol{)} \AgdaSymbol{(}\AgdaFunction{val≤} \AgdaBound{η} \AgdaBound{v}\AgdaSymbol{)}\<%
\\
%
\\
\>\AgdaFunction{nf≤} \AgdaBound{η} \AgdaSymbol{(}\AgdaInductiveConstructor{ne} \AgdaBound{m}\AgdaSymbol{)} \<[15]%
\>[15]\AgdaSymbol{=} \AgdaInductiveConstructor{ne} \AgdaSymbol{(}\AgdaFunction{nen≤} \AgdaBound{η} \AgdaBound{m}\AgdaSymbol{)}\<%
\\
\>\AgdaFunction{nf≤} \AgdaBound{η} \AgdaSymbol{(}\AgdaInductiveConstructor{lam} \AgdaBound{n}\AgdaSymbol{)} \<[15]%
\>[15]\AgdaSymbol{=} \AgdaInductiveConstructor{lam} \AgdaSymbol{(}\AgdaFunction{nf≤} \AgdaSymbol{(}\AgdaInductiveConstructor{lift} \AgdaBound{η}\AgdaSymbol{)} \AgdaBound{n}\AgdaSymbol{)}\<%
\\
%
\\
\>\AgdaFunction{nen≤} \AgdaBound{η} \AgdaSymbol{(}\AgdaInductiveConstructor{var} \AgdaBound{x}\AgdaSymbol{)} \<[17]%
\>[17]\AgdaSymbol{=} \AgdaInductiveConstructor{var} \AgdaSymbol{(}\AgdaFunction{var≤} \AgdaBound{η} \AgdaBound{x}\AgdaSymbol{)}\<%
\\
\>\AgdaFunction{nen≤} \AgdaBound{η} \AgdaSymbol{(}\AgdaInductiveConstructor{app} \AgdaBound{m} \AgdaBound{n}\AgdaSymbol{)} \AgdaSymbol{=} \AgdaInductiveConstructor{app} \AgdaSymbol{(}\AgdaFunction{nen≤} \AgdaBound{η} \AgdaBound{m}\AgdaSymbol{)} \AgdaSymbol{(}\AgdaFunction{nf≤} \AgdaBound{η} \AgdaBound{n}\AgdaSymbol{)}\<%
\end{code}}

\noindent
Having defined weakening of values by OPEs, defining the simplest form
of weakening \AgdaFunction{weakVal} that just introduces a fresh variable into the context is
easy to define:

\begin{samepage}
\begin{code}%
\>\AgdaFunction{wk} \<[9]%
\>[9]\AgdaSymbol{:} \<[12]%
\>[12]\AgdaSymbol{∀\{}\AgdaBound{Γ} \AgdaBound{a}\AgdaSymbol{\}} \AgdaSymbol{→} \AgdaSymbol{(}\AgdaBound{Γ} \AgdaInductiveConstructor{,} \AgdaBound{a}\AgdaSymbol{)} \AgdaDatatype{≤} \AgdaBound{Γ}\<%
\\
\>\AgdaFunction{wk} \<[9]%
\>[9]\AgdaSymbol{=} \<[12]%
\>[12]\AgdaInductiveConstructor{weak} \AgdaInductiveConstructor{id}\<%
\\
%
\\
\>\AgdaFunction{weakVal} \<[9]%
\>[9]\AgdaSymbol{=} \<[12]%
\>[12]\AgdaFunction{val≤} \AgdaFunction{wk}\<%
\end{code}  
\end{samepage}

\noindent
We can now define a function \AgdaFunction{readback} that turns values into
delayed normal forms, the potential delay is due to the call to the
\AgdaFunction{apply} function. The \AgdaFunction{readback} function is defined by
induction on the types. If the value is of base type then a call to
\AgdaFunction{nereadback} is made which just proceeds structurally through the
neutral term replacing values in the argument positions by normal
forms. 
If the value
is of function type then we perform eta expansion; we know the result
is a \AgdaInductiveConstructor{lam}, but the lambda body cannot be
immediately returned, since function values may be unevaluated
closures; hence, its given \AgdaInductiveConstructor{later} by
\AgdaFunction{eta}.
The function
\AgdaFunction{eta} takes the function value, weakens it, then applies it to the
fresh variable \AgdaInductiveConstructor{var zero}
yielding a delayed value at range type, which is read back recursively.

\begin{code}%
\>\AgdaFunction{readback} \<[12]%
\>[12]\AgdaSymbol{:} \AgdaSymbol{∀\{}\AgdaBound{i} \AgdaBound{Γ} \AgdaBound{a}\AgdaSymbol{\}} \<[26]%
\>[26]\AgdaSymbol{→} \AgdaDatatype{Val} \AgdaBound{Γ} \AgdaBound{a} \<[43]%
\>[43]\AgdaSymbol{→} \AgdaDatatype{Delay} \AgdaBound{i} \AgdaSymbol{(}\AgdaDatatype{Nf} \AgdaBound{Γ} \AgdaBound{a}\AgdaSymbol{)}\<%
\\
\>\AgdaFunction{nereadback} \<[12]%
\>[12]\AgdaSymbol{:} \AgdaSymbol{∀\{}\AgdaBound{i} \AgdaBound{Γ} \AgdaBound{a}\AgdaSymbol{\}} \<[26]%
\>[26]\AgdaSymbol{→} \AgdaDatatype{Ne} \AgdaDatatype{Val} \AgdaBound{Γ} \AgdaBound{a} \<[43]%
\>[43]\AgdaSymbol{→} \AgdaDatatype{Delay} \AgdaBound{i} \AgdaSymbol{(}\AgdaDatatype{Ne} \AgdaDatatype{Nf} \AgdaBound{Γ} \AgdaBound{a}\AgdaSymbol{)}\<%
\\
\>\AgdaFunction{eta} \<[12]%
\>[12]\AgdaSymbol{:} \AgdaSymbol{∀\{}\AgdaBound{i} \AgdaBound{Γ} \AgdaBound{a} \AgdaBound{b}\AgdaSymbol{\}} \<[26]%
\>[26]\AgdaSymbol{→} \AgdaDatatype{Val} \AgdaBound{Γ} \AgdaSymbol{(}\AgdaBound{a} \AgdaInductiveConstructor{⇒} \AgdaBound{b}\AgdaSymbol{)} \<[43]%
\>[43]\AgdaSymbol{→} \AgdaRecord{∞Delay} \AgdaBound{i} \AgdaSymbol{(}\AgdaDatatype{Nf} \AgdaSymbol{(}\AgdaBound{Γ} \AgdaInductiveConstructor{,} \AgdaBound{a}\AgdaSymbol{)} \AgdaBound{b}\AgdaSymbol{)}\<%
\end{code}

\begin{code}%
\>\AgdaFunction{readback} \AgdaSymbol{\{}\AgdaSymbol{a} \AgdaSymbol{=} \AgdaInductiveConstructor{★}\AgdaSymbol{\}} \<[22]%
\>[22]\AgdaSymbol{(}\AgdaInductiveConstructor{ne} \AgdaBound{w}\AgdaSymbol{)} \<[30]%
\>[30]\AgdaSymbol{=} \AgdaInductiveConstructor{ne} \<[37]%
\>[37]\AgdaFunction{\ensuremath{{<}\${>}}} \AgdaFunction{nereadback} \AgdaBound{w}\<%
\\
\>\AgdaFunction{readback} \AgdaSymbol{\{}\AgdaSymbol{a} \AgdaSymbol{=} \AgdaSymbol{\_} \AgdaInductiveConstructor{⇒} \AgdaSymbol{\_\}} \<[22]%
\>[22]\AgdaBound{v} \<[30]%
\>[30]\AgdaSymbol{=} \AgdaInductiveConstructor{lam} \<[37]%
\>[37]\AgdaFunction{\ensuremath{{<}\${>}}} \AgdaInductiveConstructor{later} \AgdaSymbol{(}\AgdaFunction{eta} \AgdaBound{v}\AgdaSymbol{)}\<%
\\
%
\\
\>\AgdaField{force} \AgdaSymbol{(}\AgdaFunction{eta} \AgdaBound{v}\AgdaSymbol{)} \<[30]%
\>[30]\AgdaSymbol{=} \AgdaFunction{readback} \AgdaFunction{\ensuremath{\mathbin{{=}\mkern-2mu{<}\mkern-8.5mu{<}}}} \AgdaFunction{apply} \AgdaSymbol{(}\AgdaFunction{weakVal} \AgdaBound{v}\AgdaSymbol{)} \AgdaSymbol{(}\AgdaInductiveConstructor{ne} \AgdaSymbol{(}\AgdaInductiveConstructor{var} \AgdaInductiveConstructor{zero}\AgdaSymbol{))}\<%
\\
%
\\
\>\AgdaFunction{nereadback} \AgdaSymbol{(}\AgdaInductiveConstructor{var} \AgdaBound{x}\AgdaSymbol{)} \<[30]%
\>[30]\AgdaSymbol{=} \AgdaInductiveConstructor{now} \AgdaSymbol{(}\AgdaInductiveConstructor{var} \AgdaBound{x}\AgdaSymbol{)}\<%
\\
\>\AgdaFunction{nereadback} \AgdaSymbol{(}\AgdaInductiveConstructor{app} \AgdaBound{w} \AgdaBound{v}\AgdaSymbol{)} \<[30]%
\>[30]\AgdaSymbol{=} \AgdaFunction{nereadback} \AgdaBound{w} \AgdaFunction{\ensuremath{\mathbin{{>}\mkern-8.5mu{>}\mkern-2mu{=}}}} \AgdaSymbol{λ} \AgdaBound{m} \AgdaSymbol{→} \AgdaInductiveConstructor{app} \AgdaBound{m} \AgdaFunction{\ensuremath{{<}\${>}}} \AgdaFunction{readback} \AgdaBound{v}\<%
\end{code} 

\noindent
The three functions are defined by an outer coinduction into the
\AgdaDatatype{Delay} monad and an inner local induction on neutral
values in \AgdaFunction{nereadback}.  Again, the sized typing of bind
and map are crucial to communicate the termination argument to Agda.

We define the identity environment by induction on the context.
\begin{code}%
\>\AgdaFunction{ide} \<[13]%
\>[13]\AgdaSymbol{:} \<[16]%
\>[16]\AgdaSymbol{∀} \AgdaBound{Γ} \AgdaSymbol{→} \AgdaDatatype{Env} \AgdaBound{Γ} \AgdaBound{Γ}\<%
\\
\>\AgdaFunction{ide} \AgdaInductiveConstructor{ε} \<[13]%
\>[13]\AgdaSymbol{=} \<[16]%
\>[16]\AgdaInductiveConstructor{ε}\<%
\\
\>\AgdaFunction{ide} \AgdaSymbol{(}\AgdaBound{Γ} \AgdaInductiveConstructor{,} \AgdaBound{a}\AgdaSymbol{)} \<[13]%
\>[13]\AgdaSymbol{=} \<[16]%
\>[16]\AgdaFunction{env≤} \AgdaFunction{wk} \AgdaSymbol{(}\AgdaFunction{ide} \AgdaBound{Γ}\AgdaSymbol{)} \AgdaInductiveConstructor{,} \AgdaInductiveConstructor{ne} \AgdaSymbol{(}\AgdaInductiveConstructor{var} \AgdaInductiveConstructor{zero}\AgdaSymbol{)}\<%
\end{code}

\noindent
Given \AgdaFunction{eval}, \AgdaFunction{ide} and \AgdaFunction{readback} we can define a
normalization function \AgdaFunction{nf} that for any term returns a delayed
normal form.

\begin{code}%
\>\AgdaFunction{nf} \<[10]%
\>[10]\AgdaSymbol{:} \<[13]%
\>[13]\AgdaSymbol{∀\{}\AgdaBound{Γ} \AgdaBound{a}\AgdaSymbol{\}(}\AgdaBound{t} \AgdaSymbol{:} \AgdaDatatype{Tm} \AgdaBound{Γ} \AgdaBound{a}\AgdaSymbol{)} \AgdaSymbol{→} \AgdaDatatype{Delay} \AgdaPostulate{∞} \AgdaSymbol{(}\AgdaDatatype{Nf} \AgdaBound{Γ} \AgdaBound{a}\AgdaSymbol{)}\<%
\\
\>\AgdaFunction{nf} \AgdaSymbol{\{}\AgdaBound{Γ}\AgdaSymbol{\}} \AgdaBound{t} \<[10]%
\>[10]\AgdaSymbol{=} \<[13]%
\>[13]\AgdaFunction{eval} \AgdaBound{t} \AgdaSymbol{(}\AgdaFunction{ide} \AgdaBound{Γ}\AgdaSymbol{)} \AgdaFunction{\ensuremath{\mathbin{{>}\mkern-8.5mu{>}\mkern-2mu{=}}}} \AgdaFunction{readback}\<%
\end{code}

\section{Termination proof}

While we have managed to define the normalizer in a way acceptable to
Agda's termination checker, we have not established that simply-typed
lambda calculus is actually normalizing, \ie, that each well-typed
term reaches its normal form after a only final number of
\AgdaInductiveConstructor{delay}s have been issued.
To this end, 
we define a logical predicate \AgdaFunction{V⟦\_⟧\_}, corresponding to strong
computability on values. It is defined by induction on the type of the
value. At base type, when the value must be neutral, the relation
states that the neutral term is strongly computable if its readback
converges. At function type it states that the function is strongly
computable if, in any weakened context (in the general OPE sense) it
takes any value which is strongly computable to a delayed value which
converges to a strongly computable value. The predicate \AgdaFunction{C⟦\_⟧\_} on
delayed values \AgdaBound{v?} is shorthand for a triple
$(\AgdaBound{v} \AgdaInductiveConstructor{,} \AgdaBound{v⇓}
  \AgdaInductiveConstructor{,} \AgdaBound{⟦v⟧})$
of a value \AgdaBound{v}, a proof \AgdaBound{v⇓} that the
delayed value converges to the value and a proof \AgdaBound{⟦v⟧} of strong
computability.

\begin{code}%
\>\AgdaFunction{V⟦\_⟧\_} \<[7]%
\>[7]\AgdaSymbol{:} \AgdaSymbol{∀\{}\AgdaBound{Γ}\AgdaSymbol{\}} \AgdaSymbol{(}\AgdaBound{a} \AgdaSymbol{:} \AgdaDatatype{Ty}\AgdaSymbol{)} \AgdaSymbol{→} \AgdaDatatype{Val} \AgdaBound{Γ} \AgdaBound{a} \<[44]%
\>[44]\AgdaSymbol{→} \AgdaPrimitiveType{Set}\<%
\\
\>\AgdaFunction{C⟦\_⟧\_} \<[7]%
\>[7]\AgdaSymbol{:} \AgdaSymbol{∀\{}\AgdaBound{Γ}\AgdaSymbol{\}} \AgdaSymbol{(}\AgdaBound{a} \AgdaSymbol{:} \AgdaDatatype{Ty}\AgdaSymbol{)} \AgdaSymbol{→} \AgdaDatatype{Delay} \AgdaPostulate{∞} \AgdaSymbol{(}\AgdaDatatype{Val} \AgdaBound{Γ} \AgdaBound{a}\AgdaSymbol{)} \<[44]%
\>[44]\AgdaSymbol{→} \AgdaPrimitiveType{Set}\<%
\\
%
\\
\>\AgdaFunction{V⟦} \AgdaInductiveConstructor{★} \<[9]%
\>[9]\AgdaFunction{⟧} \<[12]%
\>[12]\AgdaSymbol{(}\AgdaInductiveConstructor{ne} \AgdaBound{w}\AgdaSymbol{)} \<[20]%
\>[20]\AgdaSymbol{=} \AgdaFunction{nereadback} \AgdaBound{w} \AgdaFunction{⇓}\<%
\\
\>\AgdaFunction{V⟦} \AgdaBound{a} \AgdaInductiveConstructor{⇒} \AgdaBound{b} \AgdaFunction{⟧} \<[12]%
\>[12]\AgdaBound{\,f\,} \<[20]%
\>[20]\AgdaSymbol{=} \AgdaSymbol{∀\{}\AgdaBound{Δ}\AgdaSymbol{\}(}\AgdaBound{η} \AgdaSymbol{:} \AgdaBound{Δ} \AgdaDatatype{≤} \AgdaSymbol{\_)(}\AgdaBound{u} \AgdaSymbol{:} \AgdaDatatype{Val} \AgdaBound{Δ} \AgdaBound{a}\AgdaSymbol{)} \AgdaSymbol{→} \AgdaFunction{V⟦} \AgdaBound{a} \AgdaFunction{⟧} \AgdaBound{u} \AgdaSymbol{→} \AgdaFunction{C⟦} \AgdaBound{b} \AgdaFunction{⟧} \AgdaSymbol{(}\AgdaFunction{apply} \AgdaSymbol{(}\AgdaFunction{val≤} \AgdaBound{η} \AgdaBound{\,f\,}\AgdaSymbol{)} \AgdaBound{u}\AgdaSymbol{)}\<%
\\
%
\\
\>\AgdaFunction{C⟦} \AgdaBound{a} \AgdaFunction{⟧} \<[12]%
\>[12]\AgdaBound{v?} \<[20]%
\>[20]\AgdaSymbol{=} \AgdaFunction{∃} \AgdaSymbol{λ} \AgdaBound{v} \AgdaSymbol{→} \AgdaBound{v?} \AgdaDatatype{⇓} \AgdaBound{v} \AgdaFunction{×} \AgdaFunction{V⟦} \AgdaBound{a} \AgdaFunction{⟧} \AgdaBound{v}\<%
\end{code}

\noindent
The notion of strongly computable value is easily extended to environments.

\begin{samepage}
\begin{code}%
\>\AgdaFunction{E⟦\_⟧\_} \<[21]%
\>[21]\AgdaSymbol{:} \<[24]%
\>[24]\AgdaSymbol{∀\{}\AgdaBound{Δ}\AgdaSymbol{\}(}\AgdaBound{Γ} \AgdaSymbol{:} \AgdaDatatype{Cxt}\AgdaSymbol{)} \AgdaSymbol{→} \AgdaDatatype{Env} \AgdaBound{Δ} \AgdaBound{Γ} \AgdaSymbol{→} \AgdaPrimitiveType{Set}\<%
\\
\>\AgdaFunction{E⟦} \AgdaInductiveConstructor{ε} \AgdaFunction{⟧} \<[12]%
\>[12]\AgdaInductiveConstructor{ε} \<[21]%
\>[21]\AgdaSymbol{=} \<[24]%
\>[24]\AgdaRecord{⊤}\<%
\\
\>\AgdaFunction{E⟦} \AgdaBound{Γ} \AgdaInductiveConstructor{,} \AgdaBound{a} \AgdaFunction{⟧} \<[12]%
\>[12]\AgdaSymbol{(}\AgdaBound{ρ} \AgdaInductiveConstructor{,} \AgdaBound{v}\AgdaSymbol{)} \<[21]%
\>[21]\AgdaSymbol{=} \<[24]%
\>[24]\AgdaFunction{E⟦} \AgdaBound{Γ} \AgdaFunction{⟧} \AgdaBound{ρ} \AgdaFunction{×} \AgdaFunction{V⟦} \AgdaBound{a} \AgdaFunction{⟧} \AgdaBound{v}\<%
\end{code}
\end{samepage}

\noindent
Later we will require weakening (applying an OPE) variables, values,
environments, etc.\ preserve identity and composition (respect functor
laws). We state these properties now but suppress the proofs.

\begin{code}%
\>\AgdaFunction{val≤-id} \<[9]%
\>[9]\AgdaSymbol{:} \AgdaSymbol{∀\{}\AgdaBound{Δ} \AgdaBound{a}\AgdaSymbol{\}} \<[19]%
\>[19]\AgdaSymbol{(}\AgdaBound{v} \AgdaSymbol{:} \AgdaDatatype{Val} \AgdaBound{Δ} \AgdaBound{a}\AgdaSymbol{)} \<[37]%
\>[37]\AgdaSymbol{→} \AgdaFunction{val≤} \AgdaInductiveConstructor{id} \AgdaBound{v} \AgdaDatatype{≡} \AgdaBound{v}\<%
\\
%
\\
\>\AgdaFunction{env≤-id} \<[9]%
\>[9]\AgdaSymbol{:} \AgdaSymbol{∀\{}\AgdaBound{Γ} \AgdaBound{Δ}\AgdaSymbol{\}} \<[19]%
\>[19]\AgdaSymbol{(}\AgdaBound{ρ} \AgdaSymbol{:} \AgdaDatatype{Env} \AgdaBound{Δ} \AgdaBound{Γ}\AgdaSymbol{)} \<[37]%
\>[37]\AgdaSymbol{→} \AgdaFunction{env≤} \AgdaInductiveConstructor{id} \AgdaBound{ρ} \AgdaDatatype{≡} \AgdaBound{ρ}\<%
\\
%
\\
\>\AgdaFunction{nev≤-id} \<[9]%
\>[9]\AgdaSymbol{:} \AgdaSymbol{∀\{}\AgdaBound{Δ} \AgdaBound{a}\AgdaSymbol{\}} \<[19]%
\>[19]\AgdaSymbol{(}\AgdaBound{t} \AgdaSymbol{:} \AgdaDatatype{Ne} \AgdaDatatype{Val} \AgdaBound{Δ} \AgdaBound{a}\AgdaSymbol{)} \<[37]%
\>[37]\AgdaSymbol{→} \AgdaFunction{nev≤} \AgdaInductiveConstructor{id} \AgdaBound{t} \AgdaDatatype{≡} \AgdaBound{t}\<%
\end{code}

\AgdaHide{
\begin{code}%
\>\AgdaFunction{env≤-id} \AgdaInductiveConstructor{ε} \<[18]%
\>[18]\AgdaSymbol{=} \AgdaInductiveConstructor{refl}\<%
\\
\>\AgdaFunction{env≤-id} \AgdaSymbol{(}\AgdaBound{ρ} \AgdaInductiveConstructor{,} \AgdaBound{v}\AgdaSymbol{)} \<[18]%
\>[18]\AgdaSymbol{=} \AgdaFunction{cong₂} \AgdaInductiveConstructor{\_,\_} \AgdaSymbol{(}\AgdaFunction{env≤-id} \AgdaBound{ρ}\AgdaSymbol{)} \AgdaSymbol{(}\AgdaFunction{val≤-id} \AgdaBound{v}\AgdaSymbol{)}\<%
\\
%
\\
\>\AgdaFunction{val≤-id} \AgdaSymbol{(}\AgdaInductiveConstructor{ne} \AgdaBound{t}\AgdaSymbol{)} \AgdaSymbol{=} \AgdaFunction{cong} \AgdaInductiveConstructor{ne} \AgdaSymbol{(}\AgdaFunction{nev≤-id} \AgdaBound{t}\AgdaSymbol{)}\<%
\\
\>\AgdaFunction{val≤-id} \AgdaSymbol{(}\AgdaInductiveConstructor{lam} \AgdaBound{t} \AgdaBound{ρ}\AgdaSymbol{)} \AgdaSymbol{=} \AgdaFunction{cong} \AgdaSymbol{(}\AgdaInductiveConstructor{lam} \AgdaBound{t}\AgdaSymbol{)} \AgdaSymbol{(}\AgdaFunction{env≤-id} \AgdaBound{ρ}\AgdaSymbol{)}\<%
\\
%
\\
\>\AgdaFunction{nev≤-id} \AgdaSymbol{(}\AgdaInductiveConstructor{var} \AgdaBound{x}\AgdaSymbol{)} \<[18]%
\>[18]\AgdaSymbol{=} \AgdaInductiveConstructor{refl}\<%
\\
\>\AgdaFunction{nev≤-id} \AgdaSymbol{(}\AgdaInductiveConstructor{app} \AgdaBound{t} \AgdaBound{u}\AgdaSymbol{)} \AgdaSymbol{=} \AgdaFunction{cong₂} \AgdaInductiveConstructor{app} \AgdaSymbol{(}\AgdaFunction{nev≤-id} \AgdaBound{t}\AgdaSymbol{)} \AgdaSymbol{(}\AgdaFunction{val≤-id} \AgdaBound{u}\AgdaSymbol{)}\<%
\end{code}}

\begin{code}%
\>\AgdaFunction{var≤-•} \<[8]%
\>[8]\AgdaSymbol{:} \<[11]%
\>[11]\AgdaSymbol{∀\{}\AgdaBound{Δ} \AgdaBound{Δ\kern-1pt′} \AgdaBound{Δ\kern-1pt″} \AgdaBound{a}\AgdaSymbol{\}} \AgdaSymbol{(}\AgdaBound{η} \AgdaSymbol{:} \AgdaBound{Δ} \AgdaDatatype{≤} \AgdaBound{Δ\kern-1pt′}\AgdaSymbol{)} \AgdaSymbol{(}\AgdaBound{η′} \AgdaSymbol{:} \AgdaBound{Δ\kern-1pt′} \AgdaDatatype{≤} \AgdaBound{Δ\kern-1pt″}\AgdaSymbol{)} \AgdaSymbol{(}\AgdaBound{x} \AgdaSymbol{:} \AgdaDatatype{Var} \AgdaBound{Δ\kern-1pt″} \AgdaBound{a}\AgdaSymbol{)} \AgdaSymbol{→}\<%
\\
\>[2]\AgdaIndent{11}{}\<[11]%
\>[11]\AgdaFunction{var≤} \AgdaBound{η} \AgdaSymbol{(}\AgdaFunction{var≤} \AgdaBound{η′} \AgdaBound{x}\AgdaSymbol{)} \AgdaDatatype{≡} \AgdaFunction{var≤} \AgdaSymbol{(}\AgdaBound{η} \AgdaFunction{•} \AgdaBound{η′}\AgdaSymbol{)} \AgdaBound{x}\<%
\\
%
\\
\>\AgdaFunction{val≤-•} \<[8]%
\>[8]\AgdaSymbol{:} \<[11]%
\>[11]\AgdaSymbol{∀\{}\AgdaBound{Δ} \AgdaBound{Δ\kern-1pt′} \AgdaBound{Δ\kern-1pt″} \AgdaBound{a}\AgdaSymbol{\}} \AgdaSymbol{(}\AgdaBound{η} \AgdaSymbol{:} \AgdaBound{Δ} \AgdaDatatype{≤} \AgdaBound{Δ\kern-1pt′}\AgdaSymbol{)} \AgdaSymbol{(}\AgdaBound{η′} \AgdaSymbol{:} \AgdaBound{Δ\kern-1pt′} \AgdaDatatype{≤} \AgdaBound{Δ\kern-1pt″}\AgdaSymbol{)} \AgdaSymbol{(}\AgdaBound{v} \AgdaSymbol{:} \AgdaDatatype{Val} \AgdaBound{Δ\kern-1pt″} \AgdaBound{a}\AgdaSymbol{)} \AgdaSymbol{→}\<%
\\
\>[2]\AgdaIndent{11}{}\<[11]%
\>[11]\AgdaFunction{val≤} \AgdaBound{η} \AgdaSymbol{(}\AgdaFunction{val≤} \AgdaBound{η′} \AgdaBound{v}\AgdaSymbol{)} \AgdaDatatype{≡} \AgdaFunction{val≤} \AgdaSymbol{(}\AgdaBound{η} \AgdaFunction{•} \AgdaBound{η′}\AgdaSymbol{)} \AgdaBound{v}\<%
\\
%
\\
\>\AgdaFunction{env≤-•} \<[8]%
\>[8]\AgdaSymbol{:} \<[11]%
\>[11]\AgdaSymbol{∀\{}\AgdaBound{Γ} \AgdaBound{Δ} \AgdaBound{Δ\kern-1pt′} \AgdaBound{Δ\kern-1pt″}\AgdaSymbol{\}} \AgdaSymbol{(}\AgdaBound{η} \AgdaSymbol{:} \AgdaBound{Δ} \AgdaDatatype{≤} \AgdaBound{Δ\kern-1pt′}\AgdaSymbol{)} \AgdaSymbol{(}\AgdaBound{η′} \AgdaSymbol{:} \AgdaBound{Δ\kern-1pt′} \AgdaDatatype{≤} \AgdaBound{Δ\kern-1pt″}\AgdaSymbol{)} \AgdaSymbol{(}\AgdaBound{ρ} \AgdaSymbol{:} \AgdaDatatype{Env} \AgdaBound{Δ\kern-1pt″} \AgdaBound{Γ}\AgdaSymbol{)} \AgdaSymbol{→}\<%
\\
\>[2]\AgdaIndent{11}{}\<[11]%
\>[11]\AgdaFunction{env≤} \AgdaBound{η} \AgdaSymbol{(}\AgdaFunction{env≤} \AgdaBound{η′} \AgdaBound{ρ}\AgdaSymbol{)} \AgdaDatatype{≡} \AgdaFunction{env≤} \AgdaSymbol{(}\AgdaBound{η} \AgdaFunction{•} \AgdaBound{η′}\AgdaSymbol{)} \AgdaBound{ρ}\<%
\\
%
\\
\>\AgdaFunction{nev≤-•} \<[8]%
\>[8]\AgdaSymbol{:} \<[11]%
\>[11]\AgdaSymbol{∀\{}\AgdaBound{Δ} \AgdaBound{Δ\kern-1pt′} \AgdaBound{Δ\kern-1pt″} \AgdaBound{a}\AgdaSymbol{\}} \AgdaSymbol{(}\AgdaBound{η} \AgdaSymbol{:} \AgdaBound{Δ} \AgdaDatatype{≤} \AgdaBound{Δ\kern-1pt′}\AgdaSymbol{)} \AgdaSymbol{(}\AgdaBound{η′} \AgdaSymbol{:} \AgdaBound{Δ\kern-1pt′} \AgdaDatatype{≤} \AgdaBound{Δ\kern-1pt″}\AgdaSymbol{)} \AgdaSymbol{(}\AgdaBound{t} \AgdaSymbol{:} \AgdaDatatype{Ne} \AgdaDatatype{Val} \AgdaBound{Δ\kern-1pt″} \AgdaBound{a}\AgdaSymbol{)} \AgdaSymbol{→}\<%
\\
\>[2]\AgdaIndent{11}{}\<[11]%
\>[11]\AgdaFunction{nev≤} \AgdaBound{η} \AgdaSymbol{(}\AgdaFunction{nev≤} \AgdaBound{η′} \AgdaBound{t}\AgdaSymbol{)} \AgdaDatatype{≡} \AgdaFunction{nev≤} \AgdaSymbol{(}\AgdaBound{η} \AgdaFunction{•} \AgdaBound{η′}\AgdaSymbol{)} \AgdaBound{t}\<%
\end{code}

\AgdaHide{
\begin{code}%
\>\AgdaFunction{var≤-•} \AgdaInductiveConstructor{id} \<[16]%
\>[16]\AgdaBound{η′} \<[26]%
\>[26]\AgdaBound{x} \<[34]%
\>[34]\AgdaSymbol{=} \AgdaInductiveConstructor{refl}\<%
\\
\>\AgdaFunction{var≤-•} \AgdaSymbol{(}\AgdaInductiveConstructor{weak} \AgdaBound{η}\AgdaSymbol{)} \AgdaBound{η′} \<[26]%
\>[26]\AgdaBound{x} \<[34]%
\>[34]\AgdaSymbol{=} \AgdaFunction{cong} \AgdaInductiveConstructor{suc} \AgdaSymbol{(}\AgdaFunction{var≤-•} \AgdaBound{η} \AgdaBound{η′} \AgdaBound{x}\AgdaSymbol{)}\<%
\\
\>\AgdaFunction{var≤-•} \AgdaSymbol{(}\AgdaInductiveConstructor{lift} \AgdaBound{η}\AgdaSymbol{)} \AgdaInductiveConstructor{id} \<[26]%
\>[26]\AgdaBound{x} \<[34]%
\>[34]\AgdaSymbol{=} \AgdaInductiveConstructor{refl}\<%
\\
\>\AgdaFunction{var≤-•} \AgdaSymbol{(}\AgdaInductiveConstructor{lift} \AgdaBound{η}\AgdaSymbol{)} \AgdaSymbol{(}\AgdaInductiveConstructor{weak} \AgdaBound{η′}\AgdaSymbol{)} \AgdaBound{x} \<[34]%
\>[34]\AgdaSymbol{=} \AgdaFunction{cong} \AgdaInductiveConstructor{suc} \AgdaSymbol{(}\AgdaFunction{var≤-•} \AgdaBound{η} \AgdaBound{η′} \AgdaBound{x}\AgdaSymbol{)}\<%
\\
\>\AgdaFunction{var≤-•} \AgdaSymbol{(}\AgdaInductiveConstructor{lift} \AgdaBound{η}\AgdaSymbol{)} \AgdaSymbol{(}\AgdaInductiveConstructor{lift} \AgdaBound{η′}\AgdaSymbol{)} \AgdaInductiveConstructor{zero} \<[34]%
\>[34]\AgdaSymbol{=} \AgdaInductiveConstructor{refl}\<%
\\
\>\AgdaFunction{var≤-•} \AgdaSymbol{(}\AgdaInductiveConstructor{lift} \AgdaBound{η}\AgdaSymbol{)} \AgdaSymbol{(}\AgdaInductiveConstructor{lift} \AgdaBound{η′}\AgdaSymbol{)} \AgdaSymbol{(}\AgdaInductiveConstructor{suc} \AgdaBound{x}\AgdaSymbol{)} \AgdaSymbol{=} \AgdaFunction{cong} \AgdaInductiveConstructor{suc} \AgdaSymbol{(}\AgdaFunction{var≤-•} \AgdaBound{η} \AgdaBound{η′} \AgdaBound{x}\AgdaSymbol{)}\<%
\\
%
\\
\>\AgdaFunction{env≤-•} \AgdaBound{η} \AgdaBound{η′} \AgdaInductiveConstructor{ε} \<[20]%
\>[20]\AgdaSymbol{=} \AgdaInductiveConstructor{refl}\<%
\\
\>\AgdaFunction{env≤-•} \AgdaBound{η} \AgdaBound{η′} \AgdaSymbol{(}\AgdaBound{ρ} \AgdaInductiveConstructor{,} \AgdaBound{v}\AgdaSymbol{)} \AgdaSymbol{=} \AgdaFunction{cong₂} \AgdaInductiveConstructor{\_,\_} \AgdaSymbol{(}\AgdaFunction{env≤-•} \AgdaBound{η} \AgdaBound{η′} \AgdaBound{ρ}\AgdaSymbol{)} \AgdaSymbol{(}\AgdaFunction{val≤-•} \AgdaBound{η} \AgdaBound{η′} \AgdaBound{v}\AgdaSymbol{)}\<%
\\
%
\\
\>\AgdaFunction{val≤-•} \AgdaBound{η} \AgdaBound{η′} \AgdaSymbol{(}\AgdaInductiveConstructor{ne} \AgdaBound{w}\AgdaSymbol{)} \AgdaSymbol{=} \AgdaFunction{cong} \AgdaInductiveConstructor{ne} \AgdaSymbol{(}\AgdaFunction{nev≤-•} \AgdaBound{η} \AgdaBound{η′} \AgdaBound{w}\AgdaSymbol{)}\<%
\\
\>\AgdaFunction{val≤-•} \AgdaBound{η} \AgdaBound{η′} \AgdaSymbol{(}\AgdaInductiveConstructor{lam} \AgdaBound{t} \AgdaBound{ρ}\AgdaSymbol{)} \AgdaSymbol{=} \AgdaFunction{cong} \AgdaSymbol{(}\AgdaInductiveConstructor{lam} \AgdaBound{t}\AgdaSymbol{)} \AgdaSymbol{(}\AgdaFunction{env≤-•} \AgdaBound{η} \AgdaBound{η′} \AgdaBound{ρ}\AgdaSymbol{)}\<%
\\
%
\\
\>\AgdaFunction{nev≤-•} \AgdaBound{η} \AgdaBound{η′} \AgdaSymbol{(}\AgdaInductiveConstructor{var} \AgdaBound{x}\AgdaSymbol{)} \<[22]%
\>[22]\AgdaSymbol{=} \AgdaFunction{cong} \AgdaInductiveConstructor{var} \AgdaSymbol{(}\AgdaFunction{var≤-•} \AgdaBound{η} \AgdaBound{η′} \AgdaBound{x}\AgdaSymbol{)}\<%
\\
\>\AgdaFunction{nev≤-•} \AgdaBound{η} \AgdaBound{η′} \AgdaSymbol{(}\AgdaInductiveConstructor{app} \AgdaBound{w} \AgdaBound{v}\AgdaSymbol{)} \AgdaSymbol{=} \AgdaFunction{cong₂} \AgdaInductiveConstructor{app} \AgdaSymbol{(}\AgdaFunction{nev≤-•} \AgdaBound{η} \AgdaBound{η′} \AgdaBound{w}\AgdaSymbol{)} \AgdaSymbol{(}\AgdaFunction{val≤-•} \AgdaBound{η} \AgdaBound{η′} \AgdaBound{v}\AgdaSymbol{)}\<%
\end{code}}

\noindent
We also require that the operations that we introduce such as lookup,
eval, apply, readback etc.\ commute with weakening. We, again, state
these necessary properties but suppress the proofs.

\begin{code}%
\>\AgdaFunction{lookup≤} \<[9]%
\>[9]\AgdaSymbol{:} \<[12]%
\>[12]\AgdaSymbol{∀} \AgdaSymbol{\{}\AgdaBound{Γ} \AgdaBound{Δ} \AgdaBound{Δ\kern-1pt′} \AgdaBound{a}\AgdaSymbol{\}} \AgdaSymbol{(}\AgdaBound{x} \AgdaSymbol{:} \AgdaDatatype{Var} \AgdaBound{Γ} \AgdaBound{a}\AgdaSymbol{)} \AgdaSymbol{(}\AgdaBound{ρ} \AgdaSymbol{:} \AgdaDatatype{Env} \AgdaBound{Δ} \AgdaBound{Γ}\AgdaSymbol{)} \AgdaSymbol{(}\AgdaBound{η} \AgdaSymbol{:} \AgdaBound{Δ\kern-1pt′} \AgdaDatatype{≤} \AgdaBound{Δ}\AgdaSymbol{)} \AgdaSymbol{→}\<%
\\
\>[11]\AgdaIndent{12}{}\<[12]%
\>[12]\AgdaFunction{val≤} \AgdaBound{η} \AgdaSymbol{(}\AgdaFunction{lookup} \AgdaBound{x} \AgdaBound{ρ}\AgdaSymbol{)} \AgdaDatatype{≡} \AgdaFunction{lookup} \AgdaBound{x} \AgdaSymbol{(}\AgdaFunction{env≤} \AgdaBound{η} \AgdaBound{ρ}\AgdaSymbol{)}\<%
\\
%
\\
\>\AgdaFunction{eval≤} \<[9]%
\>[9]\AgdaSymbol{:} \<[12]%
\>[12]\AgdaSymbol{∀} \AgdaSymbol{\{}\AgdaBound{i} \AgdaBound{Γ} \AgdaBound{Δ} \AgdaBound{Δ\kern-1pt′} \AgdaBound{a}\AgdaSymbol{\}} \AgdaSymbol{(}\AgdaBound{t} \AgdaSymbol{:} \AgdaDatatype{Tm} \AgdaBound{Γ} \AgdaBound{a}\AgdaSymbol{)} \AgdaSymbol{(}\AgdaBound{ρ} \AgdaSymbol{:} \AgdaDatatype{Env} \AgdaBound{Δ} \AgdaBound{Γ}\AgdaSymbol{)} \AgdaSymbol{(}\AgdaBound{η} \AgdaSymbol{:} \AgdaBound{Δ\kern-1pt′} \AgdaDatatype{≤} \AgdaBound{Δ}\AgdaSymbol{)} \AgdaSymbol{→}\<%
\\
\>[11]\AgdaIndent{12}{}\<[12]%
\>[12]\AgdaSymbol{(}\AgdaFunction{val≤} \AgdaBound{η} \AgdaFunction{\ensuremath{{<}\${>}}} \AgdaSymbol{(}\AgdaFunction{eval} \AgdaBound{t} \AgdaBound{ρ}\AgdaSymbol{))} \AgdaFunction{∼⟨} \AgdaBound{i} \AgdaFunction{⟩∼} \AgdaSymbol{(}\AgdaFunction{eval} \AgdaBound{t} \AgdaSymbol{(}\AgdaFunction{env≤} \AgdaBound{η} \AgdaBound{ρ}\AgdaSymbol{))}\<%
\\
%
\\
\>\AgdaFunction{apply≤} \<[9]%
\>[9]\AgdaSymbol{:} \<[12]%
\>[12]\AgdaSymbol{∀\{}\AgdaBound{i} \AgdaBound{Γ} \AgdaBound{Δ} \AgdaBound{a} \AgdaBound{b}\AgdaSymbol{\}} \AgdaSymbol{(}\AgdaBound{\,f\,} \AgdaSymbol{:} \AgdaDatatype{Val} \AgdaBound{Γ} \AgdaSymbol{(}\AgdaBound{a} \AgdaInductiveConstructor{⇒} \AgdaBound{b}\AgdaSymbol{))(}\AgdaBound{v} \AgdaSymbol{:} \AgdaDatatype{Val} \AgdaBound{Γ} \AgdaBound{a}\AgdaSymbol{)(}\AgdaBound{η} \AgdaSymbol{:} \AgdaBound{Δ} \AgdaDatatype{≤} \AgdaBound{Γ}\AgdaSymbol{)} \AgdaSymbol{→}\<%
\\
\>[11]\AgdaIndent{12}{}\<[12]%
\>[12]\AgdaSymbol{(}\AgdaFunction{val≤} \AgdaBound{η} \AgdaFunction{\ensuremath{{<}\${>}}} \AgdaFunction{apply} \AgdaBound{\,f\,} \AgdaBound{v}\AgdaSymbol{)} \AgdaFunction{∼⟨} \AgdaBound{i} \AgdaFunction{⟩∼} \AgdaSymbol{(}\AgdaFunction{apply} \AgdaSymbol{(}\AgdaFunction{val≤} \AgdaBound{η} \AgdaBound{\,f\,}\AgdaSymbol{)} \AgdaSymbol{(}\AgdaFunction{val≤} \AgdaBound{η} \AgdaBound{v}\AgdaSymbol{))}\<%
\\
%
\\
\>\AgdaFunction{beta≤} \<[9]%
\>[9]\AgdaSymbol{:} \<[12]%
\>[12]\AgdaSymbol{∀} \AgdaSymbol{\{}\AgdaBound{i} \AgdaBound{Γ} \AgdaBound{Δ} \AgdaBound{E} \AgdaBound{a} \AgdaBound{b}\AgdaSymbol{\}} \AgdaSymbol{(}\AgdaBound{t} \AgdaSymbol{:} \AgdaDatatype{Tm} \AgdaSymbol{(}\AgdaBound{Γ} \AgdaInductiveConstructor{,} \AgdaBound{a}\AgdaSymbol{)} \AgdaBound{b}\AgdaSymbol{)(}\AgdaBound{ρ} \AgdaSymbol{:} \AgdaDatatype{Env} \AgdaBound{Δ} \AgdaBound{Γ}\AgdaSymbol{)} \AgdaSymbol{(}\AgdaBound{v} \AgdaSymbol{:} \AgdaDatatype{Val} \AgdaBound{Δ} \AgdaBound{a}\AgdaSymbol{)} \AgdaSymbol{(}\AgdaBound{η} \AgdaSymbol{:} \AgdaBound{E} \AgdaDatatype{≤} \AgdaBound{Δ}\AgdaSymbol{)} \AgdaSymbol{→}\<%
\\
\>[11]\AgdaIndent{12}{}\<[12]%
\>[12]\AgdaSymbol{(}\AgdaFunction{val≤} \AgdaBound{η} \AgdaFunction{∞\ensuremath{{<}\${>}}} \AgdaSymbol{(}\AgdaFunction{beta} \AgdaBound{t} \AgdaBound{ρ} \AgdaBound{v}\AgdaSymbol{))} \AgdaRecord{∞∼⟨} \AgdaBound{i} \AgdaRecord{⟩∼} \AgdaFunction{beta} \AgdaBound{t} \AgdaSymbol{(}\AgdaFunction{env≤} \AgdaBound{η} \AgdaBound{ρ}\AgdaSymbol{)} \AgdaSymbol{(}\AgdaFunction{val≤} \AgdaBound{η} \AgdaBound{v}\AgdaSymbol{)}\<%
\end{code} 
\AgdaHide{%
\begin{code}%
\>\AgdaFunction{lookup≤} \AgdaInductiveConstructor{zero} \<[17]%
\>[17]\AgdaSymbol{(}\AgdaBound{ρ} \AgdaInductiveConstructor{,} \AgdaBound{v}\AgdaSymbol{)} \AgdaBound{η} \AgdaSymbol{=} \AgdaInductiveConstructor{refl}\<%
\\
\>\AgdaFunction{lookup≤} \AgdaSymbol{(}\AgdaInductiveConstructor{suc} \AgdaBound{x}\AgdaSymbol{)} \<[17]%
\>[17]\AgdaSymbol{(}\AgdaBound{ρ} \AgdaInductiveConstructor{,} \AgdaBound{v}\AgdaSymbol{)} \AgdaBound{η} \AgdaSymbol{=} \AgdaFunction{lookup≤} \AgdaBound{x} \AgdaBound{ρ} \AgdaBound{η}\<%
\\
%
\\
\>\AgdaFunction{eval≤} \AgdaSymbol{(}\AgdaInductiveConstructor{var} \AgdaBound{x}\AgdaSymbol{)} \<[16]%
\>[16]\AgdaBound{ρ} \AgdaBound{η} \AgdaKeyword{rewrite} \AgdaFunction{lookup≤} \AgdaBound{x} \AgdaBound{ρ} \AgdaBound{η} \AgdaSymbol{=} \AgdaInductiveConstructor{∼now} \AgdaSymbol{\_}\<%
\\
\>\AgdaFunction{eval≤} \AgdaSymbol{(}\AgdaInductiveConstructor{abs} \AgdaBound{t}\AgdaSymbol{)} \<[16]%
\>[16]\AgdaBound{ρ} \AgdaBound{η} \AgdaSymbol{=} \AgdaInductiveConstructor{∼now} \AgdaSymbol{\_}\<%
\\
\>\AgdaFunction{eval≤} \AgdaSymbol{(}\AgdaInductiveConstructor{app} \AgdaBound{t} \AgdaBound{u}\AgdaSymbol{)} \AgdaBound{ρ} \AgdaBound{η} \AgdaSymbol{=}\<%
\\
\>[0]\AgdaIndent{2}{}\<[2]%
\>[2]\AgdaFunction{proof}\<%
\\
\>[0]\AgdaIndent{2}{}\<[2]%
\>[2]\AgdaSymbol{((}\AgdaFunction{eval} \AgdaBound{t} \AgdaBound{ρ} \AgdaFunction{\ensuremath{\mathbin{{>}\mkern-8.5mu{>}\mkern-2mu{=}}}}\<%
\\
\>[2]\AgdaIndent{4}{}\<[4]%
\>[4]\AgdaSymbol{(λ} \AgdaBound{\,f\,} \AgdaSymbol{→} \AgdaFunction{eval} \AgdaBound{u} \AgdaBound{ρ} \AgdaFunction{\ensuremath{\mathbin{{>}\mkern-8.5mu{>}\mkern-2mu{=}}}} \AgdaSymbol{(λ} \AgdaBound{v} \AgdaSymbol{→} \AgdaFunction{apply} \AgdaBound{\,f\,} \AgdaBound{v}\AgdaSymbol{)))}\<%
\\
\>[4]\AgdaIndent{6}{}\<[6]%
\>[6]\AgdaFunction{\ensuremath{\mathbin{{>}\mkern-8.5mu{>}\mkern-2mu{=}}}} \AgdaSymbol{(λ} \AgdaBound{x′} \AgdaSymbol{→} \AgdaInductiveConstructor{now} \AgdaSymbol{(}\AgdaFunction{val≤} \AgdaBound{η} \AgdaBound{x′}\AgdaSymbol{)))}\<%
\\
\>[0]\AgdaIndent{2}{}\<[2]%
\>[2]\AgdaFunction{\qquad∼⟨} \AgdaFunction{bind-assoc} \AgdaSymbol{(}\AgdaFunction{eval} \AgdaBound{t} \AgdaBound{ρ}\AgdaSymbol{)} \AgdaFunction{⟩}\<%
\\
\>[0]\AgdaIndent{2}{}\<[2]%
\>[2]\AgdaSymbol{(}\AgdaFunction{eval} \AgdaBound{t} \AgdaBound{ρ} \AgdaFunction{\ensuremath{\mathbin{{>}\mkern-8.5mu{>}\mkern-2mu{=}}}}\<%
\\
\>[2]\AgdaIndent{4}{}\<[4]%
\>[4]\AgdaSymbol{λ} \AgdaBound{\,f\,} \AgdaSymbol{→} \AgdaFunction{eval} \AgdaBound{u} \AgdaBound{ρ} \AgdaFunction{\ensuremath{\mathbin{{>}\mkern-8.5mu{>}\mkern-2mu{=}}}} \AgdaSymbol{(λ} \AgdaBound{v} \AgdaSymbol{→} \AgdaFunction{apply} \AgdaBound{\,f\,} \AgdaBound{v}\AgdaSymbol{)}\<%
\\
\>[4]\AgdaIndent{6}{}\<[6]%
\>[6]\AgdaFunction{\ensuremath{\mathbin{{>}\mkern-8.5mu{>}\mkern-2mu{=}}}} \AgdaSymbol{(λ} \AgdaBound{x′} \AgdaSymbol{→} \AgdaInductiveConstructor{now} \AgdaSymbol{(}\AgdaFunction{val≤} \AgdaBound{η} \AgdaBound{x′}\AgdaSymbol{)))}\<%
\\
\>[0]\AgdaIndent{2}{}\<[2]%
\>[2]\AgdaFunction{\qquad∼⟨} \AgdaFunction{bind-cong-r} \AgdaSymbol{(}\AgdaFunction{eval} \AgdaBound{t} \AgdaBound{ρ}\AgdaSymbol{)} \AgdaSymbol{(λ} \AgdaBound{t₁} \AgdaSymbol{→} \AgdaFunction{bind-assoc} \AgdaSymbol{(}\AgdaFunction{eval} \AgdaBound{u} \AgdaBound{ρ}\AgdaSymbol{))} \AgdaFunction{⟩}\<%
\\
\>[0]\AgdaIndent{2}{}\<[2]%
\>[2]\AgdaSymbol{(}\AgdaFunction{eval} \AgdaBound{t} \AgdaBound{ρ} \AgdaFunction{\ensuremath{\mathbin{{>}\mkern-8.5mu{>}\mkern-2mu{=}}}}\<%
\\
\>[2]\AgdaIndent{4}{}\<[4]%
\>[4]\AgdaSymbol{λ} \AgdaBound{\,f\,} \AgdaSymbol{→} \AgdaFunction{eval} \AgdaBound{u} \AgdaBound{ρ} \AgdaFunction{\ensuremath{\mathbin{{>}\mkern-8.5mu{>}\mkern-2mu{=}}}} \AgdaSymbol{λ} \AgdaBound{v} \AgdaSymbol{→} \AgdaFunction{apply} \AgdaBound{\,f\,} \AgdaBound{v}\<%
\\
\>[4]\AgdaIndent{6}{}\<[6]%
\>[6]\AgdaFunction{\ensuremath{\mathbin{{>}\mkern-8.5mu{>}\mkern-2mu{=}}}} \AgdaSymbol{(λ} \AgdaBound{x′} \AgdaSymbol{→} \AgdaInductiveConstructor{now} \AgdaSymbol{(}\AgdaFunction{val≤} \AgdaBound{η} \AgdaBound{x′}\AgdaSymbol{)))}\<%
\\
\>[0]\AgdaIndent{2}{}\<[2]%
\>[2]\AgdaFunction{\qquad∼⟨} \AgdaFunction{bind-cong-r} \AgdaSymbol{(}\AgdaFunction{eval} \AgdaBound{t} \AgdaBound{ρ}\AgdaSymbol{)}\<%
\\
\>[2]\AgdaIndent{17}{}\<[17]%
\>[17]\AgdaSymbol{(λ} \AgdaBound{t₁} \AgdaSymbol{→} \AgdaFunction{bind-cong-r} \AgdaSymbol{(}\AgdaFunction{eval} \AgdaBound{u} \AgdaBound{ρ}\AgdaSymbol{)}\<%
\\
\>[17]\AgdaIndent{37}{}\<[37]%
\>[37]\AgdaSymbol{(λ} \AgdaBound{u₁} \AgdaSymbol{→} \AgdaFunction{apply≤} \AgdaBound{t₁} \AgdaBound{u₁} \AgdaBound{η} \AgdaSymbol{))} \AgdaFunction{⟩}\<%
\\
\>[0]\AgdaIndent{2}{}\<[2]%
\>[2]\AgdaSymbol{(}\AgdaFunction{eval} \AgdaBound{t} \AgdaBound{ρ} \AgdaFunction{\ensuremath{\mathbin{{>}\mkern-8.5mu{>}\mkern-2mu{=}}}}\<%
\\
\>[2]\AgdaIndent{3}{}\<[3]%
\>[3]\AgdaSymbol{λ} \AgdaBound{x′} \AgdaSymbol{→} \AgdaFunction{eval} \AgdaBound{u} \AgdaBound{ρ} \AgdaFunction{\ensuremath{\mathbin{{>}\mkern-8.5mu{>}\mkern-2mu{=}}}} \AgdaSymbol{(λ} \AgdaBound{x′′} \AgdaSymbol{→} \AgdaFunction{apply} \AgdaSymbol{(}\AgdaFunction{val≤} \AgdaBound{η} \AgdaBound{x′}\AgdaSymbol{)} \AgdaSymbol{(}\AgdaFunction{val≤} \AgdaBound{η} \AgdaBound{x′′}\AgdaSymbol{)))}\<%
\\
\>[0]\AgdaIndent{2}{}\<[2]%
\>[2]\AgdaFunction{\qquad≡⟨⟩}\<%
\\
\>[0]\AgdaIndent{2}{}\<[2]%
\>[2]\AgdaSymbol{(}\AgdaFunction{eval} \AgdaBound{t} \AgdaBound{ρ} \AgdaFunction{\ensuremath{\mathbin{{>}\mkern-8.5mu{>}\mkern-2mu{=}}}} \AgdaSymbol{λ} \AgdaBound{x′} \AgdaSymbol{→}\<%
\\
\>[2]\AgdaIndent{6}{}\<[6]%
\>[6]\AgdaSymbol{(}\AgdaFunction{eval} \AgdaBound{u} \AgdaBound{ρ} \AgdaFunction{\ensuremath{\mathbin{{>}\mkern-8.5mu{>}\mkern-2mu{=}}}} \AgdaSymbol{λ} \AgdaBound{x′′} \AgdaSymbol{→} \AgdaInductiveConstructor{now} \AgdaSymbol{(}\AgdaFunction{val≤} \AgdaBound{η} \AgdaBound{x′′}\AgdaSymbol{)} \AgdaFunction{\ensuremath{\mathbin{{>}\mkern-8.5mu{>}\mkern-2mu{=}}}}\<%
\\
\>[6]\AgdaIndent{8}{}\<[8]%
\>[8]\AgdaSymbol{λ} \AgdaBound{v} \AgdaSymbol{→} \AgdaFunction{apply} \AgdaSymbol{(}\AgdaFunction{val≤} \AgdaBound{η} \AgdaBound{x′}\AgdaSymbol{)} \AgdaBound{v}\AgdaSymbol{))}\<%
\\
\>[0]\AgdaIndent{2}{}\<[2]%
\>[2]\AgdaFunction{\qquad∼⟨} \AgdaFunction{bind-cong-r} \AgdaSymbol{(}\AgdaFunction{eval} \AgdaBound{t} \AgdaBound{ρ}\AgdaSymbol{)} \AgdaSymbol{(λ} \AgdaBound{a} \AgdaSymbol{→} \AgdaFunction{∼sym} \AgdaSymbol{(}\AgdaFunction{bind-assoc} \AgdaSymbol{(}\AgdaFunction{eval} \AgdaBound{u} \AgdaBound{ρ}\AgdaSymbol{)))} \AgdaFunction{⟩}\<%
\\
\>[0]\AgdaIndent{2}{}\<[2]%
\>[2]\AgdaSymbol{(}\AgdaFunction{eval} \AgdaBound{t} \AgdaBound{ρ} \AgdaFunction{\ensuremath{\mathbin{{>}\mkern-8.5mu{>}\mkern-2mu{=}}}} \AgdaSymbol{λ} \AgdaBound{x′} \AgdaSymbol{→}\<%
\\
\>[2]\AgdaIndent{6}{}\<[6]%
\>[6]\AgdaSymbol{(}\AgdaFunction{eval} \AgdaBound{u} \AgdaBound{ρ} \AgdaFunction{\ensuremath{\mathbin{{>}\mkern-8.5mu{>}\mkern-2mu{=}}}} \AgdaSymbol{λ} \AgdaBound{x′′} \AgdaSymbol{→} \AgdaInductiveConstructor{now} \AgdaSymbol{(}\AgdaFunction{val≤} \AgdaBound{η} \AgdaBound{x′′}\AgdaSymbol{))} \AgdaFunction{\ensuremath{\mathbin{{>}\mkern-8.5mu{>}\mkern-2mu{=}}}}\<%
\\
\>[6]\AgdaIndent{8}{}\<[8]%
\>[8]\AgdaSymbol{(λ} \AgdaBound{v} \AgdaSymbol{→} \AgdaFunction{apply} \AgdaSymbol{(}\AgdaFunction{val≤} \AgdaBound{η} \AgdaBound{x′}\AgdaSymbol{)} \AgdaBound{v}\AgdaSymbol{))}\<%
\\
\>[0]\AgdaIndent{2}{}\<[2]%
\>[2]\AgdaFunction{\qquad∼⟨} \AgdaFunction{bind-cong-r} \AgdaSymbol{(}\AgdaFunction{eval} \AgdaBound{t} \AgdaBound{ρ}\AgdaSymbol{)} \AgdaSymbol{(λ} \AgdaBound{x′} \AgdaSymbol{→} \AgdaFunction{bind-cong-l} \<[49]%
\>[49]\AgdaSymbol{(}\AgdaFunction{eval≤} \AgdaBound{u} \AgdaBound{ρ} \AgdaBound{η}\AgdaSymbol{)} \AgdaSymbol{(λ} \AgdaBound{\_} \AgdaSymbol{→} \AgdaSymbol{\_))} \AgdaFunction{⟩}\<%
\\
\>[0]\AgdaIndent{2}{}\<[2]%
\>[2]\AgdaSymbol{(}\AgdaFunction{eval} \AgdaBound{t} \AgdaBound{ρ} \AgdaFunction{\ensuremath{\mathbin{{>}\mkern-8.5mu{>}\mkern-2mu{=}}}} \AgdaSymbol{λ} \AgdaBound{x′} \AgdaSymbol{→}\<%
\\
\>[2]\AgdaIndent{6}{}\<[6]%
\>[6]\AgdaFunction{eval} \AgdaBound{u} \AgdaSymbol{(}\AgdaFunction{env≤} \AgdaBound{η} \AgdaBound{ρ}\AgdaSymbol{)} \AgdaFunction{\ensuremath{\mathbin{{>}\mkern-8.5mu{>}\mkern-2mu{=}}}} \AgdaSymbol{(λ} \AgdaBound{v} \AgdaSymbol{→} \AgdaFunction{apply} \AgdaSymbol{(}\AgdaFunction{val≤} \AgdaBound{η} \AgdaBound{x′}\AgdaSymbol{)} \AgdaBound{v}\AgdaSymbol{))}\<%
\\
\>[0]\AgdaIndent{2}{}\<[2]%
\>[2]\AgdaFunction{\qquad≡⟨⟩}\<%
\\
\>[0]\AgdaIndent{2}{}\<[2]%
\>[2]\AgdaSymbol{(}\AgdaFunction{eval} \AgdaBound{t} \AgdaBound{ρ} \AgdaFunction{\ensuremath{\mathbin{{>}\mkern-8.5mu{>}\mkern-2mu{=}}}} \AgdaSymbol{λ} \AgdaBound{x′} \AgdaSymbol{→} \AgdaInductiveConstructor{now} \AgdaSymbol{(}\AgdaFunction{val≤} \AgdaBound{η} \AgdaBound{x′}\AgdaSymbol{)} \AgdaFunction{\ensuremath{\mathbin{{>}\mkern-8.5mu{>}\mkern-2mu{=}}}}\<%
\\
\>[2]\AgdaIndent{4}{}\<[4]%
\>[4]\AgdaSymbol{(λ} \AgdaBound{\,f\,} \AgdaSymbol{→} \AgdaFunction{eval} \AgdaBound{u} \AgdaSymbol{(}\AgdaFunction{env≤} \AgdaBound{η} \AgdaBound{ρ}\AgdaSymbol{)} \AgdaFunction{\ensuremath{\mathbin{{>}\mkern-8.5mu{>}\mkern-2mu{=}}}} \AgdaSymbol{(λ} \AgdaBound{v} \AgdaSymbol{→} \AgdaFunction{apply} \AgdaBound{\,f\,} \AgdaBound{v}\AgdaSymbol{)))}\<%
\\
\>[0]\AgdaIndent{2}{}\<[2]%
\>[2]\AgdaFunction{\qquad∼⟨} \AgdaFunction{∼sym} \AgdaSymbol{(}\AgdaFunction{bind-assoc} \AgdaSymbol{(}\AgdaFunction{eval} \AgdaBound{t} \AgdaBound{ρ}\AgdaSymbol{))} \AgdaFunction{⟩}\<%
\\
\>[0]\AgdaIndent{2}{}\<[2]%
\>[2]\AgdaSymbol{((}\AgdaFunction{eval} \AgdaBound{t} \AgdaBound{ρ} \AgdaFunction{\ensuremath{\mathbin{{>}\mkern-8.5mu{>}\mkern-2mu{=}}}} \AgdaSymbol{(λ} \AgdaBound{x′} \AgdaSymbol{→} \AgdaInductiveConstructor{now} \AgdaSymbol{(}\AgdaFunction{val≤} \AgdaBound{η} \AgdaBound{x′}\AgdaSymbol{)))} \AgdaFunction{\ensuremath{\mathbin{{>}\mkern-8.5mu{>}\mkern-2mu{=}}}}\<%
\\
\>[2]\AgdaIndent{4}{}\<[4]%
\>[4]\AgdaSymbol{(λ} \AgdaBound{\,f\,} \AgdaSymbol{→} \AgdaFunction{eval} \AgdaBound{u} \AgdaSymbol{(}\AgdaFunction{env≤} \AgdaBound{η} \AgdaBound{ρ}\AgdaSymbol{)} \AgdaFunction{\ensuremath{\mathbin{{>}\mkern-8.5mu{>}\mkern-2mu{=}}}} \AgdaSymbol{(λ} \AgdaBound{v} \AgdaSymbol{→} \AgdaFunction{apply} \AgdaBound{\,f\,} \AgdaBound{v}\AgdaSymbol{)))}\<%
\\
\>[0]\AgdaIndent{2}{}\<[2]%
\>[2]\AgdaFunction{\qquad∼⟨} \AgdaFunction{bind-cong-l} \AgdaSymbol{(}\AgdaFunction{eval≤} \AgdaBound{t} \AgdaBound{ρ} \AgdaBound{η}\AgdaSymbol{)} \AgdaSymbol{(λ} \AgdaBound{\_} \AgdaSymbol{→} \AgdaSymbol{\_)} \AgdaFunction{⟩}\<%
\\
\>[0]\AgdaIndent{2}{}\<[2]%
\>[2]\AgdaSymbol{(}\AgdaFunction{eval} \AgdaBound{t} \AgdaSymbol{(}\AgdaFunction{env≤} \AgdaBound{η} \AgdaBound{ρ}\AgdaSymbol{)} \AgdaFunction{\ensuremath{\mathbin{{>}\mkern-8.5mu{>}\mkern-2mu{=}}}}\<%
\\
\>[2]\AgdaIndent{4}{}\<[4]%
\>[4]\AgdaSymbol{(λ} \AgdaBound{\,f\,} \AgdaSymbol{→} \AgdaFunction{eval} \AgdaBound{u} \AgdaSymbol{(}\AgdaFunction{env≤} \AgdaBound{η} \AgdaBound{ρ}\AgdaSymbol{)} \AgdaFunction{\ensuremath{\mathbin{{>}\mkern-8.5mu{>}\mkern-2mu{=}}}} \AgdaSymbol{(λ} \AgdaBound{v} \AgdaSymbol{→} \AgdaFunction{apply} \AgdaBound{\,f\,} \AgdaBound{v}\AgdaSymbol{)))}\<%
\\
\>[0]\AgdaIndent{2}{}\<[2]%
\>[2]\AgdaFunction{∎}\<%
\\
\>[0]\AgdaIndent{2}{}\<[2]%
\>[2]\AgdaKeyword{where} \AgdaKeyword{open} \AgdaModule{∼-Reasoning}\<%
\\
%
\\
%
\\
\>\AgdaFunction{apply≤} \AgdaSymbol{(}\AgdaInductiveConstructor{ne} \AgdaBound{x}\AgdaSymbol{)} \<[17]%
\>[17]\AgdaBound{v} \AgdaBound{η} \AgdaSymbol{=} \AgdaFunction{∼refl} \AgdaSymbol{\_}\<%
\\
\>\AgdaFunction{apply≤} \AgdaSymbol{(}\AgdaInductiveConstructor{lam} \AgdaBound{t} \AgdaBound{ρ}\AgdaSymbol{)} \AgdaBound{v} \AgdaBound{η} \AgdaSymbol{=} \AgdaInductiveConstructor{∼later} \AgdaSymbol{(}\AgdaFunction{beta≤} \AgdaBound{t} \AgdaBound{ρ} \AgdaBound{v} \AgdaBound{η}\AgdaSymbol{)}\<%
\\
%
\\
\>\AgdaField{∼force} \AgdaSymbol{(}\AgdaFunction{beta≤} \AgdaBound{t} \AgdaBound{ρ} \AgdaBound{v} \AgdaBound{η}\AgdaSymbol{)} \AgdaSymbol{=} \AgdaFunction{eval≤} \AgdaBound{t} \AgdaSymbol{(}\AgdaBound{ρ} \AgdaInductiveConstructor{,} \AgdaBound{v}\AgdaSymbol{)} \AgdaBound{η}\<%
\end{code}}%
\begin{samepage}
\begin{code}%
\>\AgdaFunction{nereadback≤} \<[13]%
\>[13]\AgdaSymbol{:} \<[16]%
\>[16]\AgdaSymbol{∀\{}\AgdaBound{i} \AgdaBound{Γ} \AgdaBound{Δ} \AgdaBound{a}\AgdaSymbol{\}(}\AgdaBound{η} \AgdaSymbol{:} \AgdaBound{Δ} \AgdaDatatype{≤} \AgdaBound{Γ}\AgdaSymbol{)(}\AgdaBound{t} \AgdaSymbol{:} \AgdaDatatype{Ne} \AgdaDatatype{Val} \AgdaBound{Γ} \AgdaBound{a}\AgdaSymbol{)} \AgdaSymbol{→}\<%
\\
\>[2]\AgdaIndent{16}{}\<[16]%
\>[16]\AgdaSymbol{(}\AgdaFunction{nen≤} \AgdaBound{η} \AgdaFunction{\ensuremath{{<}\${>}}} \AgdaFunction{nereadback} \AgdaBound{t}\AgdaSymbol{)} \AgdaFunction{∼⟨} \AgdaBound{i} \AgdaFunction{⟩∼} \AgdaSymbol{(}\AgdaFunction{nereadback} \AgdaSymbol{(}\AgdaFunction{nev≤} \AgdaBound{η} \AgdaBound{t}\AgdaSymbol{))}\<%
\\
%
\\
\>\AgdaFunction{readback≤} \<[13]%
\>[13]\AgdaSymbol{:} \<[16]%
\>[16]\AgdaSymbol{∀\{}\AgdaBound{i} \AgdaBound{Γ} \AgdaBound{Δ}\AgdaSymbol{\}} \AgdaBound{a} \AgdaSymbol{(}\AgdaBound{η} \AgdaSymbol{:} \AgdaBound{Δ} \AgdaDatatype{≤} \AgdaBound{Γ}\AgdaSymbol{)(}\AgdaBound{v} \AgdaSymbol{:} \AgdaDatatype{Val} \AgdaBound{Γ} \AgdaBound{a}\AgdaSymbol{)} \AgdaSymbol{→}\<%
\\
\>[2]\AgdaIndent{16}{}\<[16]%
\>[16]\AgdaSymbol{(}\AgdaFunction{nf≤} \AgdaBound{η} \AgdaFunction{\ensuremath{{<}\${>}}} \AgdaFunction{readback} \AgdaBound{v}\AgdaSymbol{)} \AgdaFunction{∼⟨} \AgdaBound{i} \AgdaFunction{⟩∼} \AgdaSymbol{(}\AgdaFunction{readback} \AgdaSymbol{(}\AgdaFunction{val≤} \AgdaBound{η} \AgdaBound{v}\AgdaSymbol{))}\<%
\\
%
\\
\>\AgdaFunction{eta≤} \<[13]%
\>[13]\AgdaSymbol{:} \<[16]%
\>[16]\AgdaSymbol{∀\{}\AgdaBound{i} \AgdaBound{Γ} \AgdaBound{Δ} \AgdaBound{a} \AgdaBound{b}\AgdaSymbol{\}} \AgdaSymbol{(}\AgdaBound{η} \AgdaSymbol{:} \AgdaBound{Δ} \AgdaDatatype{≤} \AgdaBound{Γ}\AgdaSymbol{)(}\AgdaBound{v} \AgdaSymbol{:} \AgdaDatatype{Val} \AgdaBound{Γ} \AgdaSymbol{(}\AgdaBound{a} \AgdaInductiveConstructor{⇒} \AgdaBound{b}\AgdaSymbol{))} \AgdaSymbol{→}\<%
\\
\>[2]\AgdaIndent{16}{}\<[16]%
\>[16]\AgdaSymbol{(}\AgdaFunction{nf≤} \AgdaSymbol{(}\AgdaInductiveConstructor{lift} \AgdaBound{η}\AgdaSymbol{)} \AgdaFunction{∞\ensuremath{{<}\${>}}} \AgdaFunction{eta} \AgdaBound{v}\AgdaSymbol{)} \AgdaRecord{∞∼⟨} \AgdaBound{i} \AgdaRecord{⟩∼} \AgdaSymbol{(}\AgdaFunction{eta} \AgdaSymbol{(}\AgdaFunction{val≤} \AgdaBound{η} \AgdaBound{v}\AgdaSymbol{))}\<%
\end{code} 
\end{samepage}

\AgdaHide{
\begin{code}%
\>\AgdaFunction{nereadback≤} \AgdaBound{η} \AgdaSymbol{(}\AgdaInductiveConstructor{var} \AgdaBound{x}\AgdaSymbol{)} \AgdaSymbol{=} \AgdaInductiveConstructor{∼now} \AgdaSymbol{\_}\<%
\\
\>\AgdaFunction{nereadback≤} \AgdaBound{η} \AgdaSymbol{(}\AgdaInductiveConstructor{app} \AgdaBound{t} \AgdaBound{u}\AgdaSymbol{)} \AgdaSymbol{=}\<%
\\
\>[0]\AgdaIndent{2}{}\<[2]%
\>[2]\AgdaFunction{proof}\<%
\\
\>[0]\AgdaIndent{2}{}\<[2]%
\>[2]\AgdaSymbol{((}\AgdaFunction{nereadback} \AgdaBound{t} \AgdaFunction{\ensuremath{\mathbin{{>}\mkern-8.5mu{>}\mkern-2mu{=}}}}\<%
\\
\>[2]\AgdaIndent{4}{}\<[4]%
\>[4]\AgdaSymbol{(λ} \AgdaBound{t₁} \AgdaSymbol{→} \AgdaFunction{readback} \AgdaBound{u} \AgdaFunction{\ensuremath{\mathbin{{>}\mkern-8.5mu{>}\mkern-2mu{=}}}} \AgdaSymbol{(λ} \AgdaBound{n} \AgdaSymbol{→} \AgdaInductiveConstructor{now} \AgdaSymbol{(}\AgdaInductiveConstructor{app} \AgdaBound{t₁} \AgdaBound{n}\AgdaSymbol{))))}\<%
\\
\>[4]\AgdaIndent{33}{}\<[33]%
\>[33]\AgdaFunction{\ensuremath{\mathbin{{>}\mkern-8.5mu{>}\mkern-2mu{=}}}} \AgdaSymbol{(λ} \AgdaBound{x′} \AgdaSymbol{→} \AgdaInductiveConstructor{now} \AgdaSymbol{(}\AgdaFunction{nen≤} \AgdaBound{η} \AgdaBound{x′}\AgdaSymbol{)))}\<%
\\
\>[0]\AgdaIndent{2}{}\<[2]%
\>[2]\AgdaFunction{\qquad∼⟨} \AgdaFunction{bind-assoc} \AgdaSymbol{(}\AgdaFunction{nereadback} \AgdaBound{t}\AgdaSymbol{)} \AgdaFunction{⟩}\<%
\\
\>[0]\AgdaIndent{2}{}\<[2]%
\>[2]\AgdaSymbol{(}\AgdaFunction{nereadback} \AgdaBound{t} \AgdaFunction{\ensuremath{\mathbin{{>}\mkern-8.5mu{>}\mkern-2mu{=}}}} \AgdaSymbol{(λ} \AgdaBound{x} \AgdaSymbol{→}\<%
\\
\>[2]\AgdaIndent{4}{}\<[4]%
\>[4]\AgdaSymbol{(}\AgdaFunction{readback} \AgdaBound{u} \AgdaFunction{\ensuremath{\mathbin{{>}\mkern-8.5mu{>}\mkern-2mu{=}}}} \AgdaSymbol{(λ} \AgdaBound{n} \AgdaSymbol{→} \AgdaInductiveConstructor{now} \AgdaSymbol{(}\AgdaInductiveConstructor{app} \AgdaBound{x} \AgdaBound{n}\AgdaSymbol{)))}\<%
\\
\>[4]\AgdaIndent{33}{}\<[33]%
\>[33]\AgdaFunction{\ensuremath{\mathbin{{>}\mkern-8.5mu{>}\mkern-2mu{=}}}} \AgdaSymbol{(λ} \AgdaBound{x′} \AgdaSymbol{→} \AgdaInductiveConstructor{now} \AgdaSymbol{(}\AgdaFunction{nen≤} \AgdaBound{η} \AgdaBound{x′}\AgdaSymbol{))))}\<%
\\
\>[0]\AgdaIndent{2}{}\<[2]%
\>[2]\AgdaFunction{\qquad∼⟨} \AgdaFunction{bind-cong-r} \AgdaSymbol{(}\AgdaFunction{nereadback} \AgdaBound{t}\AgdaSymbol{)} \AgdaSymbol{(λ} \AgdaBound{x} \AgdaSymbol{→} \AgdaFunction{bind-assoc} \AgdaSymbol{(}\AgdaFunction{readback} \AgdaBound{u}\AgdaSymbol{))} \AgdaFunction{⟩}\<%
\\
\>[0]\AgdaIndent{2}{}\<[2]%
\>[2]\AgdaSymbol{(}\AgdaFunction{nereadback} \AgdaBound{t} \AgdaFunction{\ensuremath{\mathbin{{>}\mkern-8.5mu{>}\mkern-2mu{=}}}} \AgdaSymbol{(λ} \AgdaBound{x} \AgdaSymbol{→}\<%
\\
\>[2]\AgdaIndent{5}{}\<[5]%
\>[5]\AgdaFunction{readback} \AgdaBound{u} \AgdaFunction{\ensuremath{\mathbin{{>}\mkern-8.5mu{>}\mkern-2mu{=}}}} \AgdaSymbol{(λ} \AgdaBound{y} \AgdaSymbol{→} \AgdaInductiveConstructor{now} \AgdaSymbol{(}\AgdaInductiveConstructor{app} \AgdaBound{x} \AgdaBound{y}\AgdaSymbol{)} \AgdaFunction{\ensuremath{\mathbin{{>}\mkern-8.5mu{>}\mkern-2mu{=}}}} \AgdaSymbol{(λ} \AgdaBound{x′} \AgdaSymbol{→} \AgdaInductiveConstructor{now} \AgdaSymbol{(}\AgdaFunction{nen≤} \AgdaBound{η} \AgdaBound{x′}\AgdaSymbol{)))))}\<%
\\
\>[0]\AgdaIndent{2}{}\<[2]%
\>[2]\AgdaFunction{\qquad≡⟨⟩}\<%
\\
\>[0]\AgdaIndent{2}{}\<[2]%
\>[2]\AgdaSymbol{(}\AgdaFunction{nereadback} \AgdaBound{t} \AgdaFunction{\ensuremath{\mathbin{{>}\mkern-8.5mu{>}\mkern-2mu{=}}}}\<%
\\
\>[2]\AgdaIndent{4}{}\<[4]%
\>[4]\AgdaSymbol{(λ} \AgdaBound{x} \AgdaSymbol{→} \AgdaSymbol{(}\AgdaFunction{readback} \AgdaBound{u} \AgdaFunction{\ensuremath{\mathbin{{>}\mkern-8.5mu{>}\mkern-2mu{=}}}} \AgdaSymbol{(λ} \AgdaBound{y} \AgdaSymbol{→} \AgdaInductiveConstructor{now} \AgdaSymbol{(}\AgdaInductiveConstructor{app} \AgdaSymbol{(}\AgdaFunction{nen≤} \AgdaBound{η} \AgdaBound{x}\AgdaSymbol{)} \AgdaSymbol{(}\AgdaFunction{nf≤} \AgdaBound{η} \AgdaBound{y}\AgdaSymbol{))))))}\<%
\\
\>[0]\AgdaIndent{2}{}\<[2]%
\>[2]\AgdaFunction{\qquad≡⟨⟩}\<%
\\
\>[0]\AgdaIndent{2}{}\<[2]%
\>[2]\AgdaSymbol{(}\AgdaFunction{nereadback} \AgdaBound{t} \AgdaFunction{\ensuremath{\mathbin{{>}\mkern-8.5mu{>}\mkern-2mu{=}}}}\<%
\\
\>[2]\AgdaIndent{9}{}\<[9]%
\>[9]\AgdaSymbol{(λ} \AgdaBound{x} \AgdaSymbol{→} \AgdaSymbol{(}\AgdaFunction{readback} \AgdaBound{u} \AgdaFunction{\ensuremath{\mathbin{{>}\mkern-8.5mu{>}\mkern-2mu{=}}}} \AgdaSymbol{(λ} \AgdaBound{x′} \AgdaSymbol{→} \AgdaInductiveConstructor{now} \AgdaSymbol{(}\AgdaFunction{nf≤} \AgdaBound{η} \AgdaBound{x′}\AgdaSymbol{)} \AgdaFunction{\ensuremath{\mathbin{{>}\mkern-8.5mu{>}\mkern-2mu{=}}}}\<%
\\
\>[9]\AgdaIndent{13}{}\<[13]%
\>[13]\AgdaSymbol{(λ} \AgdaBound{n} \AgdaSymbol{→} \AgdaInductiveConstructor{now} \AgdaSymbol{(}\AgdaInductiveConstructor{app} \AgdaSymbol{(}\AgdaFunction{nen≤} \AgdaBound{η} \AgdaBound{x}\AgdaSymbol{)} \AgdaBound{n}\AgdaSymbol{))))))}\<%
\\
\>[0]\AgdaIndent{2}{}\<[2]%
\>[2]\AgdaFunction{\qquad∼⟨} \AgdaFunction{bind-cong-r} \AgdaSymbol{(}\AgdaFunction{nereadback} \AgdaBound{t}\AgdaSymbol{)} \AgdaSymbol{(λ} \AgdaBound{x} \AgdaSymbol{→} \AgdaFunction{∼sym} \AgdaSymbol{(}\AgdaFunction{bind-assoc} \AgdaSymbol{(}\AgdaFunction{readback} \AgdaBound{u}\AgdaSymbol{)))} \AgdaFunction{⟩}\<%
\\
\>[0]\AgdaIndent{2}{}\<[2]%
\>[2]\AgdaSymbol{(}\AgdaFunction{nereadback} \AgdaBound{t} \AgdaFunction{\ensuremath{\mathbin{{>}\mkern-8.5mu{>}\mkern-2mu{=}}}}\<%
\\
\>[2]\AgdaIndent{9}{}\<[9]%
\>[9]\AgdaSymbol{(λ} \AgdaBound{x} \AgdaSymbol{→} \AgdaSymbol{((}\AgdaFunction{readback} \AgdaBound{u} \AgdaFunction{\ensuremath{\mathbin{{>}\mkern-8.5mu{>}\mkern-2mu{=}}}} \AgdaSymbol{(λ} \AgdaBound{x′} \AgdaSymbol{→} \AgdaInductiveConstructor{now} \AgdaSymbol{(}\AgdaFunction{nf≤} \AgdaBound{η} \AgdaBound{x′}\AgdaSymbol{)))} \AgdaFunction{\ensuremath{\mathbin{{>}\mkern-8.5mu{>}\mkern-2mu{=}}}}\<%
\\
\>[9]\AgdaIndent{13}{}\<[13]%
\>[13]\AgdaSymbol{(λ} \AgdaBound{n} \AgdaSymbol{→} \AgdaInductiveConstructor{now} \AgdaSymbol{(}\AgdaInductiveConstructor{app} \AgdaSymbol{(}\AgdaFunction{nen≤} \AgdaBound{η} \AgdaBound{x}\AgdaSymbol{)} \AgdaBound{n}\AgdaSymbol{)))))}\<%
\\
\>[0]\AgdaIndent{2}{}\<[2]%
\>[2]\AgdaFunction{\qquad≡⟨⟩}\<%
\\
\>[0]\AgdaIndent{2}{}\<[2]%
\>[2]\AgdaSymbol{(}\AgdaFunction{nereadback} \AgdaBound{t} \AgdaFunction{\ensuremath{\mathbin{{>}\mkern-8.5mu{>}\mkern-2mu{=}}}} \AgdaSymbol{(λ} \AgdaBound{x} \AgdaSymbol{→} \AgdaInductiveConstructor{now} \AgdaSymbol{(}\AgdaFunction{nen≤} \AgdaBound{η} \AgdaBound{x}\AgdaSymbol{)} \AgdaFunction{\ensuremath{\mathbin{{>}\mkern-8.5mu{>}\mkern-2mu{=}}}}\<%
\\
\>[2]\AgdaIndent{4}{}\<[4]%
\>[4]\AgdaSymbol{(λ} \AgdaBound{t₁} \AgdaSymbol{→} \AgdaSymbol{((}\AgdaFunction{readback} \AgdaBound{u} \AgdaFunction{\ensuremath{\mathbin{{>}\mkern-8.5mu{>}\mkern-2mu{=}}}} \AgdaSymbol{(λ} \AgdaBound{x′} \AgdaSymbol{→} \AgdaInductiveConstructor{now} \AgdaSymbol{(}\AgdaFunction{nf≤} \AgdaBound{η} \AgdaBound{x′}\AgdaSymbol{)))} \AgdaFunction{\ensuremath{\mathbin{{>}\mkern-8.5mu{>}\mkern-2mu{=}}}}\<%
\\
\>[4]\AgdaIndent{8}{}\<[8]%
\>[8]\AgdaSymbol{(λ} \AgdaBound{n} \AgdaSymbol{→} \AgdaInductiveConstructor{now} \AgdaSymbol{(}\AgdaInductiveConstructor{app} \AgdaBound{t₁} \AgdaBound{n}\AgdaSymbol{))))))}\<%
\\
\>[0]\AgdaIndent{2}{}\<[2]%
\>[2]\AgdaFunction{\qquad∼⟨} \AgdaFunction{∼sym} \AgdaSymbol{(}\AgdaFunction{bind-assoc} \AgdaSymbol{(}\AgdaFunction{nereadback} \AgdaBound{t}\AgdaSymbol{))} \AgdaFunction{⟩}\<%
\\
\>[0]\AgdaIndent{2}{}\<[2]%
\>[2]\AgdaSymbol{((}\AgdaFunction{nereadback} \AgdaBound{t} \AgdaFunction{\ensuremath{\mathbin{{>}\mkern-8.5mu{>}\mkern-2mu{=}}}} \AgdaSymbol{(λ} \AgdaBound{x′} \AgdaSymbol{→} \AgdaInductiveConstructor{now} \AgdaSymbol{(}\AgdaFunction{nen≤} \AgdaBound{η} \AgdaBound{x′}\AgdaSymbol{)))} \AgdaFunction{\ensuremath{\mathbin{{>}\mkern-8.5mu{>}\mkern-2mu{=}}}}\<%
\\
\>[2]\AgdaIndent{4}{}\<[4]%
\>[4]\AgdaSymbol{(λ} \AgdaBound{t₁} \AgdaSymbol{→} \AgdaSymbol{((}\AgdaFunction{readback} \AgdaBound{u} \AgdaFunction{\ensuremath{\mathbin{{>}\mkern-8.5mu{>}\mkern-2mu{=}}}} \AgdaSymbol{(λ} \AgdaBound{x′} \AgdaSymbol{→} \AgdaInductiveConstructor{now} \AgdaSymbol{(}\AgdaFunction{nf≤} \AgdaBound{η} \AgdaBound{x′}\AgdaSymbol{)))} \AgdaFunction{\ensuremath{\mathbin{{>}\mkern-8.5mu{>}\mkern-2mu{=}}}}\<%
\\
\>[4]\AgdaIndent{8}{}\<[8]%
\>[8]\AgdaSymbol{(λ} \AgdaBound{n} \AgdaSymbol{→} \AgdaInductiveConstructor{now} \AgdaSymbol{(}\AgdaInductiveConstructor{app} \AgdaBound{t₁} \AgdaBound{n}\AgdaSymbol{)))))}\<%
\\
\>[0]\AgdaIndent{2}{}\<[2]%
\>[2]\AgdaFunction{\qquad≡⟨⟩}\<%
\\
\>[0]\AgdaIndent{2}{}\<[2]%
\>[2]\AgdaSymbol{(}\AgdaFunction{nen≤} \AgdaBound{η} \AgdaFunction{\ensuremath{{<}\${>}}} \AgdaFunction{nereadback} \AgdaBound{t} \AgdaFunction{\ensuremath{\mathbin{{>}\mkern-8.5mu{>}\mkern-2mu{=}}}}\<%
\\
\>[2]\AgdaIndent{5}{}\<[5]%
\>[5]\AgdaSymbol{(λ} \AgdaBound{t₁} \AgdaSymbol{→} \AgdaFunction{nf≤} \AgdaBound{η} \AgdaFunction{\ensuremath{{<}\${>}}} \AgdaFunction{readback} \AgdaBound{u} \AgdaFunction{\ensuremath{\mathbin{{>}\mkern-8.5mu{>}\mkern-2mu{=}}}} \AgdaSymbol{(λ} \AgdaBound{n} \AgdaSymbol{→} \AgdaInductiveConstructor{now} \AgdaSymbol{(}\AgdaInductiveConstructor{app} \AgdaBound{t₁} \AgdaBound{n}\AgdaSymbol{))))}\<%
\\
\>[0]\AgdaIndent{2}{}\<[2]%
\>[2]\AgdaFunction{\qquad∼⟨} \AgdaFunction{bind-cong-r} \AgdaSymbol{(}\AgdaFunction{nen≤} \AgdaBound{η} \AgdaFunction{\ensuremath{{<}\${>}}} \AgdaFunction{nereadback} \AgdaBound{t}\AgdaSymbol{)}\<%
\\
\>[2]\AgdaIndent{17}{}\<[17]%
\>[17]\AgdaSymbol{(λ} \AgdaBound{x} \AgdaSymbol{→} \AgdaFunction{bind-cong-l} \AgdaSymbol{(}\AgdaFunction{readback≤} \AgdaSymbol{\_} \AgdaBound{η} \AgdaBound{u}\AgdaSymbol{)}\<%
\\
\>[17]\AgdaIndent{36}{}\<[36]%
\>[36]\AgdaSymbol{(λ} \AgdaBound{x} \AgdaSymbol{→} \AgdaSymbol{\_))} \AgdaFunction{⟩}\<%
\\
\>[0]\AgdaIndent{2}{}\<[2]%
\>[2]\AgdaSymbol{(}\AgdaFunction{nen≤} \AgdaBound{η} \AgdaFunction{\ensuremath{{<}\${>}}} \AgdaFunction{nereadback} \AgdaBound{t} \AgdaFunction{\ensuremath{\mathbin{{>}\mkern-8.5mu{>}\mkern-2mu{=}}}}\<%
\\
\>[2]\AgdaIndent{5}{}\<[5]%
\>[5]\AgdaSymbol{(λ} \AgdaBound{t₁} \AgdaSymbol{→} \AgdaFunction{readback} \AgdaSymbol{(}\AgdaFunction{val≤} \AgdaBound{η} \AgdaBound{u}\AgdaSymbol{)} \AgdaFunction{\ensuremath{\mathbin{{>}\mkern-8.5mu{>}\mkern-2mu{=}}}} \AgdaSymbol{(λ} \AgdaBound{n} \AgdaSymbol{→} \AgdaInductiveConstructor{now} \AgdaSymbol{(}\AgdaInductiveConstructor{app} \AgdaBound{t₁} \AgdaBound{n}\AgdaSymbol{))))}\<%
\\
\>[0]\AgdaIndent{2}{}\<[2]%
\>[2]\AgdaFunction{\qquad∼⟨} \<[6]%
\>[6]\AgdaFunction{bind-cong-l} \AgdaSymbol{(}\AgdaFunction{nereadback≤} \AgdaBound{η} \AgdaBound{t}\AgdaSymbol{)} \AgdaSymbol{(λ} \AgdaBound{x} \AgdaSymbol{→} \AgdaSymbol{\_)} \AgdaFunction{⟩}\<%
\\
\>[0]\AgdaIndent{2}{}\<[2]%
\>[2]\AgdaSymbol{(}\AgdaFunction{nereadback} \AgdaSymbol{(}\AgdaFunction{nev≤} \AgdaBound{η} \AgdaBound{t}\AgdaSymbol{)} \AgdaFunction{\ensuremath{\mathbin{{>}\mkern-8.5mu{>}\mkern-2mu{=}}}}\<%
\\
\>[2]\AgdaIndent{5}{}\<[5]%
\>[5]\AgdaSymbol{(λ} \AgdaBound{t₁} \AgdaSymbol{→} \AgdaFunction{readback} \AgdaSymbol{(}\AgdaFunction{val≤} \AgdaBound{η} \AgdaBound{u}\AgdaSymbol{)} \AgdaFunction{\ensuremath{\mathbin{{>}\mkern-8.5mu{>}\mkern-2mu{=}}}} \AgdaSymbol{(λ} \AgdaBound{n} \AgdaSymbol{→} \AgdaInductiveConstructor{now} \AgdaSymbol{(}\AgdaInductiveConstructor{app} \AgdaBound{t₁} \AgdaBound{n}\AgdaSymbol{))))}\<%
\\
\>[0]\AgdaIndent{2}{}\<[2]%
\>[2]\AgdaFunction{∎}\<%
\\
\>[0]\AgdaIndent{2}{}\<[2]%
\>[2]\AgdaKeyword{where} \AgdaKeyword{open} \AgdaModule{∼-Reasoning}\<%
\\
\>\<%
\end{code}}

\noindent
As an example of a commutivity lemma, we show the proofs of the base case (type \AgdaInductiveConstructor{★}) for
\AgdaFunction{readback≤}.  The proof is a chain of bisimulation equations (in
relation \presizedbisim{i}), and we use the preorder reasoning package of
Agda's standard library which provides nice syntax for equality
chains, following an idea of Augustsson~\cite{augustsson:equalityProofs}.
Justification for each step is provided in angle brackets, some steps
(\AgdaFunction{≡⟨⟩}) hold directly by definition.

\begin{code}%
\>\AgdaFunction{readback≤} \AgdaInductiveConstructor{★} \AgdaBound{η} \AgdaSymbol{(}\AgdaInductiveConstructor{ne} \AgdaBound{w}\AgdaSymbol{)} \AgdaSymbol{=}\<%
\\
\>[0]\AgdaIndent{2}{}\<[2]%
\>[2]\AgdaFunction{proof}\<%
\\
\>[2]\AgdaIndent{4}{}\<[4]%
\>[4]\AgdaFunction{nf≤} \AgdaBound{η} \<[11]%
\>[11]\AgdaFunction{\ensuremath{{<}\${>}}} \<[16]%
\>[16]\AgdaSymbol{(}\AgdaInductiveConstructor{ne} \<[22]%
\>[22]\AgdaFunction{\ensuremath{{<}\${>}}} \AgdaFunction{nereadback} \AgdaBound{w}\AgdaSymbol{)} \<[42]%
\>[42]\AgdaFunction{\qquad∼⟨} \AgdaFunction{map-compose} \AgdaSymbol{(}\AgdaFunction{nereadback} \AgdaBound{w}\AgdaSymbol{)} \AgdaFunction{⟩}\<%
\\
\>[2]\AgdaIndent{4}{}\<[4]%
\>[4]\AgdaSymbol{(}\AgdaFunction{nf≤} \AgdaBound{η} \AgdaFunction{∘} \AgdaInductiveConstructor{ne}\AgdaSymbol{)} \<[22]%
\>[22]\AgdaFunction{\ensuremath{{<}\${>}}} \AgdaFunction{nereadback} \AgdaBound{w} \<[43]%
\>[43]\AgdaFunction{\qquad≡⟨⟩}\<%
\\
\>[2]\AgdaIndent{4}{}\<[4]%
\>[4]\AgdaSymbol{(}\AgdaInductiveConstructor{Nf.ne} \AgdaFunction{∘} \AgdaFunction{nen≤} \AgdaBound{η}\AgdaSymbol{)} \<[22]%
\>[22]\AgdaFunction{\ensuremath{{<}\${>}}} \AgdaFunction{nereadback} \AgdaBound{w} \<[43]%
\>[43]\AgdaFunction{\qquad∼⟨} \AgdaFunction{∼sym} \AgdaSymbol{(}\AgdaFunction{map-compose} \AgdaSymbol{(}\AgdaFunction{nereadback} \AgdaBound{w}\AgdaSymbol{))} \AgdaFunction{⟩}\<%
\\
\>[2]\AgdaIndent{4}{}\<[4]%
\>[4]\AgdaInductiveConstructor{ne} \AgdaFunction{\ensuremath{{<}\${>}}} \<[12]%
\>[12]\AgdaSymbol{(}\AgdaFunction{nen≤} \AgdaBound{η} \<[22]%
\>[22]\AgdaFunction{\ensuremath{{<}\${>}}} \AgdaFunction{nereadback} \AgdaBound{w}\AgdaSymbol{)} \<[43]%
\>[43]\AgdaFunction{\qquad∼⟨} \AgdaFunction{map-cong} \AgdaInductiveConstructor{ne} \AgdaSymbol{(}\AgdaFunction{nereadback≤} \AgdaBound{η} \AgdaBound{w}\AgdaSymbol{)} \AgdaFunction{⟩}\<%
\\
\>[2]\AgdaIndent{4}{}\<[4]%
\>[4]\AgdaInductiveConstructor{ne} \AgdaFunction{\ensuremath{{<}\${>}}} \<[13]%
\>[13]\AgdaFunction{nereadback} \AgdaSymbol{(}\AgdaFunction{nev≤} \AgdaBound{η} \AgdaBound{w}\AgdaSymbol{)}\<%
\\
\>[0]\AgdaIndent{2}{}\<[2]%
\>[2]\AgdaFunction{∎}\<%
\\
\>[0]\AgdaIndent{2}{}\<[2]%
\>[2]\AgdaKeyword{where} \AgdaKeyword{open} \AgdaModule{∼-Reasoning}\<%
\end{code} 
\AgdaHide{%
\begin{code}%
\>\AgdaFunction{readback≤} \AgdaSymbol{(}\AgdaBound{a} \AgdaInductiveConstructor{⇒} \AgdaBound{b}\AgdaSymbol{)} \AgdaBound{η} \AgdaBound{\,f\,} \<[27]%
\>[27]\AgdaSymbol{=} \AgdaInductiveConstructor{∼later} \AgdaSymbol{(}\<%
\\
\>[0]\AgdaIndent{2}{}\<[2]%
\>[2]\AgdaFunction{proof}\<%
\\
\>[0]\AgdaIndent{2}{}\<[2]%
\>[2]\AgdaSymbol{(}\AgdaFunction{eta} \AgdaBound{\,f\,} \AgdaFunction{\ensuremath{\mathbin{\infty\mkern-3mu{>}\mkern-8.5mu{>}\mkern-2mu{=}}}} \AgdaSymbol{(λ} \AgdaBound{a₁} \AgdaSymbol{→} \AgdaInductiveConstructor{now} \AgdaSymbol{(}\AgdaInductiveConstructor{lam} \AgdaBound{a₁}\AgdaSymbol{)))} \AgdaFunction{\ensuremath{\mathbin{\infty\mkern-3mu{>}\mkern-8.5mu{>}\mkern-2mu{=}}}} \AgdaSymbol{(λ} \AgdaBound{x'} \AgdaSymbol{→} \AgdaInductiveConstructor{now} \AgdaSymbol{(}\AgdaFunction{nf≤} \AgdaBound{η} \AgdaBound{x'}\AgdaSymbol{))}\<%
\\
\>[0]\AgdaIndent{2}{}\<[2]%
\>[2]\AgdaFunction{∞∼⟨} \AgdaFunction{∞bind-assoc} \AgdaSymbol{(}\AgdaFunction{eta} \AgdaBound{\,f\,}\AgdaSymbol{)} \AgdaFunction{⟩}\<%
\\
\>[0]\AgdaIndent{2}{}\<[2]%
\>[2]\AgdaSymbol{(}\AgdaFunction{eta} \AgdaBound{\,f\,} \AgdaFunction{\ensuremath{\mathbin{\infty\mkern-3mu{>}\mkern-8.5mu{>}\mkern-2mu{=}}}} \AgdaSymbol{λ} \AgdaBound{a₁} \AgdaSymbol{→} \AgdaInductiveConstructor{now} \AgdaSymbol{(}\AgdaInductiveConstructor{lam} \AgdaBound{a₁}\AgdaSymbol{)} \AgdaFunction{\ensuremath{\mathbin{{>}\mkern-8.5mu{>}\mkern-2mu{=}}}} \AgdaSymbol{λ} \AgdaBound{x'} \AgdaSymbol{→} \AgdaInductiveConstructor{now} \AgdaSymbol{(}\AgdaFunction{nf≤} \AgdaBound{η} \AgdaBound{x'}\AgdaSymbol{))}\<%
\\
\>[0]\AgdaIndent{2}{}\<[2]%
\>[2]\AgdaFunction{\qquad≡⟨⟩}\<%
\\
\>[0]\AgdaIndent{2}{}\<[2]%
\>[2]\AgdaSymbol{(}\AgdaFunction{eta} \AgdaBound{\,f\,} \AgdaFunction{\ensuremath{\mathbin{\infty\mkern-3mu{>}\mkern-8.5mu{>}\mkern-2mu{=}}}} \AgdaSymbol{(λ} \AgdaBound{a₁} \AgdaSymbol{→} \AgdaInductiveConstructor{now} \AgdaSymbol{(}\AgdaInductiveConstructor{lam} \AgdaSymbol{(}\AgdaFunction{nf≤} \AgdaSymbol{(}\AgdaInductiveConstructor{lift} \AgdaBound{η}\AgdaSymbol{)} \AgdaBound{a₁}\AgdaSymbol{))))}\<%
\\
\>[0]\AgdaIndent{2}{}\<[2]%
\>[2]\AgdaFunction{\qquad≡⟨⟩}\<%
\\
\>[0]\AgdaIndent{2}{}\<[2]%
\>[2]\AgdaSymbol{(}\AgdaFunction{eta} \AgdaBound{\,f\,} \AgdaFunction{\ensuremath{\mathbin{\infty\mkern-3mu{>}\mkern-8.5mu{>}\mkern-2mu{=}}}} \AgdaSymbol{λ} \AgdaBound{a₁} \AgdaSymbol{→} \AgdaInductiveConstructor{now} \AgdaSymbol{(}\AgdaFunction{nf≤} \AgdaSymbol{(}\AgdaInductiveConstructor{lift} \AgdaBound{η}\AgdaSymbol{)} \AgdaBound{a₁}\AgdaSymbol{)} \AgdaFunction{\ensuremath{\mathbin{{>}\mkern-8.5mu{>}\mkern-2mu{=}}}} \AgdaSymbol{λ} \AgdaBound{a₁} \AgdaSymbol{→} \AgdaInductiveConstructor{now} \AgdaSymbol{(}\AgdaInductiveConstructor{lam} \AgdaBound{a₁}\AgdaSymbol{))}\<%
\\
\>[0]\AgdaIndent{2}{}\<[2]%
\>[2]\AgdaFunction{∞∼⟨} \AgdaFunction{∞∼sym} \AgdaSymbol{(}\AgdaFunction{∞bind-assoc} \AgdaSymbol{(}\AgdaFunction{eta} \AgdaBound{\,f\,}\AgdaSymbol{))} \AgdaFunction{⟩}\<%
\\
\>[0]\AgdaIndent{2}{}\<[2]%
\>[2]\AgdaSymbol{(}\AgdaFunction{eta} \AgdaBound{\,f\,} \AgdaFunction{\ensuremath{\mathbin{\infty\mkern-3mu{>}\mkern-8.5mu{>}\mkern-2mu{=}}}} \AgdaSymbol{(λ} \AgdaBound{a₁} \AgdaSymbol{→} \AgdaInductiveConstructor{now} \AgdaSymbol{(}\AgdaFunction{nf≤} \AgdaSymbol{(}\AgdaInductiveConstructor{lift} \AgdaBound{η}\AgdaSymbol{)} \AgdaBound{a₁}\AgdaSymbol{)))} \AgdaFunction{\ensuremath{\mathbin{\infty\mkern-3mu{>}\mkern-8.5mu{>}\mkern-2mu{=}}}} \AgdaSymbol{(λ} \AgdaBound{a₁} \AgdaSymbol{→} \AgdaInductiveConstructor{now} \AgdaSymbol{(}\AgdaInductiveConstructor{lam} \AgdaBound{a₁}\AgdaSymbol{))}\<%
\\
\>[0]\AgdaIndent{2}{}\<[2]%
\>[2]\AgdaFunction{∞∼⟨} \AgdaFunction{∞bind-cong-l} \AgdaSymbol{(}\AgdaFunction{eta≤} \AgdaBound{η} \AgdaBound{\,f\,}\AgdaSymbol{)} \AgdaSymbol{(λ} \AgdaBound{a} \AgdaSymbol{→} \AgdaInductiveConstructor{now} \AgdaSymbol{(}\AgdaInductiveConstructor{lam} \AgdaBound{a}\AgdaSymbol{))} \AgdaFunction{⟩}\<%
\\
\>[0]\AgdaIndent{2}{}\<[2]%
\>[2]\AgdaFunction{eta} \AgdaSymbol{(}\AgdaFunction{val≤} \AgdaBound{η} \AgdaBound{\,f\,}\AgdaSymbol{)} \AgdaFunction{\ensuremath{\mathbin{\infty\mkern-3mu{>}\mkern-8.5mu{>}\mkern-2mu{=}}}} \AgdaSymbol{(λ} \AgdaBound{a₁} \AgdaSymbol{→} \AgdaInductiveConstructor{now} \AgdaSymbol{(}\AgdaInductiveConstructor{lam} \AgdaBound{a₁}\AgdaSymbol{))}\<%
\\
\>[0]\AgdaIndent{2}{}\<[2]%
\>[2]\AgdaFunction{∎}\AgdaSymbol{)}\<%
\\
\>[0]\AgdaIndent{2}{}\<[2]%
\>[2]\AgdaKeyword{where} \AgdaKeyword{open} \AgdaModule{∞∼-Reasoning}\<%
\\
%
\\
\>\AgdaField{∼force} \AgdaSymbol{(}\AgdaFunction{eta≤} \AgdaBound{η} \AgdaBound{\,f\,}\AgdaSymbol{)} \AgdaSymbol{=}\<%
\\
\>[0]\AgdaIndent{2}{}\<[2]%
\>[2]\AgdaFunction{proof}\<%
\\
\>[0]\AgdaIndent{2}{}\<[2]%
\>[2]\AgdaSymbol{((}\AgdaFunction{apply} \AgdaSymbol{(}\AgdaFunction{weakVal} \AgdaBound{\,f\,}\AgdaSymbol{)} \AgdaSymbol{(}\AgdaInductiveConstructor{ne} \AgdaSymbol{(}\AgdaInductiveConstructor{var} \AgdaInductiveConstructor{zero}\AgdaSymbol{))} \AgdaFunction{\ensuremath{\mathbin{{>}\mkern-8.5mu{>}\mkern-2mu{=}}}} \AgdaFunction{readback}\AgdaSymbol{)}\<%
\\
\>[2]\AgdaIndent{4}{}\<[4]%
\>[4]\AgdaFunction{\ensuremath{\mathbin{{>}\mkern-8.5mu{>}\mkern-2mu{=}}}} \AgdaSymbol{(λ} \AgdaBound{a} \AgdaSymbol{→} \AgdaInductiveConstructor{now} \AgdaSymbol{(}\AgdaFunction{nf≤} \AgdaSymbol{(}\AgdaInductiveConstructor{lift} \AgdaBound{η}\AgdaSymbol{)} \AgdaBound{a}\AgdaSymbol{)))}\<%
\\
\>[0]\AgdaIndent{2}{}\<[2]%
\>[2]\AgdaFunction{\qquad∼⟨} \AgdaFunction{bind-assoc} \AgdaSymbol{(}\AgdaFunction{apply} \AgdaSymbol{(}\AgdaFunction{weakVal} \AgdaBound{\,f\,}\AgdaSymbol{)} \AgdaSymbol{(}\AgdaInductiveConstructor{ne} \AgdaSymbol{(}\AgdaInductiveConstructor{var} \AgdaInductiveConstructor{zero}\AgdaSymbol{)))} \AgdaFunction{⟩}\<%
\\
\>[0]\AgdaIndent{2}{}\<[2]%
\>[2]\AgdaSymbol{(}\AgdaFunction{apply} \AgdaSymbol{(}\AgdaFunction{weakVal} \AgdaBound{\,f\,}\AgdaSymbol{)} \AgdaSymbol{(}\AgdaInductiveConstructor{ne} \AgdaSymbol{(}\AgdaInductiveConstructor{var} \AgdaInductiveConstructor{zero}\AgdaSymbol{))} \AgdaFunction{\ensuremath{\mathbin{{>}\mkern-8.5mu{>}\mkern-2mu{=}}}}\<%
\\
\>[2]\AgdaIndent{7}{}\<[7]%
\>[7]\AgdaSymbol{(λ} \AgdaBound{z} \AgdaSymbol{→} \AgdaFunction{readback} \AgdaBound{z} \AgdaFunction{\ensuremath{\mathbin{{>}\mkern-8.5mu{>}\mkern-2mu{=}}}} \AgdaSymbol{(λ} \AgdaBound{x'} \AgdaSymbol{→} \AgdaInductiveConstructor{now} \AgdaSymbol{(}\AgdaFunction{nf≤} \AgdaSymbol{(}\AgdaInductiveConstructor{lift} \AgdaBound{η}\AgdaSymbol{)} \AgdaBound{x'}\AgdaSymbol{))))}\<%
\\
\>[0]\AgdaIndent{2}{}\<[2]%
\>[2]\AgdaFunction{\qquad∼⟨} \AgdaFunction{bind-cong-r} \AgdaSymbol{(}\AgdaFunction{apply} \AgdaSymbol{(}\AgdaFunction{weakVal} \AgdaBound{\,f\,}\AgdaSymbol{)} \AgdaSymbol{(}\AgdaInductiveConstructor{ne} \AgdaSymbol{(}\AgdaInductiveConstructor{var} \AgdaInductiveConstructor{zero}\AgdaSymbol{)))}\<%
\\
\>[2]\AgdaIndent{17}{}\<[17]%
\>[17]\AgdaSymbol{(λ} \AgdaBound{x} \AgdaSymbol{→} \AgdaFunction{readback≤} \AgdaSymbol{\_} \AgdaSymbol{(}\AgdaInductiveConstructor{lift} \AgdaBound{η}\AgdaSymbol{)} \AgdaBound{x}\AgdaSymbol{)} \AgdaFunction{⟩}\<%
\\
\>[0]\AgdaIndent{2}{}\<[2]%
\>[2]\AgdaSymbol{(}\AgdaFunction{apply} \AgdaSymbol{(}\AgdaFunction{weakVal} \AgdaBound{\,f\,}\AgdaSymbol{)} \AgdaSymbol{(}\AgdaInductiveConstructor{ne} \AgdaSymbol{(}\AgdaInductiveConstructor{var} \AgdaInductiveConstructor{zero}\AgdaSymbol{))} \AgdaFunction{\ensuremath{\mathbin{{>}\mkern-8.5mu{>}\mkern-2mu{=}}}}\<%
\\
\>[2]\AgdaIndent{4}{}\<[4]%
\>[4]\AgdaSymbol{(λ} \AgdaBound{x'} \AgdaSymbol{→} \AgdaFunction{readback} \AgdaSymbol{(}\AgdaFunction{val≤} \AgdaSymbol{(}\AgdaInductiveConstructor{lift} \AgdaBound{η}\AgdaSymbol{)} \AgdaBound{x'}\AgdaSymbol{)))}\<%
\\
\>[0]\AgdaIndent{2}{}\<[2]%
\>[2]\AgdaFunction{\qquad≡⟨⟩}\<%
\\
\>[0]\AgdaIndent{2}{}\<[2]%
\>[2]\AgdaSymbol{(}\AgdaFunction{apply} \AgdaSymbol{(}\AgdaFunction{weakVal} \AgdaBound{\,f\,}\AgdaSymbol{)} \AgdaSymbol{(}\AgdaInductiveConstructor{ne} \AgdaSymbol{(}\AgdaInductiveConstructor{var} \AgdaInductiveConstructor{zero}\AgdaSymbol{))} \AgdaFunction{\ensuremath{\mathbin{{>}\mkern-8.5mu{>}\mkern-2mu{=}}}}\<%
\\
\>[2]\AgdaIndent{8}{}\<[8]%
\>[8]\AgdaSymbol{(λ} \AgdaBound{x'} \AgdaSymbol{→} \AgdaInductiveConstructor{now} \AgdaSymbol{(}\AgdaFunction{val≤} \AgdaSymbol{(}\AgdaInductiveConstructor{lift} \AgdaBound{η}\AgdaSymbol{)} \AgdaBound{x'}\AgdaSymbol{)} \AgdaFunction{\ensuremath{\mathbin{{>}\mkern-8.5mu{>}\mkern-2mu{=}}}} \AgdaFunction{readback}\AgdaSymbol{))}\<%
\\
\>[0]\AgdaIndent{2}{}\<[2]%
\>[2]\AgdaFunction{\qquad∼⟨} \AgdaFunction{∼sym} \AgdaSymbol{(}\AgdaFunction{bind-assoc} \AgdaSymbol{(}\AgdaFunction{apply} \AgdaSymbol{(}\AgdaFunction{weakVal} \AgdaBound{\,f\,}\AgdaSymbol{)} \AgdaSymbol{(}\AgdaInductiveConstructor{ne} \AgdaSymbol{(}\AgdaInductiveConstructor{var} \AgdaInductiveConstructor{zero}\AgdaSymbol{))))} \<[60]%
\>[60]\AgdaFunction{⟩}\<%
\\
\>[0]\AgdaIndent{2}{}\<[2]%
\>[2]\AgdaSymbol{((}\AgdaFunction{apply} \AgdaSymbol{(}\AgdaFunction{weakVal} \AgdaBound{\,f\,}\AgdaSymbol{)} \AgdaSymbol{(}\AgdaInductiveConstructor{ne} \AgdaSymbol{(}\AgdaInductiveConstructor{var} \AgdaInductiveConstructor{zero}\AgdaSymbol{))} \AgdaFunction{\ensuremath{\mathbin{{>}\mkern-8.5mu{>}\mkern-2mu{=}}}}\<%
\\
\>[2]\AgdaIndent{8}{}\<[8]%
\>[8]\AgdaSymbol{(λ} \AgdaBound{x'} \AgdaSymbol{→} \AgdaInductiveConstructor{now} \AgdaSymbol{(}\AgdaFunction{val≤} \AgdaSymbol{(}\AgdaInductiveConstructor{lift} \AgdaBound{η}\AgdaSymbol{)} \AgdaBound{x'}\AgdaSymbol{)))}\<%
\\
\>[0]\AgdaIndent{7}{}\<[7]%
\>[7]\AgdaFunction{\ensuremath{\mathbin{{>}\mkern-8.5mu{>}\mkern-2mu{=}}}} \AgdaFunction{readback}\AgdaSymbol{)}\<%
\\
\>[0]\AgdaIndent{2}{}\<[2]%
\>[2]\AgdaFunction{\qquad∼⟨} \AgdaFunction{bind-cong-l} \AgdaSymbol{(}\AgdaFunction{apply≤} \AgdaSymbol{(}\AgdaFunction{weakVal} \AgdaBound{\,f\,}\AgdaSymbol{)} \AgdaSymbol{(}\AgdaInductiveConstructor{ne} \AgdaSymbol{(}\AgdaInductiveConstructor{var} \AgdaInductiveConstructor{zero}\AgdaSymbol{))} \AgdaSymbol{(}\AgdaInductiveConstructor{lift} \AgdaBound{η}\AgdaSymbol{))} \AgdaFunction{readback} \AgdaFunction{⟩}\<%
\\
\>[0]\AgdaIndent{2}{}\<[2]%
\>[2]\AgdaSymbol{(}\AgdaFunction{apply} \AgdaSymbol{(}\AgdaFunction{val≤} \AgdaSymbol{(}\AgdaInductiveConstructor{lift} \AgdaBound{η}\AgdaSymbol{)} \AgdaSymbol{(}\AgdaFunction{val≤} \AgdaFunction{wk} \AgdaBound{\,f\,}\AgdaSymbol{))} \AgdaSymbol{(}\AgdaInductiveConstructor{ne} \AgdaSymbol{(}\AgdaInductiveConstructor{var} \AgdaInductiveConstructor{zero}\AgdaSymbol{))} \AgdaFunction{\ensuremath{\mathbin{{>}\mkern-8.5mu{>}\mkern-2mu{=}}}} \AgdaFunction{readback}\AgdaSymbol{)}\<%
\\
\>[0]\AgdaIndent{2}{}\<[2]%
\>[2]\AgdaFunction{≡⟨} \AgdaFunction{cong} \AgdaSymbol{(λ} \AgdaBound{f₁} \AgdaSymbol{→} \AgdaFunction{apply} \AgdaBound{f₁} \AgdaSymbol{(}\AgdaInductiveConstructor{ne} \AgdaSymbol{(}\AgdaInductiveConstructor{var} \AgdaInductiveConstructor{zero}\AgdaSymbol{))} \AgdaFunction{\ensuremath{\mathbin{{>}\mkern-8.5mu{>}\mkern-2mu{=}}}} \AgdaFunction{readback}\AgdaSymbol{)}\<%
\\
\>[2]\AgdaIndent{11}{}\<[11]%
\>[11]\AgdaSymbol{(}\AgdaFunction{val≤-•} \AgdaSymbol{(}\AgdaInductiveConstructor{lift} \AgdaBound{η}\AgdaSymbol{)} \AgdaFunction{wk} \AgdaBound{\,f\,}\AgdaSymbol{)} \AgdaFunction{⟩}\<%
\\
\>[0]\AgdaIndent{2}{}\<[2]%
\>[2]\AgdaSymbol{(}\AgdaFunction{apply} \AgdaSymbol{(}\AgdaFunction{val≤} \AgdaSymbol{(}\AgdaInductiveConstructor{weak} \AgdaSymbol{(}\AgdaBound{η} \AgdaFunction{•} \AgdaInductiveConstructor{id}\AgdaSymbol{))} \AgdaBound{\,f\,}\AgdaSymbol{)} \AgdaSymbol{(}\AgdaInductiveConstructor{ne} \AgdaSymbol{(}\AgdaInductiveConstructor{var} \AgdaInductiveConstructor{zero}\AgdaSymbol{))} \AgdaFunction{\ensuremath{\mathbin{{>}\mkern-8.5mu{>}\mkern-2mu{=}}}} \AgdaFunction{readback}\AgdaSymbol{)}\<%
\\
\>[0]\AgdaIndent{2}{}\<[2]%
\>[2]\AgdaFunction{≡⟨} \AgdaFunction{cong} \AgdaSymbol{(λ} \AgdaBound{η₁} \AgdaSymbol{→} \AgdaFunction{apply} \AgdaSymbol{(}\AgdaFunction{val≤} \AgdaSymbol{(}\AgdaInductiveConstructor{weak} \AgdaBound{η₁}\AgdaSymbol{)} \AgdaBound{\,f\,}\AgdaSymbol{)} \AgdaSymbol{(}\AgdaInductiveConstructor{ne} \AgdaSymbol{(}\AgdaInductiveConstructor{var} \AgdaInductiveConstructor{zero}\AgdaSymbol{))} \AgdaFunction{\ensuremath{\mathbin{{>}\mkern-8.5mu{>}\mkern-2mu{=}}}} \AgdaFunction{readback}\AgdaSymbol{)}\<%
\\
\>[2]\AgdaIndent{10}{}\<[10]%
\>[10]\AgdaSymbol{(}\AgdaFunction{η•id} \AgdaBound{η}\AgdaSymbol{)} \AgdaFunction{⟩}\<%
\\
\>[0]\AgdaIndent{2}{}\<[2]%
\>[2]\AgdaSymbol{(}\AgdaFunction{apply} \AgdaSymbol{(}\AgdaFunction{val≤} \AgdaSymbol{(}\AgdaInductiveConstructor{weak} \AgdaBound{η}\AgdaSymbol{)} \AgdaBound{\,f\,}\AgdaSymbol{)} \AgdaSymbol{(}\AgdaInductiveConstructor{ne} \AgdaSymbol{(}\AgdaInductiveConstructor{var} \AgdaInductiveConstructor{zero}\AgdaSymbol{))} \AgdaFunction{\ensuremath{\mathbin{{>}\mkern-8.5mu{>}\mkern-2mu{=}}}} \AgdaFunction{readback}\AgdaSymbol{)}\<%
\\
\>[0]\AgdaIndent{2}{}\<[2]%
\>[2]\AgdaFunction{≡⟨} \AgdaFunction{cong} \AgdaSymbol{(λ} \AgdaBound{f₁} \AgdaSymbol{→} \AgdaFunction{apply} \AgdaBound{f₁} \AgdaSymbol{(}\AgdaInductiveConstructor{ne} \AgdaSymbol{(}\AgdaInductiveConstructor{var} \AgdaInductiveConstructor{zero}\AgdaSymbol{))} \AgdaFunction{\ensuremath{\mathbin{{>}\mkern-8.5mu{>}\mkern-2mu{=}}}} \AgdaFunction{readback}\AgdaSymbol{)}\<%
\\
\>[2]\AgdaIndent{10}{}\<[10]%
\>[10]\AgdaSymbol{(}\AgdaFunction{sym} \AgdaSymbol{(}\AgdaFunction{val≤-•} \AgdaFunction{wk} \AgdaBound{η} \AgdaBound{\,f\,}\AgdaSymbol{))} \AgdaFunction{⟩}\<%
\\
\>[0]\AgdaIndent{2}{}\<[2]%
\>[2]\AgdaSymbol{(}\AgdaFunction{apply} \AgdaSymbol{(}\AgdaFunction{weakVal} \AgdaSymbol{(}\AgdaFunction{val≤} \AgdaBound{η} \AgdaBound{\,f\,}\AgdaSymbol{))} \AgdaSymbol{(}\AgdaInductiveConstructor{ne} \AgdaSymbol{(}\AgdaInductiveConstructor{var} \AgdaInductiveConstructor{zero}\AgdaSymbol{))} \AgdaFunction{\ensuremath{\mathbin{{>}\mkern-8.5mu{>}\mkern-2mu{=}}}} \AgdaFunction{readback}\AgdaSymbol{)}\<%
\\
\>[0]\AgdaIndent{2}{}\<[2]%
\>[2]\AgdaFunction{∎}\<%
\\
\>[0]\AgdaIndent{2}{}\<[2]%
\>[2]\AgdaKeyword{where} \AgdaKeyword{open} \AgdaModule{∼-Reasoning}\<%
\end{code}}

\noindent
We must also be able to weaken proofs of strong computability. Again
we skip the proofs.

\begin{code}%
\>\AgdaFunction{nereadback≤⇓} \<[14]%
\>[14]\AgdaSymbol{:} \<[17]%
\>[17]\AgdaSymbol{∀\{}\AgdaBound{Γ} \AgdaBound{Δ} \AgdaBound{a}\AgdaSymbol{\}} \AgdaSymbol{(}\AgdaBound{η} \AgdaSymbol{:} \AgdaBound{Δ} \AgdaDatatype{≤} \AgdaBound{Γ}\AgdaSymbol{)} \AgdaSymbol{(}\AgdaBound{t} \AgdaSymbol{:} \AgdaDatatype{Ne} \AgdaDatatype{Val} \AgdaBound{Γ} \AgdaBound{a}\AgdaSymbol{)} \AgdaSymbol{\{}\AgdaBound{n} \AgdaSymbol{:} \AgdaDatatype{Ne} \AgdaDatatype{Nf} \AgdaBound{Γ} \AgdaBound{a}\AgdaSymbol{\}} \AgdaSymbol{→}\<%
\\
\>[2]\AgdaIndent{17}{}\<[17]%
\>[17]\AgdaFunction{nereadback} \AgdaBound{t} \AgdaDatatype{⇓} \AgdaBound{n} \AgdaSymbol{→} \AgdaFunction{nereadback} \AgdaSymbol{(}\AgdaFunction{nev≤} \AgdaBound{η} \AgdaBound{t}\AgdaSymbol{)} \AgdaDatatype{⇓} \AgdaFunction{nen≤} \AgdaBound{η} \AgdaBound{n}\<%
\\
%
\\
\>\AgdaFunction{V⟦⟧≤} \<[14]%
\>[14]\AgdaSymbol{:} \<[17]%
\>[17]\AgdaSymbol{∀\{}\AgdaBound{Δ} \AgdaBound{Δ\kern-1pt′}\AgdaSymbol{\}} \AgdaBound{a} \<[28]%
\>[28]\AgdaSymbol{(}\AgdaBound{η} \AgdaSymbol{:} \AgdaBound{Δ\kern-1pt′} \AgdaDatatype{≤} \AgdaBound{Δ}\AgdaSymbol{)} \<[42]%
\>[42]\AgdaSymbol{(}\AgdaBound{v} \AgdaSymbol{:} \AgdaDatatype{Val} \AgdaBound{Δ} \AgdaBound{a}\AgdaSymbol{)} \<[57]%
\>[57]\AgdaSymbol{→} \AgdaFunction{V⟦} \AgdaBound{a} \AgdaFunction{⟧} \AgdaBound{v} \<[69]%
\>[69]\AgdaSymbol{→} \AgdaFunction{V⟦} \AgdaBound{a} \AgdaFunction{⟧} \AgdaSymbol{(}\AgdaFunction{val≤} \AgdaBound{η} \AgdaBound{v}\AgdaSymbol{)}\<%
\\
\>\AgdaFunction{E⟦⟧≤} \<[14]%
\>[14]\AgdaSymbol{:} \<[17]%
\>[17]\AgdaSymbol{∀\{}\AgdaBound{Γ} \AgdaBound{Δ} \AgdaBound{Δ\kern-1pt′}\AgdaSymbol{\}} \<[28]%
\>[28]\AgdaSymbol{(}\AgdaBound{η} \AgdaSymbol{:} \AgdaBound{Δ\kern-1pt′} \AgdaDatatype{≤} \AgdaBound{Δ}\AgdaSymbol{)} \<[42]%
\>[42]\AgdaSymbol{(}\AgdaBound{ρ} \AgdaSymbol{:} \AgdaDatatype{Env} \AgdaBound{Δ} \AgdaBound{Γ}\AgdaSymbol{)} \<[57]%
\>[57]\AgdaSymbol{→} \AgdaFunction{E⟦} \AgdaBound{Γ} \AgdaFunction{⟧} \AgdaBound{ρ} \<[69]%
\>[69]\AgdaSymbol{→} \AgdaFunction{E⟦} \AgdaBound{Γ} \AgdaFunction{⟧} \AgdaSymbol{(}\AgdaFunction{env≤} \AgdaBound{η} \AgdaBound{ρ}\AgdaSymbol{)}\<%
\end{code}

\AgdaHide{
\begin{code}%
\>\AgdaFunction{nereadback≤⇓} \AgdaBound{η} \AgdaBound{t} \AgdaSymbol{\{}\AgdaBound{n}\AgdaSymbol{\}} \AgdaBound{p} \AgdaSymbol{=} \AgdaFunction{subst∼⇓} \AgdaSymbol{(}\AgdaFunction{map⇓} \AgdaSymbol{(}\AgdaFunction{nen≤} \AgdaBound{η}\AgdaSymbol{)} \AgdaBound{p}\AgdaSymbol{)} \AgdaSymbol{(}\AgdaFunction{nereadback≤} \AgdaBound{η} \AgdaBound{t}\AgdaSymbol{)}\<%
\\
%
\\
\>\AgdaFunction{V⟦⟧≤} \AgdaInductiveConstructor{★} \<[13]%
\>[13]\AgdaBound{η} \AgdaSymbol{(}\AgdaInductiveConstructor{ne} \AgdaBound{t}\AgdaSymbol{)} \AgdaSymbol{(}\AgdaBound{n} \AgdaInductiveConstructor{,} \AgdaBound{p}\AgdaSymbol{)} \<[37]%
\>[37]\AgdaSymbol{=} \AgdaFunction{nen≤} \AgdaBound{η} \AgdaBound{n} \AgdaInductiveConstructor{,} \AgdaFunction{nereadback≤⇓} \AgdaBound{η} \AgdaBound{t} \AgdaBound{p}\<%
\\
\>\AgdaFunction{V⟦⟧≤} \AgdaSymbol{(}\AgdaBound{a} \AgdaInductiveConstructor{⇒} \AgdaBound{b}\AgdaSymbol{)} \AgdaBound{η} \AgdaBound{v} \<[22]%
\>[22]\AgdaBound{p} \<[30]%
\>[30]\AgdaBound{ρ} \AgdaBound{u} \AgdaBound{u⇓} \AgdaSymbol{=}\<%
\\
\>[0]\AgdaIndent{2}{}\<[2]%
\>[2]\AgdaKeyword{let} \AgdaBound{v′} \AgdaInductiveConstructor{,} \AgdaBound{p′} \AgdaInductiveConstructor{,} \AgdaBound{p′′} \AgdaSymbol{=} \AgdaBound{p} \AgdaSymbol{(}\AgdaBound{ρ} \AgdaFunction{•} \AgdaBound{η}\AgdaSymbol{)} \AgdaBound{u} \AgdaBound{u⇓} \AgdaKeyword{in}\<%
\\
\>[2]\AgdaIndent{6}{}\<[6]%
\>[6]\AgdaBound{v′} \AgdaInductiveConstructor{,} \AgdaFunction{subst} \AgdaSymbol{(λ} \AgdaBound{\,f\,} \AgdaSymbol{→} \AgdaFunction{apply} \AgdaBound{\,f\,} \AgdaBound{u} \AgdaDatatype{⇓} \AgdaField{fst} \AgdaSymbol{(}\AgdaBound{p} \AgdaSymbol{(}\AgdaBound{ρ} \AgdaFunction{•} \AgdaBound{η}\AgdaSymbol{)} \AgdaBound{u} \AgdaBound{u⇓}\AgdaSymbol{))}\<%
\\
\>[6]\AgdaIndent{17}{}\<[17]%
\>[17]\AgdaSymbol{((}\AgdaFunction{sym} \AgdaSymbol{(}\AgdaFunction{val≤-•} \AgdaBound{ρ} \AgdaBound{η} \AgdaBound{v}\AgdaSymbol{)))}\<%
\\
\>[6]\AgdaIndent{17}{}\<[17]%
\>[17]\AgdaBound{p′}\<%
\\
\>[0]\AgdaIndent{9}{}\<[9]%
\>[9]\AgdaInductiveConstructor{,} \AgdaBound{p′′}\<%
\\
%
\\
\>\AgdaFunction{E⟦⟧≤} \AgdaBound{η} \AgdaInductiveConstructor{ε} \<[15]%
\>[15]\AgdaBound{θ} \<[24]%
\>[24]\AgdaSymbol{=} \AgdaSymbol{\_}\<%
\\
\>\AgdaFunction{E⟦⟧≤} \AgdaBound{η} \AgdaSymbol{(}\AgdaBound{ρ} \AgdaInductiveConstructor{,} \AgdaBound{v}\AgdaSymbol{)} \AgdaSymbol{(}\AgdaBound{θ} \AgdaInductiveConstructor{,} \AgdaBound{⟦v⟧}\AgdaSymbol{)} \AgdaSymbol{=} \AgdaFunction{E⟦⟧≤} \AgdaBound{η} \AgdaBound{ρ} \AgdaBound{θ} \AgdaInductiveConstructor{,} \AgdaFunction{V⟦⟧≤} \AgdaSymbol{\_} \AgdaBound{η} \AgdaBound{v} \AgdaBound{⟦v⟧}\<%
\end{code}
}

\noindent
Finally, we can work our way up towards the fundamental theorem of
logical relations (called \AgdaFunction{term} for \emph{termination} below).
In our case, it is just a logical predicate, namely, strong
computability \AgdaFunction{C⟦\_⟧\_}, but the proof technique is the same:
induction on well-typed terms.
To this end, we establish lemmas for each case, calling them
\AgdaFunction{⟦var⟧},
\AgdaFunction{⟦abs⟧}, and
\AgdaFunction{⟦app⟧}.
To start, soundness of variable evaluation is a consequence of a sound
(\AgdaBound{θ}) environment \AgdaBound{ρ}:
\begin{code}%
\>\AgdaFunction{⟦var⟧} \AgdaSymbol{:} \AgdaSymbol{∀\{}\AgdaBound{Δ} \AgdaBound{Γ} \AgdaBound{a}\AgdaSymbol{\}} \AgdaSymbol{(}\AgdaBound{x} \AgdaSymbol{:} \AgdaDatatype{Var} \AgdaBound{Γ} \AgdaBound{a}\AgdaSymbol{)} \AgdaSymbol{(}\AgdaBound{ρ} \AgdaSymbol{:} \AgdaDatatype{Env} \AgdaBound{Δ} \AgdaBound{Γ}\AgdaSymbol{)} \AgdaSymbol{→} \AgdaFunction{E⟦} \AgdaBound{Γ} \AgdaFunction{⟧} \AgdaBound{ρ} \AgdaSymbol{→} \AgdaFunction{C⟦} \AgdaBound{a} \AgdaFunction{⟧} \AgdaSymbol{(}\AgdaInductiveConstructor{now} \AgdaSymbol{(}\AgdaFunction{lookup} \AgdaBound{x} \AgdaBound{ρ}\AgdaSymbol{))}\<%
\\
\>\AgdaFunction{⟦var⟧} \AgdaInductiveConstructor{zero} \<[14]%
\>[14]\AgdaSymbol{(\_} \AgdaInductiveConstructor{,} \AgdaBound{v}\AgdaSymbol{)} \<[23]%
\>[23]\AgdaSymbol{(\_} \AgdaInductiveConstructor{,} \AgdaBound{v⇓}\AgdaSymbol{)} \<[33]%
\>[33]\AgdaSymbol{=} \AgdaBound{v} \AgdaInductiveConstructor{,} \AgdaInductiveConstructor{now⇓} \AgdaInductiveConstructor{,} \AgdaBound{v⇓}\<%
\\
\>\AgdaFunction{⟦var⟧}\AgdaSymbol{(}\AgdaInductiveConstructor{suc} \AgdaBound{x}\AgdaSymbol{)} \<[14]%
\>[14]\AgdaSymbol{(}\AgdaBound{ρ} \AgdaInductiveConstructor{,} \AgdaSymbol{\_)} \<[23]%
\>[23]\AgdaSymbol{(}\AgdaBound{θ} \AgdaInductiveConstructor{,} \AgdaSymbol{\_} \AgdaSymbol{)} \<[33]%
\>[33]\AgdaSymbol{=} \AgdaFunction{⟦var⟧} \AgdaBound{x} \AgdaBound{ρ} \AgdaBound{θ}\<%
\end{code}

\noindent
The abstraction case requires another, albeit trivial lemma:
\AgdaFunction{sound-β}, which states the semantic soundness of
$\beta$-expansion.

\begin{samepage}
\begin{code}%
\>\AgdaFunction{sound-β} \<[9]%
\>[9]\AgdaSymbol{:} \<[12]%
\>[12]\AgdaSymbol{∀} \AgdaSymbol{\{}\AgdaBound{Δ} \AgdaBound{Γ} \AgdaBound{a} \AgdaBound{b}\AgdaSymbol{\}} \AgdaSymbol{(}\AgdaBound{t} \AgdaSymbol{:} \AgdaDatatype{Tm} \AgdaSymbol{(}\AgdaBound{Γ} \AgdaInductiveConstructor{,} \AgdaBound{a}\AgdaSymbol{)} \AgdaBound{b}\AgdaSymbol{)} \AgdaSymbol{(}\AgdaBound{ρ} \AgdaSymbol{:} \AgdaDatatype{Env} \AgdaBound{Δ} \AgdaBound{Γ}\AgdaSymbol{)} \AgdaSymbol{(}\AgdaBound{u} \AgdaSymbol{:} \AgdaDatatype{Val} \AgdaBound{Δ} \AgdaBound{a}\AgdaSymbol{)} \AgdaSymbol{→}\<%
\\
\>[9]\AgdaIndent{12}{}\<[12]%
\>[12]\AgdaFunction{C⟦} \AgdaBound{b} \AgdaFunction{⟧} \AgdaSymbol{(}\AgdaFunction{eval} \AgdaBound{t} \<[28]%
\>[28]\AgdaSymbol{(}\AgdaBound{ρ} \AgdaInductiveConstructor{,} \AgdaBound{u}\AgdaSymbol{))} \AgdaSymbol{→} \AgdaFunction{C⟦} \AgdaBound{b} \AgdaFunction{⟧} \AgdaSymbol{(}\AgdaFunction{apply} \AgdaSymbol{(}\AgdaInductiveConstructor{lam} \AgdaBound{t} \AgdaBound{ρ}\AgdaSymbol{)} \AgdaBound{u}\AgdaSymbol{)}\<%
\\
\>\AgdaFunction{sound-β} \AgdaBound{t} \AgdaBound{ρ} \AgdaBound{u} \AgdaSymbol{(}\AgdaBound{v} \AgdaInductiveConstructor{,} \AgdaBound{v⇓} \AgdaInductiveConstructor{,} \AgdaBound{⟦v⟧}\AgdaSymbol{)} \AgdaSymbol{=} \AgdaBound{v} \AgdaInductiveConstructor{,} \AgdaInductiveConstructor{later⇓} \AgdaBound{v⇓} \AgdaInductiveConstructor{,} \AgdaBound{⟦v⟧}\<%
\\
%
\\
\>\AgdaFunction{⟦abs⟧} \<[9]%
\>[9]\AgdaSymbol{:} \<[12]%
\>[12]\AgdaSymbol{∀} \AgdaSymbol{\{}\AgdaBound{Δ} \AgdaBound{Γ} \AgdaBound{a} \AgdaBound{b}\AgdaSymbol{\}} \AgdaSymbol{(}\AgdaBound{t} \AgdaSymbol{:} \AgdaDatatype{Tm} \AgdaSymbol{(}\AgdaBound{Γ} \AgdaInductiveConstructor{,} \AgdaBound{a}\AgdaSymbol{)} \AgdaBound{b}\AgdaSymbol{)} \AgdaSymbol{(}\AgdaBound{ρ} \AgdaSymbol{:} \AgdaDatatype{Env} \AgdaBound{Δ} \AgdaBound{Γ}\AgdaSymbol{)} \AgdaSymbol{(}\AgdaBound{θ} \AgdaSymbol{:} \AgdaFunction{E⟦} \AgdaBound{Γ} \AgdaFunction{⟧} \AgdaBound{ρ}\AgdaSymbol{)} \AgdaSymbol{→}\<%
\\
\>[9]\AgdaIndent{12}{}\<[12]%
\>[12]\AgdaSymbol{(∀\{}\AgdaBound{Δ\kern-1pt′}\AgdaSymbol{\}(}\AgdaBound{η} \AgdaSymbol{:} \AgdaBound{Δ\kern-1pt′} \AgdaDatatype{≤} \AgdaBound{Δ}\AgdaSymbol{)(}\AgdaBound{u} \AgdaSymbol{:} \AgdaDatatype{Val} \AgdaBound{Δ\kern-1pt′} \AgdaBound{a}\AgdaSymbol{)(}\AgdaBound{u⇓} \AgdaSymbol{:} \AgdaFunction{V⟦} \AgdaBound{a} \AgdaFunction{⟧} \AgdaBound{u}\AgdaSymbol{)} \AgdaSymbol{→} \AgdaFunction{C⟦} \AgdaBound{b} \AgdaFunction{⟧} \AgdaSymbol{(}\AgdaFunction{eval} \AgdaBound{t} \AgdaSymbol{(}\AgdaFunction{env≤} \AgdaBound{η} \AgdaBound{ρ} \AgdaInductiveConstructor{,} \AgdaBound{u}\AgdaSymbol{)))} \AgdaSymbol{→}\<%
\\
\>[9]\AgdaIndent{12}{}\<[12]%
\>[12]\AgdaFunction{C⟦} \AgdaBound{a} \AgdaInductiveConstructor{⇒} \AgdaBound{b} \AgdaFunction{⟧} \AgdaSymbol{(}\AgdaInductiveConstructor{now} \AgdaSymbol{(}\AgdaInductiveConstructor{lam} \AgdaBound{t} \AgdaBound{ρ}\AgdaSymbol{))}\<%
\\
\>\AgdaFunction{⟦abs⟧} \AgdaBound{t} \AgdaBound{ρ} \AgdaBound{θ} \AgdaBound{ih} \AgdaSymbol{=} \AgdaInductiveConstructor{lam} \AgdaBound{t} \AgdaBound{ρ} \AgdaInductiveConstructor{,} \AgdaInductiveConstructor{now⇓} \AgdaInductiveConstructor{,} \AgdaSymbol{(λ} \AgdaBound{η} \AgdaBound{u} \AgdaBound{p} \AgdaSymbol{→} \AgdaFunction{sound-β} \AgdaBound{t} \AgdaSymbol{(}\AgdaFunction{env≤} \AgdaBound{η} \AgdaBound{ρ}\AgdaSymbol{)} \AgdaBound{u} \AgdaSymbol{(}\AgdaBound{ih} \AgdaBound{η} \AgdaBound{u} \AgdaBound{p}\AgdaSymbol{))}\<%
\end{code}
\end{samepage}

\noindent
The lemma for application is straightforward, the proof term is just a
bit bloated by the need to apply the first functor law \AgdaFunction{val≤-id}
to fix the types.

\begin{code}%
\>\AgdaFunction{⟦app⟧} \<[7]%
\>[7]\AgdaSymbol{:} \<[10]%
\>[10]\AgdaSymbol{∀\{}\AgdaBound{Δ} \AgdaBound{a} \AgdaBound{b}\AgdaSymbol{\}} \AgdaSymbol{\{}\AgdaBound{f?} \AgdaSymbol{:} \AgdaDatatype{Delay} \AgdaSymbol{\_} \AgdaSymbol{(}\AgdaDatatype{Val} \AgdaBound{Δ} \AgdaSymbol{(}\AgdaBound{a} \AgdaInductiveConstructor{⇒} \AgdaBound{b}\AgdaSymbol{))\}} \AgdaSymbol{\{}\AgdaBound{u?} \AgdaSymbol{:} \AgdaDatatype{Delay} \AgdaSymbol{\_} \AgdaSymbol{(}\AgdaDatatype{Val} \AgdaBound{Δ} \AgdaBound{a}\AgdaSymbol{)\}} \AgdaSymbol{→}\<%
\\
\>[0]\AgdaIndent{10}{}\<[10]%
\>[10]\AgdaFunction{C⟦} \AgdaBound{a} \AgdaInductiveConstructor{⇒} \AgdaBound{b} \AgdaFunction{⟧} \AgdaBound{f?} \AgdaSymbol{→} \AgdaFunction{C⟦} \AgdaBound{a} \AgdaFunction{⟧} \AgdaBound{u?} \AgdaSymbol{→} \AgdaFunction{C⟦} \AgdaBound{b} \AgdaFunction{⟧} \AgdaSymbol{(}\AgdaBound{f?} \AgdaFunction{\ensuremath{\mathbin{{>}\mkern-8.5mu{>}\mkern-2mu{=}}}} \AgdaSymbol{λ} \AgdaBound{\,f\,} \AgdaSymbol{→} \AgdaBound{u?} \AgdaFunction{\ensuremath{\mathbin{{>}\mkern-8.5mu{>}\mkern-2mu{=}}}} \AgdaFunction{apply} \AgdaBound{\,f\,}\AgdaSymbol{)}\<%
\\
\>\AgdaFunction{⟦app⟧} \AgdaSymbol{\{}\AgdaSymbol{u?} \AgdaSymbol{=} \AgdaBound{u?}\AgdaSymbol{\}} \AgdaSymbol{(}\AgdaBound{\,f\,} \AgdaInductiveConstructor{,} \AgdaBound{f⇓} \AgdaInductiveConstructor{,} \AgdaBound{⟦\,f\,⟧}\AgdaSymbol{)} \AgdaSymbol{(}\AgdaBound{u} \AgdaInductiveConstructor{,} \AgdaBound{u⇓} \AgdaInductiveConstructor{,} \AgdaBound{⟦u⟧}\AgdaSymbol{)} \AgdaSymbol{=}\<%
\\
\>[0]\AgdaIndent{2}{}\<[2]%
\>[2]\AgdaKeyword{let} \<[7]%
\>[7]\AgdaBound{v} \AgdaInductiveConstructor{,} \AgdaBound{v⇓} \AgdaInductiveConstructor{,} \AgdaBound{⟦v⟧} \<[21]%
\>[21]\AgdaSymbol{=} \<[24]%
\>[24]\AgdaBound{⟦\,f\,⟧} \AgdaInductiveConstructor{id} \AgdaBound{u} \AgdaBound{⟦u⟧}\<%
\\
\>[2]\AgdaIndent{7}{}\<[7]%
\>[7]\AgdaBound{v⇓′} \<[21]%
\>[21]\AgdaSymbol{=} \<[24]%
\>[24]\AgdaFunction{bind⇓} \<[31]%
\>[31]\AgdaSymbol{(λ} \AgdaBound{\,f\,′} \AgdaSymbol{→} \AgdaBound{u?} \AgdaFunction{\ensuremath{\mathbin{{>}\mkern-8.5mu{>}\mkern-2mu{=}}}} \AgdaFunction{apply} \AgdaBound{\,f\,′}\AgdaSymbol{)}\<%
\\
\>[7]\AgdaIndent{31}{}\<[31]%
\>[31]\AgdaBound{f⇓}\<%
\\
\>[7]\AgdaIndent{31}{}\<[31]%
\>[31]\AgdaSymbol{(}\AgdaFunction{bind⇓} \<[39]%
\>[39]\AgdaSymbol{(}\AgdaFunction{apply} \AgdaBound{\,f\,}\AgdaSymbol{)}\<%
\\
\>[31]\AgdaIndent{39}{}\<[39]%
\>[39]\AgdaBound{u⇓}\<%
\\
\>[31]\AgdaIndent{39}{}\<[39]%
\>[39]\AgdaSymbol{(}\AgdaFunction{subst} \<[47]%
\>[47]\AgdaSymbol{(λ} \AgdaBound{\,f\,′} \AgdaSymbol{→} \AgdaFunction{apply} \AgdaBound{\,f\,′} \AgdaBound{u} \AgdaDatatype{⇓} \AgdaBound{v}\AgdaSymbol{)}\<%
\\
\>[39]\AgdaIndent{47}{}\<[47]%
\>[47]\AgdaSymbol{(}\AgdaFunction{val≤-id} \AgdaBound{\,f\,}\AgdaSymbol{)}\<%
\\
\>[39]\AgdaIndent{47}{}\<[47]%
\>[47]\AgdaBound{v⇓}\AgdaSymbol{))}\<%
\\
\>[0]\AgdaIndent{2}{}\<[2]%
\>[2]\AgdaKeyword{in} \<[7]%
\>[7]\AgdaBound{v} \AgdaInductiveConstructor{,} \AgdaBound{v⇓′} \AgdaInductiveConstructor{,} \AgdaBound{⟦v⟧}\<%
\end{code}

\noindent
Evaluation is sound, in particular, it terminates.
The proof of \AgdaFunction{term} proceeds by induction on the terms and is
straightforward after our preparations.

\begin{code}%
\>\AgdaFunction{term} \<[21]%
\>[21]\AgdaSymbol{:} \<[24]%
\>[24]\AgdaSymbol{∀} \AgdaSymbol{\{}\AgdaBound{Δ} \AgdaBound{Γ} \AgdaBound{a}\AgdaSymbol{\}} \AgdaSymbol{(}\AgdaBound{t} \AgdaSymbol{:} \AgdaDatatype{Tm} \AgdaBound{Γ} \AgdaBound{a}\AgdaSymbol{)} \AgdaSymbol{(}\AgdaBound{ρ} \AgdaSymbol{:} \AgdaDatatype{Env} \AgdaBound{Δ} \AgdaBound{Γ}\AgdaSymbol{)} \AgdaSymbol{(}\AgdaBound{θ} \AgdaSymbol{:} \AgdaFunction{E⟦} \AgdaBound{Γ} \AgdaFunction{⟧} \AgdaBound{ρ}\AgdaSymbol{)} \AgdaSymbol{→} \AgdaFunction{C⟦} \AgdaBound{a} \AgdaFunction{⟧} \AgdaSymbol{(}\AgdaFunction{eval} \AgdaBound{t} \AgdaBound{ρ}\AgdaSymbol{)}\<%
\\
\>\AgdaFunction{term} \AgdaSymbol{(}\AgdaInductiveConstructor{var} \AgdaBound{x}\AgdaSymbol{)} \<[16]%
\>[16]\AgdaBound{ρ} \AgdaBound{θ} \<[21]%
\>[21]\AgdaSymbol{=} \<[24]%
\>[24]\AgdaFunction{⟦var⟧} \AgdaBound{x} \AgdaBound{ρ} \AgdaBound{θ}\<%
\\
\>\AgdaFunction{term} \AgdaSymbol{(}\AgdaInductiveConstructor{abs} \AgdaBound{t}\AgdaSymbol{)} \<[16]%
\>[16]\AgdaBound{ρ} \AgdaBound{θ} \<[21]%
\>[21]\AgdaSymbol{=} \<[24]%
\>[24]\AgdaFunction{⟦abs⟧} \AgdaBound{t} \AgdaBound{ρ} \AgdaBound{θ} \AgdaSymbol{(λ} \AgdaBound{η} \AgdaBound{u} \AgdaBound{p} \AgdaSymbol{→} \AgdaFunction{term} \AgdaBound{t} \AgdaSymbol{(}\AgdaFunction{env≤} \AgdaBound{η} \AgdaBound{ρ} \AgdaInductiveConstructor{,} \AgdaBound{u}\AgdaSymbol{)} \AgdaSymbol{(}\AgdaFunction{E⟦⟧≤} \AgdaBound{η} \AgdaBound{ρ} \AgdaBound{θ} \AgdaInductiveConstructor{,} \AgdaBound{p}\AgdaSymbol{))}\<%
\\
\>\AgdaFunction{term} \AgdaSymbol{(}\AgdaInductiveConstructor{app} \AgdaBound{t} \AgdaBound{u}\AgdaSymbol{)} \<[16]%
\>[16]\AgdaBound{ρ} \AgdaBound{θ} \<[21]%
\>[21]\AgdaSymbol{=} \<[24]%
\>[24]\AgdaFunction{⟦app⟧} \AgdaSymbol{(}\AgdaFunction{term} \AgdaBound{t} \AgdaBound{ρ} \AgdaBound{θ}\AgdaSymbol{)} \AgdaSymbol{(}\AgdaFunction{term} \AgdaBound{u} \AgdaBound{ρ} \AgdaBound{θ}\AgdaSymbol{)}\<%
\end{code}

\noindent
Termination of readback for strongly computable values follows from
the following two mutually defined lemmas. They are proved mutually by
induction on types.

To reify a functional value \AgdaBound{\,f\,}, we need to reflect the fresh
variable $\AgdaInductiveConstructor{var}\;\AgdaInductiveConstructor{zero}$ to obtain a value \AgdaBound{u} with semantics
\AgdaBound{⟦u⟧}.  We can then apply the semantic function \AgdaBound{⟦\,f\,⟧} to
\AgdaBound{u} and recursively reify the returned value~\AgdaBound{v}.

\begin{code}%
\>\AgdaKeyword{mutual}\<%
\\
\>[0]\AgdaIndent{2}{}\<[2]%
\>[2]\AgdaFunction{reify} \AgdaSymbol{:} \AgdaSymbol{∀\{}\AgdaBound{Γ}\AgdaSymbol{\}} \AgdaBound{a} \AgdaSymbol{(}\AgdaBound{v} \AgdaSymbol{:} \AgdaDatatype{Val} \AgdaBound{Γ} \AgdaBound{a}\AgdaSymbol{)} \AgdaSymbol{→} \AgdaFunction{V⟦} \AgdaBound{a} \AgdaFunction{⟧} \AgdaBound{v} \AgdaSymbol{→} \AgdaFunction{readback} \AgdaBound{v} \AgdaFunction{⇓}\<%
\\
\>[0]\AgdaIndent{2}{}\<[2]%
\>[2]\AgdaFunction{reify} \AgdaInductiveConstructor{★} \<[17]%
\>[17]\AgdaSymbol{(}\AgdaInductiveConstructor{ne} \AgdaSymbol{\_)} \<[25]%
\>[25]\AgdaSymbol{(}\AgdaBound{m} \AgdaInductiveConstructor{,} \AgdaBound{⇓m}\AgdaSymbol{)} \<[35]%
\>[35]\AgdaSymbol{=} \AgdaInductiveConstructor{ne} \AgdaBound{m} \AgdaInductiveConstructor{,} \AgdaFunction{map⇓} \AgdaInductiveConstructor{ne} \AgdaBound{⇓m}\<%
\\
\>[0]\AgdaIndent{2}{}\<[2]%
\>[2]\AgdaFunction{reify} \AgdaSymbol{(}\AgdaBound{a} \AgdaInductiveConstructor{⇒} \AgdaBound{b}\AgdaSymbol{)} \<[17]%
\>[17]\AgdaBound{\,f\,} \<[25]%
\>[25]\AgdaBound{⟦\,f\,⟧} \<[35]%
\>[35]\AgdaSymbol{=}\<%
\\
\>[2]\AgdaIndent{4}{}\<[4]%
\>[4]\AgdaKeyword{let} \AgdaBound{u} \<[22]%
\>[22]\AgdaSymbol{=} \AgdaInductiveConstructor{ne} \AgdaSymbol{(}\AgdaInductiveConstructor{var} \AgdaInductiveConstructor{zero}\AgdaSymbol{)}\<%
\\
\>[4]\AgdaIndent{8}{}\<[8]%
\>[8]\AgdaBound{⟦u⟧} \<[22]%
\>[22]\AgdaSymbol{=} \AgdaFunction{reflect} \AgdaBound{a} \AgdaSymbol{(}\AgdaInductiveConstructor{var} \AgdaInductiveConstructor{zero}\AgdaSymbol{)} \AgdaSymbol{(}\AgdaInductiveConstructor{var} \AgdaInductiveConstructor{zero} \AgdaInductiveConstructor{,} \AgdaInductiveConstructor{now⇓}\AgdaSymbol{)}\<%
\\
\>[4]\AgdaIndent{8}{}\<[8]%
\>[8]\AgdaBound{v} \AgdaInductiveConstructor{,} \AgdaBound{v⇓} \AgdaInductiveConstructor{,} \AgdaBound{⟦v⟧} \<[22]%
\>[22]\AgdaSymbol{=} \AgdaBound{⟦\,f\,⟧} \AgdaFunction{wk} \AgdaBound{u} \AgdaBound{⟦u⟧}\<%
\\
\>[4]\AgdaIndent{8}{}\<[8]%
\>[8]\AgdaBound{n} \AgdaInductiveConstructor{,} \AgdaBound{⇓n} \<[22]%
\>[22]\AgdaSymbol{=} \AgdaFunction{reify} \AgdaBound{b} \AgdaBound{v} \AgdaBound{⟦v⟧}\<%
\\
\>[4]\AgdaIndent{8}{}\<[8]%
\>[8]\AgdaBound{⇓λn} \<[22]%
\>[22]\AgdaSymbol{=} \AgdaInductiveConstructor{later⇓} \AgdaSymbol{(}\AgdaFunction{bind⇓} \<[39]%
\>[39]\AgdaSymbol{(λ} \AgdaBound{x} \AgdaSymbol{→} \AgdaInductiveConstructor{now} \AgdaSymbol{(}\AgdaInductiveConstructor{lam} \AgdaBound{x}\AgdaSymbol{))}\<%
\\
\>[8]\AgdaIndent{39}{}\<[39]%
\>[39]\AgdaSymbol{(}\AgdaFunction{bind⇓} \AgdaFunction{readback} \AgdaBound{v⇓} \AgdaBound{⇓n}\AgdaSymbol{)}\<%
\\
\>[8]\AgdaIndent{39}{}\<[39]%
\>[39]\AgdaInductiveConstructor{now⇓}\AgdaSymbol{)}\<%
\\
\>[0]\AgdaIndent{4}{}\<[4]%
\>[4]\AgdaKeyword{in} \<[8]%
\>[8]\AgdaInductiveConstructor{lam} \AgdaBound{n} \AgdaInductiveConstructor{,} \AgdaBound{⇓λn}\<%
\end{code}

\noindent
Reflecting a neutral value \AgdaBound{w} at function type $\AgdaBound{a}\;\AgdaInductiveConstructor{⇒}\;\AgdaBound{b}$ returns
a semantic function, which, if applied to a value \AgdaBound{u} of type
\AgdaBound{a} and its semantics \AgdaBound{⟦u⟧}, in essence reflects recursively
the application of \AgdaBound{w} to \AgdaBound{u}, which is again neutral, at
type \AgdaBound{b}.  A little more has to be done, though, e.g.,
we also show that this application can be read back.

\begin{code}%
\>[0]\AgdaIndent{2}{}\<[2]%
\>[2]\AgdaFunction{reflect} \AgdaSymbol{:} \AgdaSymbol{∀\{}\AgdaBound{Γ}\AgdaSymbol{\}} \AgdaBound{a} \AgdaSymbol{(}\AgdaBound{w} \AgdaSymbol{:} \AgdaDatatype{Ne} \AgdaDatatype{Val} \AgdaBound{Γ} \AgdaBound{a}\AgdaSymbol{)} \AgdaSymbol{→} \AgdaFunction{nereadback} \AgdaBound{w} \AgdaFunction{⇓} \AgdaSymbol{→} \AgdaFunction{V⟦} \AgdaBound{a} \AgdaFunction{⟧} \AgdaSymbol{(}\AgdaInductiveConstructor{ne} \AgdaBound{w}\AgdaSymbol{)}\<%
\\
\>[0]\AgdaIndent{2}{}\<[2]%
\>[2]\AgdaFunction{reflect} \AgdaInductiveConstructor{★} \<[19]%
\>[19]\AgdaBound{w} \<[22]%
\>[22]\AgdaBound{w⇓} \<[41]%
\>[41]\AgdaSymbol{=} \AgdaBound{w⇓}\<%
\\
\>[0]\AgdaIndent{2}{}\<[2]%
\>[2]\AgdaFunction{reflect} \AgdaSymbol{(}\AgdaBound{a} \AgdaInductiveConstructor{⇒} \AgdaBound{b}\AgdaSymbol{)} \<[19]%
\>[19]\AgdaBound{w} \<[22]%
\>[22]\AgdaSymbol{(}\AgdaBound{m} \AgdaInductiveConstructor{,} \AgdaBound{w⇓m}\AgdaSymbol{)} \AgdaBound{η} \AgdaBound{u} \AgdaBound{⟦u⟧} \<[41]%
\>[41]\AgdaSymbol{=}\<%
\\
\>[2]\AgdaIndent{4}{}\<[4]%
\>[4]\AgdaKeyword{let} \<[9]%
\>[9]\AgdaBound{n} \AgdaInductiveConstructor{,} \AgdaBound{⇓n} \<[17]%
\>[17]\AgdaSymbol{=} \AgdaFunction{reify} \AgdaBound{a} \AgdaBound{u} \AgdaBound{⟦u⟧}\<%
\\
\>[4]\AgdaIndent{9}{}\<[9]%
\>[9]\AgdaBound{m′} \<[17]%
\>[17]\AgdaSymbol{=} \AgdaFunction{nen≤} \AgdaBound{η} \AgdaBound{m}\<%
\\
\>[4]\AgdaIndent{9}{}\<[9]%
\>[9]\AgdaBound{⇓m} \<[17]%
\>[17]\AgdaSymbol{=} \AgdaFunction{nereadback≤⇓} \AgdaBound{η} \AgdaBound{w} \AgdaBound{w⇓m}\<%
\\
\>[4]\AgdaIndent{9}{}\<[9]%
\>[9]\AgdaBound{wu} \<[17]%
\>[17]\AgdaSymbol{=} \AgdaInductiveConstructor{app} \AgdaSymbol{(}\AgdaFunction{nev≤} \AgdaBound{η} \AgdaBound{w}\AgdaSymbol{)} \AgdaBound{u}\<%
\\
\>[4]\AgdaIndent{9}{}\<[9]%
\>[9]\AgdaBound{⟦wu⟧} \<[17]%
\>[17]\AgdaSymbol{=} \AgdaFunction{reflect} \AgdaBound{b} \AgdaBound{wu} \<[33]%
\>[33]\AgdaSymbol{(}\AgdaInductiveConstructor{app} \AgdaBound{m′} \AgdaBound{n} \AgdaInductiveConstructor{,}\<%
\\
\>[9]\AgdaIndent{33}{}\<[33]%
\>[33]\AgdaFunction{bind⇓} \<[40]%
\>[40]\AgdaSymbol{(λ} \AgdaBound{m} \AgdaSymbol{→} \AgdaInductiveConstructor{app} \AgdaBound{m} \AgdaFunction{\ensuremath{{<}\${>}}} \AgdaFunction{readback} \AgdaBound{u}\AgdaSymbol{)}\<%
\\
\>[33]\AgdaIndent{40}{}\<[40]%
\>[40]\AgdaBound{⇓m}\<%
\\
\>[33]\AgdaIndent{40}{}\<[40]%
\>[40]\AgdaSymbol{(}\AgdaFunction{bind⇓} \AgdaSymbol{(λ} \AgdaBound{n} \AgdaSymbol{→} \AgdaInductiveConstructor{now} \AgdaSymbol{(}\AgdaInductiveConstructor{app} \AgdaBound{m′} \AgdaBound{n}\AgdaSymbol{))} \AgdaBound{⇓n} \AgdaInductiveConstructor{now⇓}\AgdaSymbol{))}\<%
\\
\>[0]\AgdaIndent{4}{}\<[4]%
\>[4]\AgdaKeyword{in} \<[9]%
\>[9]\AgdaInductiveConstructor{ne} \AgdaBound{wu} \AgdaInductiveConstructor{,} \AgdaInductiveConstructor{now⇓} \AgdaInductiveConstructor{,} \AgdaBound{⟦wu⟧}\<%
\\
\>\<%
\end{code} 

\noindent
As immediate corollaries we get that all variables are strongly
computable and that the identity environment is strongly computable.

\begin{code}%
\>\AgdaFunction{var↑} \<[15]%
\>[15]\AgdaSymbol{:} \<[18]%
\>[18]\AgdaSymbol{∀\{}\AgdaBound{Γ} \AgdaBound{a}\AgdaSymbol{\}(}\AgdaBound{x} \AgdaSymbol{:} \AgdaDatatype{Var} \AgdaBound{Γ} \AgdaBound{a}\AgdaSymbol{)} \AgdaSymbol{→} \AgdaFunction{V⟦} \AgdaBound{a} \AgdaFunction{⟧} \AgdaInductiveConstructor{ne} \AgdaSymbol{(}\AgdaInductiveConstructor{var} \AgdaBound{x}\AgdaSymbol{)}\<%
\\
\>\AgdaFunction{var↑} \AgdaBound{x} \<[15]%
\>[15]\AgdaSymbol{=} \<[18]%
\>[18]\AgdaFunction{reflect} \AgdaSymbol{\_} \AgdaSymbol{(}\AgdaInductiveConstructor{var} \AgdaBound{x}\AgdaSymbol{)} \AgdaSymbol{(}\AgdaInductiveConstructor{var} \AgdaBound{x} \AgdaInductiveConstructor{,} \AgdaInductiveConstructor{now⇓}\AgdaSymbol{)}\<%
\\
%
\\
\>\AgdaFunction{⟦ide⟧} \<[15]%
\>[15]\AgdaSymbol{:} \<[18]%
\>[18]\AgdaSymbol{∀} \AgdaBound{Γ} \AgdaSymbol{→} \AgdaFunction{E⟦} \AgdaBound{Γ} \AgdaFunction{⟧} \AgdaSymbol{(}\AgdaFunction{ide} \AgdaBound{Γ}\AgdaSymbol{)}\<%
\\
\>\AgdaFunction{⟦ide⟧} \AgdaInductiveConstructor{ε} \<[15]%
\>[15]\AgdaSymbol{=} \<[18]%
\>[18]\AgdaSymbol{\_}\<%
\\
\>\AgdaFunction{⟦ide⟧} \AgdaSymbol{(}\AgdaBound{Γ} \AgdaInductiveConstructor{,} \AgdaBound{a}\AgdaSymbol{)} \<[15]%
\>[15]\AgdaSymbol{=} \<[18]%
\>[18]\AgdaFunction{E⟦⟧≤} \AgdaFunction{wk} \AgdaSymbol{(}\AgdaFunction{ide} \AgdaBound{Γ}\AgdaSymbol{)} \AgdaSymbol{(}\AgdaFunction{⟦ide⟧} \AgdaBound{Γ}\AgdaSymbol{)} \AgdaInductiveConstructor{,} \AgdaFunction{var↑} \AgdaInductiveConstructor{zero}\<%
\end{code}

\noindent
Finally we can plug the termination of \AgdaFunction{eval} in the identity
environment to yield a strongly computable value and the termination
of \AgdaFunction{readback} give a strongly computable value to yield the
termination of \AgdaFunction{nf}.

\begin{code}%
\>\AgdaFunction{normalize} \<[17]%
\>[17]\AgdaSymbol{:} \<[20]%
\>[20]\AgdaSymbol{∀} \AgdaBound{Γ} \AgdaBound{a} \AgdaSymbol{(}\AgdaBound{t} \AgdaSymbol{:} \AgdaDatatype{Tm} \AgdaBound{Γ} \AgdaBound{a}\AgdaSymbol{)} \AgdaSymbol{→} \AgdaFunction{∃} \AgdaSymbol{λ} \AgdaBound{n} \AgdaSymbol{→} \AgdaFunction{nf} \AgdaBound{t} \AgdaDatatype{⇓} \AgdaBound{n}\<%
\\
\>\AgdaFunction{normalize} \AgdaBound{Γ} \AgdaBound{a} \AgdaBound{t} \<[17]%
\>[17]\AgdaSymbol{=} \<[20]%
\>[20]\AgdaKeyword{let} \<[25]%
\>[25]\AgdaBound{v} \AgdaInductiveConstructor{,} \AgdaBound{v⇓} \AgdaInductiveConstructor{,} \AgdaBound{⟦v⟧} \<[39]%
\>[39]\AgdaSymbol{=} \AgdaFunction{term} \AgdaBound{t} \AgdaSymbol{(}\AgdaFunction{ide} \AgdaBound{Γ}\AgdaSymbol{)} \AgdaSymbol{(}\AgdaFunction{⟦ide⟧} \AgdaBound{Γ}\AgdaSymbol{)}\<%
\\
\>[4]\AgdaIndent{25}{}\<[25]%
\>[25]\AgdaBound{n} \AgdaInductiveConstructor{,} \AgdaBound{⇓n} \<[39]%
\>[39]\AgdaSymbol{=} \AgdaFunction{reify} \AgdaBound{a} \AgdaBound{v} \AgdaBound{⟦v⟧}\<%
\\
\>[0]\AgdaIndent{20}{}\<[20]%
\>[20]\AgdaKeyword{in} \<[25]%
\>[25]\AgdaBound{n} \AgdaInductiveConstructor{,} \AgdaFunction{bind⇓} \AgdaFunction{readback} \AgdaBound{v⇓} \AgdaBound{⇓n}\<%
\end{code}

\section{Conclusions}

We have presented a coinductive normalizer for
simply typed lambda calculus and proved that it terminates. The
combination of the coinductive normalizer and termination proof yield
a terminating normalizer function in type theory.

The successful formalization serves as a proof-of-concept for
coinductive programming and proving using sized types and copatterns,
a new and presently experimental feature of Agda. The approach we have
taken lifts easily to extensions such as G\"odel's System T.

\paragraph*{Acknowledgments}

The authors are grateful to Nils Anders Danielsson for discussions and
his talk at \emph{Shonan Meeting 026: Coinduction for computation
structures and programming} in October 2013 which inspired this
work. We also thank the anonymous referees for their helpful comments.

Andreas Abel has been supported by framework grant 254820104
of Vetenskapsr\aa{}det
to the Chalmers ProgLog group, held by Thierry Coquand.
James Chapman has been supported by the ERDF funded Estonian CoE project EXCS
and ICT National Programme project “Coinduction”, the Estonian
Ministry of Research and Education target-financed research theme
no.~0140007s12 and the Estonian Science Foundation grant no.~9219.

This article has been type set with the Stevan Andjelkovic's LaTeX backend for Agda.

\bibliographystyle{eptcs}
\bibliography{auto-eptcs14,byhand}

\end{document}